\documentclass[12pt,fleqn]{report}

\usepackage{etoolbox}
\newbool{ACM}
\makeatletter
\@ifclassloaded{acmsmallmod}
{\booltrue{ACM}}
{\boolfalse{ACM}}
\makeatother

\ifbool{ACM}
{
\usepackage{fullpage}
\usepackage{setspace}
\setstretch{1.5}
\newcommand{\tgnstretch}{}
\renewenvironment{example}{
\begin{exam}}{\end{exam}}
}
{

\usepackage{fullpage}
\usepackage{mathptmx}
\usepackage[12pt]{moresize}
\usepackage{setspace}

\newbool{tgndoublespace}
\boolfalse{tgndoublespace}
\ifbool{tgndoublespace}{\setstretch{2}}{}
\newcommand{\tgnstretch}{
\ifbool{tgndoublespace}{}
{\renewcommand{\arraystretch}{1.5}}
}

\pagestyle{headings}
\usepackage[headsep=5ex,top=1.25in, bottom=1.25in, left=1in, right=1in]{geometry}

\usepackage[sort,nocompress]{cite}

\newtheorem{theorem}{Theorem}[chapter]

\newtheorem{lemma}[theorem]{Lemma}
\newtheorem{corollary}[theorem]{Corollary}
\newtheorem{rem}[theorem]{Remark}
\newtheorem{exam}[theorem]{Example}

\newtheorem{defi}[theorem]{Definition}
\newenvironment{definition}[1]{\begin{defi}
#1}
{\end{defi}}

\newenvironment{example}[1]{\begin{exam}\rm #1}
{\end{exam}}

\newenvironment{remark}[1]{\begin{rem}\rm #1}
{\end{rem}}

\newcommand{\qed}{\hfill\kern 10pt{\unitlength1pt\linethickness{.4pt}\framebox(5,5){}}}

\newenvironment{proof}[1][Proof]{
\begin{trivlist}
\item[\hskip \labelsep {\itshape #1.}]
}
{\qed\end{trivlist}}

}

\usepackage{amsmath,amssymb,blkarray,color,enumerate,listings,mathtools,tabularx,url,xspace,array,amsfonts,xcolor}

\usepackage
[justification=justified,singlelinecheck=true]{caption}

\usepackage{mathptmx}       
\usepackage{helvet}         
\usepackage{courier}        
\usepackage{type1cm}        

\usepackage{makeidx}         
\makeindex
\usepackage{graphicx}        
\usepackage{multicol,multirow}        
\usepackage[bottom]{footmisc}

\usepackage{latexsym}

\usepackage
  {hyperref}
  \hypersetup{
  colorlinks=true,
  linkcolor=red,
  citecolor=blue,
  filecolor=magenta,
  urlcolor=cyan}

\usepackage{bookmark}
\usepackage[titletoc]{appendix}
\usepackage[normalem]{ulem}

\makeatletter
\def\rf#1{(\@rf#1,.)}
\def\@rf#1,{\ref{eq:#1}\@ifnextchar . {\@endrf}{, \@rf}}
\def\@endrf.{}
\makeatother

\usepackage{fancybox}
\usepackage[noprefix]{nomencl}
\DeclareMathAlphabet{\mathcal}{OMS}{cmsy}{b}{n}
\DeclareMathAlphabet{\mathcal}{OMS}{cmsy}{m}{n}

\renewcommand{\~}[1]{\mathbf{#1}}

\newcommand{\setbg}[1]{\bigl\{\,#1\,\bigr\}}

\newcommand{\casemod}[2]{\left\{ \begin{array}{#1} #2 \end{array}\right. }

{\sf
\begin{tabbing}{\bf Algorithm (#1).}\\%
\hspace*{2em}\=\hspace*{2em}\=\hspace*{2em}\=\hspace*{2em}\=\hspace*{2em}\=\hspace*{2em}\=\hspace*{2em}\=\hspace*{2em}\=\kill\!\!}
{\end{tabbing}}
\newcommand{\li}[1]{\lstinline[basicstyle=\tt]{#1}}

\newcommand{\pendL}{L}
\newcommand{\pendG}{g}

\newcommand{\Sigmeth}{$\Sigma$-method\xspace}
\newcommand{\mc}[3]{\multicolumn{#1}{#2}{#3}}
\newcommand{\s}[1]{\text{\scriptsize$#1$}}
\newcommand{\valSig}[2]{\val{#1}=#2}
\newcommand{\detJac}[2]{\det (#1)=#2}
\newcommand{\mcdetJac}[2]{\mc{#1}{c}{$#2$}}
\newcommand{\nud}{\nu_d} 
\newcommand{\nuS}{\nu_S} 

\newcommand{\sij}[2]{\sigma_{#1#2}}

\newcommand{\Sig}{\Sigma}
\newcommand{\Jac}{\~J}
\newcommand{\Jij}[2]{\~J_{#1#2}}

\newcommand{\compSig}{\widetilde\Sigma}
\newcommand{\compJ}{\~{\widetilde J}}
\newcommand{\compJij}[2]{\~{\widetilde J}_{#1#2}}
\newcommand{\compsij}[2]{\widetilde\sigma_{#1#2}}

\newcommand{\compc}{\~{\widetilde c}}
\newcommand{\compd}{\~{\widetilde d}}

\newcommand{\newsig}{\overline\sigma}
\newcommand{\newsij}[2]{\newsig_{#1#2}}
\newcommand{\newsicj}[2]{\newsig_{#1,#2}}
\newcommand{\newSig}{\overline\Sigma}

\newcommand{\newJ}{\~{\overline J}}
\newcommand{\newJac}{\~{\overline J}}

\newcommand{\newT}{\overline T}
\newcommand{\newf}{\overline f}

\newcommand{\newc}{\overline c}
\newcommand{\newd}{\overline d}
\newcommand{\daeF}{\mathcal{F}}
\newcommand{\newdaeF}{\overline\daeF}
\newcommand{\newx}{\overline{\~x}}
\newcommand{\newy}{\overline{\~y}}

\newcommand{\Aij}[2]{A_{#1#2}}
\newcommand{\Sess}{S_{\text{ess}}}
\newcommand{\rnge}[2]{#1:#2}
\newcommand{\setrnge}[2]{\{#1,\ldots,#2\}}
\newcommand{\rngeset}[2]{\{#1:#2\}}
\newcommand{\oneton}{\rnge{1}{n}}
\newcommand{\onetos}{\rnge{1}{s}}
\newcommand{\inlinefrac}[2]{(#1/#2)}

\newcommand{\clsp}{\phantom{\ }}
\def\mod2pend{{\sc Mod2Pend}}

\newcommand{\neginf}{-\infty}

\newcommand{\calX}{\xi}

\newcommand{\derset}{\text{derset}}
\newcommand{\xderset}[1]{\~x_{\derset(#1)}}

\newcommand{\daesa}{{\sc daesa}\xspace}
\newcommand{\matlab}{{\sc matlab}\xspace}
\newcommand{\daets}{{\sc daets}\xspace}

\newcommand{\lam}{\lambda}

\newcommand{\indxk}{l}

\newcommand{\nzset}{L}
\newcommand{\zset}{\overline \nzset}
\newcommand{\eqsetI}{I}
\newcommand{\neqsetI}{\overline \eqsetI}

\newcommand{\cm}[1]{C(#1)}

\newcommand{\hoder}[2]{\sigma\left(#1,#2\right)}
\newcommand{\comphoder}[2]{\widetilde\sigma\left(#1,#2\right)}

\newcommand{\jone}{j}
\newcommand{\jtwo}{r}

\newcommand{\LargeLess}{\parbox{12pt}{\Large $<$}}
\newcommand{\HugeLess}{\parbox{12pt}{\Large $<$}}
\newcommand{\HugeLe}{\parbox{12pt}{\Large $\le$}}
\newcommand{\HugeInf}{\parbox{12pt}{\Large $-\infty$}}

\newcommand{\val}[1]{\text{Val}(#1)}

\newcommand{\pp}[2]{\frac{\partial #1}{\partial #2}}
\newcommand{\ppin}[2]{\partial #1/\partial #2}
\newcommand{\dd}[2]{\frac{\text{d} #1}{\text{d} #2}}

\definecolor{lgray}{rgb}{0.6,0.6,0.6}

\newcommand{\lgo}{\OK{0}}
\definecolor{dgreen}{RGB}{1,120,1}
\definecolor{purple}{RGB}{191,0,95}

\definecolor{palegray}{rgb}{0.7,0.7,0.7}
\definecolor{palepink}{rgb}{1,0.8,0.8}
\definecolor{paleblue}{rgb}{0.8,0.8,1}
\definecolor{palegreen}{rgb}{0.8,1,0.8}

\newcommand{\OK}[1]{\colorbox{palegray}{\ensuremath{#1}}}

\newcommand{\pend}{{\sc Pend}\xspace}
\newcommand{\modpenda}{{\sc ModPendA}\xspace}
\newcommand{\modpendb}{{\sc ModPendB}\xspace}

\newcommand{\rrf}[2]{(\ref{eq:#1}--\ref{eq:#2})}
\newcommand{\chref}[1]{\ref{ch:#1}}

\newcommand{\exref}[1]{\ref{ex:#1}}

\newcommand{\CHref}[1]{\S\ref{ch:#1}}
\newcommand{\SSCref}[1]{\S\ref{ssc:#1}}
\newcommand{\SCref}[1]{\S\ref{sc:#1}}
\newcommand{\EXref}[1]{Example~\ref{ex:#1}}
\newcommand{\RMref}[1]{Remark~\ref{rm:#1}}

\newcommand{\TBref}[1]{Table~\ref{tb:#1}}
\newcommand{\DFref}[1]{Definition~\ref{df:#1}}
\newcommand{\LEref}[1]{Lemma~\ref{le:#1}}
\newcommand{\FGref}[1]{Figure~\ref{fg:#1}}
\newcommand{\COref}[1]{Corollary~\ref{co:#1}}
\newcommand{\THref}[1]{Theorem~\ref{th:#1}}

\newcommand{\uiter}[1]{u^#1}

\newcommand{\Iiter}[1]{\eqsetI^#1}
\newcommand{\thiter}[1]{\theta^#1}
\newcommand{\Siter}[1]{\Sigma^#1}
\newcommand{\Jiter}[1]{\Jac^#1}
\newcommand{\JiterT}[1]{(\Jiter{#1})^T}
\newcommand{\Kiter}[1]{\nzset^#1}
\newcommand{\daeiter}[1]{\daeF^#1}
\newcommand{\kiter}[1]{\indxk^#1}

\newcommand{\indJ}{\mathcal{J}}

\newcommand{\intvI}{\mathbb{I}}
\newcommand{\bbR}{\mathbb{R}}
\newcommand{\bbN}{\mathbb{N}}

\newcommand{\xjl}[2]{x_{#1}^{(#2)}}

\newcommand\bbmx{\begin{bmatrix}}
\newcommand\ebmx{\end{bmatrix}}

\begin{document}
\title{Symbolic-Numeric Methods for 
Improving Structural Analysis of 
Differential-Algebraic Equation Systems}

\markboth{Tan, Nedialkov, Pryce}{Symbolic-Numeric Methods for Improving Structural Analysis of Differential-Algebraic Equation Systems}

\ifbool{ACM}
{\author{
GUANGNING TAN
\affil{McMaster University}
NEDIALKO S. NEDIALKOV
\affil{McMaster University} \vspace{3pt}
JOHN D. PRYCE
\affil{Cardiff University} \vspace{3pt}
}

\begin{abstract}
Systems of differential-algebraic equations (DAEs) are generated routinely by simulation and modeling environments such as {\sc Modelica} and {\sc MapleSim}. Before a simulation starts and a numerical solution method is applied, some kind of structural analysis is performed to determine the structure and the index of a DAE. Structural analysis methods serve as a necessary preprocessing stage, and among them, Pantelides's algorithm is widely used. 
Recently Pryce's $\Sigma$-method is becoming increasingly popular, owing to its straightforward approach and capability of analyzing high-order systems. Both methods are equivalent in the sense that when one succeeds, producing a nonsingular system Jacobian, the other also succeeds, and the two give the same structural index.

Although provably successful on fairly many problems of interest, the structural analysis methods can fail on some simple, solvable DAEs and give incorrect structural information including the index.
In this report, we focus on the $\Sigma$-method. We investigate its failures, and develop two symbolic-numeric conversion methods for converting a DAE, on which the $\Sigma$-method fails, to an equivalent problem on which this method succeeds. Aimed at making structural analysis methods more reliable, our conversion methods exploit structural information of a DAE, and require a symbolic tool for their implementation.
\end{abstract}

\maketitle
}
{
\author{
Guangning Tan, Nedialko S. Nedialkov\\
McMaster University\\[1ex]
John D. Pryce\\
Cardiff University\\[1ex]
}
\date{\today}
\maketitle

}

\tableofcontents
\numberwithin{equation}{chapter}

\chapter{Introduction}\label{ch:intro}
We are interested in solving initial value problems in DAEs of the general form
\begin{align}\label{eq:maineq}
  f_i(\, t,\, \text{the $x_j$ and derivatives of them}\,) = 0, \quad i=\oneton,
\end{align}
where the $x_j(t)$ are $n$ state variables, and $t$ is the time variable.  
The formulation \rf{maineq} includes high-order systems and systems that are jointly nonlinear in leading derivatives. 
Moreover, \rf{maineq} includes ordinary differential equations (ODEs) and purely algebraic systems.

An important characteristic of a DAE is its {\em index}. Generally, the index measures the difficulty of solving a DAE numerically. If a DAE is of index-1, then a general index-1 solver can be used, e.g., {\sc dassl}\cite{BrenanCampbelPetzold}, {\sc ida} of {\sc sundials}\cite{Hindmarsh}, and \matlab's \li{ode15s} and \li{ode23t}. If a DAE is of high index, that is, index $\ge 2$, then we need a high-index DAE solver, e.g., {\sc radau5} for DAEs of index $\le 3$\cite{HairerII} or \daets for DAEs of any index\cite{nedialkov2008solving}. We can also use index reduction techniques to convert the original DAE to an index-1 problem\cite{Matt93a, scholz2013combined, DDandSA2015}, and then apply an index-1 solver.

{\em Structural analysis} (SA) methods serve as a preprocessing stage to help determine the index. Among them is the Pantelides's method\cite{Pant88b}, which is a graph-based algorithm that finds how many times each equation needs to be differentiated. Pryce's structural analysis---the {\em Signature method} or {\em $\Sigma$-method}---is essentially equivalent to that of Pantelides \cite{Pryce2001a}, and in particular computes the same {\em structural index} when both methods succeed. However, Pantelides's algorithm can only handle first-order systems, while Pryce's can be applied to \rf{maineq} of any order and is generally easier to apply.

This SA determines the structural index,  which is often the same as the {\em differentiation index}, the number of degrees of freedom, the variables and derivatives that need to be initialized, and the constraints of the DAE. We give the definition of 
the differentiation index in \CHref{background}
and that of 
the structural index in \CHref{SA}.

Nedialkov and Pryce \cite{NedialkovPryce05a,nedialkov2007solving,nedialkov2008solving} use the $\Sigma$-method to analyze a DAE of the form \rf{maineq}, and solve it numerically using Taylor series.
On each integration step, Taylor coefficients (TCs) for the solution are computed up to some order. These coefficients are computed in a stage-wise manner. This stage by stage solution scheme, also derived from the SA, indicates at each stage which equations need to be solved and for which variables\cite{nedialkov2014a}. In 
 \cite{Barrio05b,griewank2006efficient,HoefkensPhD}, 
 the \Sigmeth is also applied to perform structural analysis, and the resulting offset vectors are used to prescribe the computation of TCs.





Although the $\Sigma$-method provably gives correct structural information (including index) on many  DAEs of practical interest \cite{Pryce2001a}, it can fail---whence also Pantelides's algorithm and other SA methods \cite{LPT-1995-05,soares2012structural} can fail---to find a DAE's true structure, producing an identically singular system Jacobian. (See \CHref{SA} for the definition of system Jacobian.)

Scholz et al. \cite{scholz2013combined} show that several simulation environments such as {\sc Dymola, OpenModelica} and {\sc SimulationX} all fail on a simple, solvable $4\times 4$ linear constant coefficient DAE; we discuss this DAE in \EXref{lenaintro}.
Other examples where SA fails are the Campbell-Griepentrog Robot Arm \cite{Campbell:1995:SGD:203046.203047} and the Ring Modulator \cite{TestSetIVP}. When SA fails, the structural index usually underestimates the differentiation index.
In other cases, when SA produces a nonsingular system Jacobian, the structural index may overestimate the differentiation index \cite{Reissig1999a}. 
We review in Appendix \ref{ch:Fixes} how these DAEs in the early literature are handled so that SA reports the correct index.

SA can fail if there are hidden symbolic cancellations in a DAE; this is the simplest case among SA's failures. However, SA can fail in a more obscure way. In this case, it is difficult to understand the causes of such failures and to provide fixes to the formulation of the problem.
%
Such deficiencies can pose limitations to the application of SA, as it becomes unreliable. Our goal is to construct methods that convert automatically a system on which SA fails into an equivalent form on which it succeeds. This report is devoted to developing such methods.

\medskip

It is organized as follows. Chapter~\chref{background} overviews work that has been done to date. Chapter~\chref{SA} summarizes the $\Sigma$-method and gives definitions and tools that are needed for our theoretical development. The problem of SA's failures on some DAEs is described in Chapter~\chref{fail}. In Chapters~\chref{LCmethod} and \chref{ESmethod}, we develop two methods, the {\em linear combination method} and the {\em expression substitution method}, respectively.
We show in Chapter~\chref{example} how to apply our methods on several examples. Chapter~\chref{conclu} gives conclusions and indicates several research directions.

\chapter{Background}\label{ch:background}
The index of a DAE is an important concept in DAE theory. There are various definitions of an index: differentiation index\cite{Camp95a, gear1988differential, gear1990differential}, geometric index\cite{rabier1991, rheinboldt1984differential}, structural index\cite{Pryce2001a, Duff86b, Pant88b}, perturbation index\cite{HairerII}, tractability index\cite{griepentrogmarz1986}, and strangeness index\cite{KunM06}.

%
%

The most commonly used index is the {\em differentiation index}; we refer to it as d-index or $\nud$. The following definition is from \cite[p. 236]{AscherPetzold}. 

\begin{definition}\label{df:dindex}
Consider a general form of a first-order DAE
\begin{align}\label{eq:orderoneform}
\~F(t,\~x,\~x') = \~0,
\end{align}
where $\partial \~F/\partial \~x'$ may be singular. The {\em differentiation index}  along a solution $\~x(t)$ is the minimum number of differentiations of the system that would be required to solve $\~x'$ uniquely in terms of $\~x$ and $t$, that is, to define an ODE for $\~x$. Thus this index is defined in terms of the overdetermined system
\begin{equation*}
\begin{aligned}
\~F\left(t,\~x,\~x'\right) &= \~0, \\
\dd{\~F}{t}\left(t,\~x,\~x',\~x''\right) &= \~0,\\
\vdots \\
\frac{\text{d}^p\~F}{\text{d}t^p}\left(t,\~x,\~x',\cdots,\~x^{(p+1)}\right) &= \~0
\end{aligned}
\end{equation*}
to be the smallest integer $p$ so that $\~x'$ in \rf{orderoneform} can be solved for in terms of $\~x$ and $t$.
\end{definition}

If a DAE \rf{maineq} is of high-order, then one can introduce additional variables to reduce the order of the system so that it is still in the general form \rf{orderoneform}. 

We give a definition for solution of a DAE.
\begin{definition}\label{df:solution}
An $n$-vector valued function $\~x(t)$, defined on a time interval $\intvI\subset\bbR$, is a {\em solution} of \rf{maineq}, if $(t,\~x(t))$ satisfies $f_i=0$, $i=\rnge{1}{n}$, pointwise for all $t\in\intvI$: that is, every $f_i$ {\em vanishes} on $\intvI$.
\end{definition}


Rei\ss ig et al. \cite{Reissig1999a} claim that a DAE of d-index 1 may have arbitrarily high structural index. They construct a class of linear constant coefficient DAEs in some specific form. On these DAEs of d-index 1, Pantelides's algorithm performs a high number of iterations and differentiations, and obtains a high structural index that far exceeds the d-index 1.
 A simple $3\times 3$ linear electrical circuit example is also presented: choosing a specific node as the ground node results in a DAE of d-index 1, but of structural index 2. 

Pryce \cite{Pryce2001a} shows that, if the \Sigmeth succeeds, then the structural index $\nu_S$
 is always an upper bound on the d-index. This implies that, if the structural index computed by the \Sigmeth is smaller than the d-index,  then the method {\em must} fail; otherwise we would have a statement that contradicts to the above \DFref{dindex}. Pryce also shows that the \Sigmeth succeeds on one of Rei\ss ig's DAEs and produces a nonsingular system Jacobian\cite{Pryce2001a}. His method also produces the same high structural index as does Pantelides's.

In \cite{Pryce98}, Pryce shows that the \Sigmeth fails on the index-5 Campbell-Griepentrog Robot Arm DAE---the SA produces an identically singular Jacobian.
He then provides a remedy: identify the common subexpressions in the problem, introduce extra variables, and substitute them for those subexpressions. The resulting equivalent problem is an enlarged one, where the $\Sigma$-method succeeds and reports the correct structural index 5. Pryce introduces the term {\em structure-revealing} to conjecture that a nonsingular system Jacobian might be an effect of DAE formulation, but not of DAE's inherent nature. 

Choudhry et al. \cite{chowdhry2004symbolic} propose a method called symbolic numeric index analysis (SNIA). Their method can accurately detect symbolic cancellation of variables that appear linearly in equations, and therefore can deal with linear constant coefficient systems. For general nonlinear DAEs, SNIA provides a correct result in some cases, but not all. Furthermore, it is limited to order-1 systems, and it cannot handle complex expression substitution and symbolic cancellations, such as $(x\cos y)'-x'\cos y$. For the general case, their method does not derive from the original problem an equivalent one that has the correct index.

Scholz et al. \cite{scholz2013combined} are interested in a class of DAEs called coupled systems. In their case, a coupled system is composed by coupling two semi-explicit d-index 1 systems. They show that the \Sigmeth succeeds if and only if the coupled system is again of d-index 1. As a consequence, if the coupled system is of high index, SA methods {\em must} fail.
%
They develop a structural-algebraic approach to deal with such coupled systems. They differentiate a linear combination of certain algebraic equations that contribute to singularity, append the resulting equations, and replace certain derivatives with newly introduced variables.  
They use this {\em regularization} process to convert the regular coupled system to a d-index 1 problem, on which SA succeeds with nonsingular Jacobian.
\chapter{Summary of Pryce's structural analysis}\label{ch:SA}
We call this SA \cite{Pryce2001a} the \Sigmeth, because it constructs for \rf{maineq} an $n\times n$ {\em signature matrix} $\Sigma=(\sij{i}{j})$ such that
\begin{equation}\label{eq:sigmx}
\sij{i}{j} =
\left\{
\begin{array}{ll}  
\text{
the order of the highest order derivative to which $x_j$ occurs in $f_i$; or}\\
-\infty\  \text{ if $x_j$ does not occur in $f_i$.}
\end{array}
\right.
\end{equation}

A {\em transversal} $T$ is a set of $n$ positions $(i,j)$ with one entry in each row and each column. The sum of entries $\sij{i}{j}$ over $T$, or $\sum_{(i,j)\in T} \sij{i}{j}$, is called the {\em value $T$}, written $\val{T}$. We seek a {\em highest-value transversal} (HVT) that gives this sum the largest value. 
%
We call this number the {\em value of the signature matrix}, written Val$(\Sigma)$. 

We give a definition for a DAE's structural posed-ness.
\begin{definition}\label{df:SWPSIP}
We say that a DAE is {\em structurally well-posed} (SWP) if its $\val{\Sigma}$ is finite. That is, all entries in a HVT are finite, or equivalently, there exists some finite transversal. Otherwise, if $\val{\Sigma}=\neginf$, then we say a DAE is {\em structurally ill-posed} (SIP).
\end{definition}

For a SWP DAE, we find {\em equation and variable offsets} $\~c$ and $\~d$, respectively, which are non-negative integer $n$-vectors satisfying 
\begin{equation}\label{eq:cidj}
c_i\ge 0;\quad d_j-c_i\geq \sij{i}{j} \quad \text{for all $i,j$ with equality on a HVT.}
\end{equation}
An equality $d_j-c_i=\sij{i}{j}$ on some HVT also holds on all HVTs\cite{NedialkovPryce2012a}.
We refer to $\~c$ and $\~d$ satisfying \rf{cidj} as {\em valid offsets}. They are not unique, but there exists unique $\~c$ and $\~d$ that are the smallest component-wise valid offsets. We refer to them as {\em canonical offsets}.

The {\em structural index} is defined by
\begin{equation*}\label{eq:saindex}
\nu_S =
\Biggl\{ 
\begin{array}{ll}
\max_i c_i+1 \qquad &\text{if $d_j=0$ for some $j$, or}\\[1ex]
\max_i c_i   \qquad &\text{otherwise.}
\end{array}
\end{equation*}


Critical to the success of this method is the nonsingularity of
 the DAE's $n\times n$ {\em system Jacobian} matrix $\Jac=(\Jij{i}{j})$, where
\begin{equation}\label{eq:sysjac}
\Jij{i}{j} =\frac{\partial f_i}{\partial x_j^{(d_j-c_i)}} =
\left\{ 
\begin{array}{ll}
\partial f_i/ \partial x_j^{(\sij{i}{j})} &\text{if $d_j-c_i=\sij{i}{j}$, and}\\[1ex]
 0  &\text{otherwise.}
\end{array}
\right.
\end{equation}

Note that $\Jac=\Jac(\~c,\~d)$ depends on the choice of valid offsets $\~c,\~d$, which satisfy \rf{cidj}. That is, using different valid offsets, one may obtain different system Jacobians. However, they all have the same determinant; see \THref{SigJac}. For all the examples in this report, we shall use canonical offsets and the system Jacobian derived from them.

We can use $\Sigma$ and $\~c,\~d$ to determine a {\em solution scheme} for computing derivatives of the solution to \rf{maineq}. They are computed in stages 
\[
k=k_d,k_d+1,\ldots,0,1,\ldots \quad\text{where $k_d=-\max_j{d_j}$}. 
\]
At each stage we solve equations
\begin{equation}\label{eq:solvefi}
0 = f_i^{(c_i+k)} \quad \text{for all $i$ such that $c_i+k\geq 0$}
\end{equation}
for derivatives
\begin{equation}\label{eq:forxj}
x_j^{(d_j+k)} \qquad\,\,\,\,\,\, \text{for all $j$ such that $d_j+k\geq 0$}
\end{equation}
using the previously found
\begin{equation*}
x_j^{(r)} \qquad\qquad \text{for all $j$ such that $0\le r< d_j+k$}.
\end{equation*}
We refer to \cite{nedialkov2014a} for more details on this solution scheme; see also \EXref{simplepend}.

Throughout this report, for brevity, we write ``derivatives of $x_j$'' instead of ``$x_j$ and derivatives of it''---derivatives $v^{(l)}$ of a variable $v$ include $v$ itself as the case $l=0$.

If the solution scheme \rrf{solvefi}{forxj} can be carried out up to stage $k=0$, and the derivatives of each variable $x_j$ can be uniquely determined up to order $d_j$, then we say the solution scheme and the SA {\em succeed}. The system Jacobian is nonsingular at a point
\begin{equation}\label{eq:consPointNL}
\left( t;\ x_1,\ldots,x_1^{(d_1)};\ x_2,\ldots,x_2^{(d_2)}; \ \ldots ; \ x_n,\ldots,x_n^{(d_n)} \right),
\end{equation}
and there exists a unique solution through this point \cite{Pryce2001a,NedialkovPryce05a,NedialkovPryce2012a}. We say the DAE is {\em locally solvable}, and call \rf{consPointNL} a {\em consistent point}, if derivatives $x_j^{(d_j)}$ do not occur jointly linearly in $f_i^{(c_i)}$. 
In the linear case, a consistent point is
\begin{equation}\label{eq:consPointL}
\left( t;\ x_1,\ldots,x_1^{(d_1-1)};\ x_2,\ldots,x_2^{(d_2-1)}; \ \ldots ; \ x_n,\ldots,x_n^{(d_n-1)} \right).
\end{equation} 
For a more rigorous discussion of a consistent point, we refer the readers to \cite{NedialkovPryce05a,NedialkovPryce2012a,nedialkov2014a}.
 
To perform a numerical check for SA's success, or a {\em success check} for short, we attempt to compute numerically a consistent point at which $\Jac$ is nonsingular up to roundoff: we provide an appropriate set of derivatives of $x_j$'s and follow the solution scheme \rrf{solvefi}{forxj} for stages $k=\rnge{k_d}{0}$. This set of derivatives is the set of {\em initial values} for a DAE initial value problem, and a minimal set of derivatives required for initial values is discussed in \cite{NedialkovPryce2012a}.
 

When SA succeeds, the structural index is an upper bound for the differentiation index, and often they are the same: $\nud\le \nuS$\cite{Pryce2001a}. Also, the {\em number of degrees of freedom} (DOF) is
\begin{equation*}\label{eq:dof}
\text{DOF} = \val{\Sigma} = \sum_j d_j - \sum_i c_i = \sum_{(i,j)\in T} \sij{i}{j}.
\end{equation*}

We say the solution scheme and SA {\em fails}, if we cannot determine uniquely a consistent point using the solution scheme defined by \rrf{solvefi}{forxj}---otherwise said, we cannot follow the solution scheme up to stage $k=0$ and find a consistent point at which $\Jac$ is nonsingular.
In our experience, in the failure case usually $\nud>\nuS$, but not always, and the true number of DOF is overestimated by $\val{\Sigma}$. This is discussed in Examples~\exref{pendsing1},~\exref{pendsing2},~\exref{lenaintro},~\exref{correctindexbutfails}.

\medskip

We illustrate the above concepts using the following example.
\begin{example}\label{ex:simplepend}
The simple pendulum DAE (\pend) in Cartesian coordinates is
\begin{equation}\label{eq:pend}
\begin{aligned}
0 = f_1 &= x''+x\lam\\
0 = f_2 &= y''+y\lam -\pendG\\
0 = f_3 &= x^2+y^2-\pendL^2.
\end{aligned}
\end{equation}
Here the state variables are $x,y,\lam$; $\pendG$ is gravity, and $\pendL>0$ is the length of the pendulum.

The signature matrix and system Jacobian of this DAE are
\begin{align*}\tgnstretch
\Sigma = 
\begin{blockarray}{rccc cc}
& \clsp x\clsp &  \clsp y \clsp&\clsp \lam \clsp & \s{c_i} & \\
\begin{block}{r @{\hspace{10pt}}[ccc]cc}
f_1 & 2^\bullet  & -  & 0 & \s0 \\
f_2 & -  & 2  & 0^\bullet & \s0 \\
f_3 & 0  & 0^\bullet  & - & \s2  \\
\end{block}
 \s{d_j}& \s2 &\s2 &\s0  \\
 \end{blockarray}\quad\text{and}\quad
\Jac = \begin{blockarray}{rccc cc}
&\clsp x\clsp & \clsp y\clsp &\clsp \lam\clsp \\
\begin{block}{r @{\hspace{10pt}}[ccc]cc}
f_1 & 1  &0    &x  \\
f_2 & 0  &1    &y  \\
f_3 & 2x &2y   &0  \\
\end{block}
\phantom{\s{d_j}} \\
\end{blockarray}.
\end{align*}
We write $\Sigma$ in a {\em signature tableau}: a HVT is marked by $\bullet$; $-$ denotes $-\infty$; the canonical offsets $\~c,\;\~d$ are annotated on the right of $\Sigma$ and at the bottom of it, respectively. 

The structural index is 
\begin{align*}
\nuS &=\max_i{c_i}+1=c_3+1=3,
\end{align*}
which is the same as the d-index.
The number of degrees of freedom is
\begin{align*}
\text{DOF} &= \sum_j d_j - \sum_i c_i  = 2.
\end{align*}

Since the derivatives $x_j^{(d_j)}$, $j=1,2,3$, that is, $x'',y'',\lam$, occur jointly linearly in \rf{pend}, a consistent point is given by $(t,x,x',y,y')$. If we evaluate $\Jac$ at this point, then
\[
\det(\Jac)=-2(x^2+y^2)=-2\pendL^2\neq 0
\] 
(because $x^2+y^2=\pendL^2$ by $f_3=0$) and SA succeeds\cite{Pryce2001a}. The solution scheme is in \TBref{bltriPend}. The notation $z^{(<r)}$ is short for $z,z',\ldots,z^{(r-1)}$.

\begin{table}
  \[
  \tgnstretch
  \begin{array}{ll@{\hspace{5mm}}l@{\hspace{5mm}}l}
  \text{stage $k$}&  \text{solve}  &\text{for} &\text{using previously found}\\
   \hline
   -2 &0=f_3     & x,y    &-   \\  
   -1 &0=f'_3    & x',y'  &x,y   \\
    \ge 0 &0=f_1^{(k)},f^{(k)}_2,f^{(k+2)}_3 & x^{(k+2)},y^{(k+2)},\lam^{(k)}
& x^{(<k+2)}, y^{(<k+2)}, \lam^{(<k)} \\
  \end{array}
  \]
  \caption{Solution scheme for \protect\rf{pend}}
  \label{tb:bltriPend}
 \end{table}
\end{example}

For brevity, in the following chapters, when we give a system of equations, we  write down 
\begin{itemize}
\item the signature matrix, 
\item a HVT in it (marked by $\bullet$), 
\item the canonical offsets $\~c,\;\~d$, 
\item positions $(i,j)$ where $d_j-c_i>\sij{i}{j}\ge 0$ (marked by $\OK{\phantom{\times}}$), and 
\item the accompanying system Jacobian.
\end{itemize}
When we present a SA result, we omit the words 
\[\text{ ``the signature matrix and system Jacobian are in the following.''}
\] 
 Provided there is a finite HVT in $\Sigma$, we also show the value of the signature matrix and the determinant of the system Jacobian---$\val{\Sigma}$ and $\det(\Jac)$. For instance, after giving \rf{pend}, we simply put $\Sigma$ with $\val{\Sigma}$ attached, and $\Jac$ with $\det(\Jac)$ at the bottom.
\begin{align*}\tgnstretch
\Sigma = 
\begin{blockarray}{rccc ll}
& \clsp x\clsp &  \clsp y \clsp&\clsp \lam \clsp & \s{c_i} & \\
\begin{block}{r @{\hspace{10pt}}[ccc]ll}
f_1 & 2^\bullet  & -  & 0 & \s0 \\
f_2 & -  & 2  & 0^\bullet & \s0 \\
f_3 & 0  & 0^\bullet  & - & \s2  \\
\end{block}
 \s{d_j}& \s2 &\s2 &\s0  &\valSig{\Sigma}{2}\\
 \end{blockarray}
\Jac = \begin{blockarray}{rccc cc}
&\clsp x\clsp & \clsp y\clsp &\clsp \lam\clsp \\
\begin{block}{r @{\hspace{10pt}}[ccc]cc}
f_1 & 1  &0    &x  \\
f_2 & 0  &1    &y  \\
f_3 & 2x &2y   &0  \\
\end{block}
& \mcdetJac{4}{\detJac{\Jac}{-2\pendL^2}}
\end{blockarray}
\end{align*}

Similarly, if we write the signature matrix of a system as $\newSig$, then we write correspondingly the canonical offsets as $\~\newc,\;\~\newd$, and the Jacobian as $\newJ$. Throughout this report, we shall show DAE problems for which our conversion methods are suitable. These methods critically depend on the SA results. 

\chapter{Structural analysis's failure}\label{ch:fail}
In this chapter, we investigate how SA fails on some DAEs. That is, SA produces a singular system Jacobian, and the problem is solvable. 
In \SCref{success}, we give definitions for (a) a structural zero in the system Jacobian, and (b) a structurally singular DAE, where the system Jacobian is identically singular. In \SCref{idfailure} we identify two types of SA's failure.

\section{Success check}\label{sc:success}
To perform a success check for SA on a SWP DAE, we attempt to
evaluate the system Jacobian $\Jac$ in \rf{sysjac}. If a point \rf{consPointNL} satisfies the solution scheme \rrf{solvefi}{forxj} at stages $k=k_d,k_d+1,\ldots,0$, and $\Jac$ is nonsingular, then SA succeeds.

\bigskip

In the definitions that follow, we let $A$ be an $n\times n$ matrix function.

\begin{definition}
An $(i,j)$ position is a {\em structural zero} of $A$ if $\Aij{i}{j}$ is identically $0$; otherwise it is a {\em structural nonzero}.
\end{definition}

\begin{definition}\label{df:structsing}\cite{NedialkovPryce05a}
Matrix $A$ is {\em structurally singular} if every $B\in \bbR^{n\times n}$, with $B_{ij}=0$ in $A$'s structural zero positions, is singular---equivalently, if every transversal of $A$ contains a structural zero. Otherwise $A$ is {\em structurally nonsingular} .
\end{definition}
\begin{definition}\label{df:idsing}
Matrix $A$ is {\em identically singular}, if its determinant is identically $0$; otherwise it is {\em generically nonsingular}. 
\end{definition}

For a matrix function, being structurally singular is a special case of being identically singular; see \EXref{matex} below.
\begin{example}\label{ex:matex}
Consider the following three matrix functions of variables $x$ and $y$:

\[
A_1=
\bbmx
x & x\\[.5ex] 
0 & 0
\ebmx,\quad
A_2=
\bbmx
x & x\\[.5ex] 
y & y
\ebmx,\quad\text{and}\quad
A_3=
\bbmx
x & y\\[.5ex] 
y & x
\ebmx.
\]

$A_1$ is identically singular because $\det(A_1)=0$. It is also structurally singular, since every $B\in \bbR^{2\times 2}$ with $B_{21}=B_{22}=0$ is singular. Here, $(2,1)$ and $(2,2)$ are structural zero positions of $A$, and each transversal in $A$ contains a structural zero.

$A_2$ is also identically singular, as $\det(A_2)=xy-xy=0$.
It is structurally nonsingular, since a transversal does not contain a structural zero.

$A_3$ is structurally nonsingular. It is generically nonsingular, since $\det(A_3)=x^2-y^2$ is not identically zero. $A_3$ is singular only when $x=\pm y$.
\end{example}

In the following, we denote \rf{maineq} by $\daeF$ and define two concepts for it:
\begin{itemize}
\item a {\em structural zero in the system Jacobian} $\Jac$, and 
\item a {\em structurally singular DAE}.
\end{itemize}

Let $\indJ$ be the set of index-pairs
\begin{equation}\label{eq:indJ}
\indJ = \setbg{(j,l) \mid j=\rnge{1}{n}\,, \ l\in\bbN }.
\end{equation}
Given an $n$-vector function $\~x=\~x(t)$ that is sufficiently smooth (but not necessarily a solution of $\daeF$), let 
\[
\~x_\indJ = \setbg{x_j^{(l)}\mid (j,l)\in\indJ}.
\]
For a finite subset $J$ of $\indJ$, we define a $|J|$-vector $\~x_J$
  whose components are $x_j^{(l)}$ as $(j,l)$ ranges over $J$. (The ordering of these components does not matter.)
%

Now we denote a DAE as $\daeF$. We define the {\em derivative set} of $\daeF$
as 
\begin{equation}\label{eq:derset}
\derset(\daeF)
= \setbg{ (j,l)  \mid \xjl{j}{l} \text{ occurs in  $\daeF$}}.
\end{equation}
Then the derivatives occurring in $\daeF$ can be denoted concisely as $\xderset{\daeF}$.

By a {\em value point} we mean a $\calX\in \bbR\times \bbR^{|\derset(\daeF)|}$ that contains values for $t$ and  values for the derivative symbols in $\xderset{\daeF}$.

\begin{example}
In the simple pendulum DAE \rf{pend}, the state variables $x,y,\lam$ are $x_1,x_2,x_3$. Let $\pendL=5$ and $\pendG=9.8$. 
Then 
\[
\derset(\daeF) = \setbg{ (1,0),\, (1,2),\, (2,0),\, (2,2),\, (3,0)  }.
\]
A possible value point  can be
\[
\calX = (t, x_1,x''_1, x_2, x''_2, x_3) = (2, 3,-3,4,1.6,1),
\]
which satisfies $f_1$ and $f_3$ but not $f_2$.
\end{example}

Similarly, we define the {\em derivative set of $\Jac$}:
\begin{equation*}
\derset(\Jac) = \setbg{ (j,l)   \mid x_j^{(l)} \text{ occurs in }\Jac}.
\end{equation*}
From \rf{sysjac}, a derivative occurring in $\Jac$ must also occur in $\daeF$, but not vice versa. For example, in \pend, $x'',y'',\lam$ do not appear in $\Jac$, and $\derset(\Jac) = \setbg{(1,0),(2,0)}$; cf. \EXref{simplepend}.
The derivative set of $\Jac$ is a subset of that of $\daeF$: $\derset(\Jac)\subseteq \derset(\daeF)$. 

\begin{definition}\label{df:structzero}
An $(i,j)$ position is a {\em structural zero} of $\Jac$, if $\Jij{i}{j}$ is identically zero at all
value points  $\calX\in \bbR\times \bbR^{|\derset(\daeF)|}$
that satisfy 0 or more equations from
\begin{equation}\label{eq:fizero}
0=f_i^{(m)},\quad m\ge 0,\quad i=\rnge{1}{n}.
\end{equation}
Otherwise, $(i,j)$ is a {\em structural nonzero}.
\end{definition}

For the present purpose, we do not require the DAE to have a unique solution, or even {\em any} solution. That is, we do not consider existence and uniqueness of the DAE at this stage, while identifying structural zeros of $\Jac$ and the singularity of $\Jac$ discussed below.

Recall \rf{sysjac} that defines $\Jac$. If  $d_j-c_i>\sij{i}{j}$, then $\Jij{i}{j}=0$ and thus position $(i,j)$ is a structural zero in $\Jac$. The converse is not true; see \EXref{pendsing1}.

\begin{example}\label{ex:pendsing1}
Consider an artificially modified simple pendulum DAE. We multiply the first equation $f_1$ by $x^2+y^2-\pendL^2$ and obtain

\begin{equation}\label{eq:pendmess1} 
\begin{aligned}
0 = f_1 &= (x''+x\lam )(x^2+y^2-\pendL^2)\\
0 = f_2 &= y''+y\lam -\pendG\\
0 = f_3 &= x^2+y^2-\pendL^2.
\end{aligned}
\end{equation}
\begin{align*}\tgnstretch
\Sigma = 
\begin{blockarray}{rccc ll}
& \clsp x\clsp &  \clsp y \clsp&\clsp \lam \clsp & \s{c_i} & \\
\begin{block}{r @{\hspace{10pt}}[ccc]ll}
f_1 & 2^\bullet  & \lgo  & 0 & \s0 \\
f_2 & -  & 2  & 0^\bullet & \s0 \\
f_3 & 0  & 0^\bullet  & - & \s2  \\
\end{block}
 \s{d_j}& \s2 &\s2 &\s0 & \valSig{\Sigma}{2} \\
 \end{blockarray}
\Jac = \begin{blockarray}{rccc cc}
&\clsp x\clsp & \clsp y\clsp &\clsp \lam\clsp \\
\begin{block}{r @{\hspace{10pt}}[ccc]cc}
f_1 & \mu  &0    &x\mu  \\
f_2 & 0  &1    &y  \\
f_3 & 2x &2y   &0  \\
\end{block}
\mcdetJac{5}{\detJac{\Jac}{-2\mu (x^2+y^2)}}
\end{blockarray}
\end{align*}
In $\Jac$, $\mu=x^2+y^2-\pendL^2$. 
To decide which entries in $\Jac$ are structural zeros, we notice the following.
\begin{itemize}
\item If we evaluate $\Jac$ at some random  $\calX$, then $\mu$ is not identically equal zero. Hence positions $(f_1,x)$ and $(f_1,\lam)$ are not identical zeros.
\item If we evaluate $\Jac$ at some $\calX$ that satisfies 
\[
\mu=f_3=x^2+y^2-\pendL^2=0,
\]
then according to \DFref{structzero}, positions $(f_1,x)$ and $(f_1,\lam)$ are structural zeros of $\Jac$.
\end{itemize}
\end{example}

We give a definition for {\em structural regularity} of a DAE.

\begin{definition}\label{df:DAEreg}
A DAE is {\em structurally singular} if $\Jac$ is identically singular at all value points $\calX\in \bbR\times \bbR^{|\derset(\daeF)|}$ that satisfy 0 or more equations from \rf{fizero}. 
Otherwise the DAE is {\em structurally nonsingular}, or {\em structurally regular}. 
\end{definition}

\begin{example}\label{ex:pendsing2}
In the previous example, positions $(f_1,x)$ and $(f_1,\lam)$ are structural zeros of $\Jac$ at any point that satisfies $f_3=0$. 
By \DFref{DAEreg},  \rf{pendmess1} is structurally singular.

In fact, it can be shown that a solution of \pend is a solution to \rf{pendmess1}, but not vice versa.
\end{example}

\begin{example}
Consider the DAE in \cite[p. 235, Example 9.2]{AscherPetzold}, written in \rf{maineq} form:
\begin{equation}\label{eq:ascherpetzoldex} 
\begin{aligned}
0 = f_1 &= -y'_1 + y_3\\
0 = f_2 &= y_2(1-y_2)\\
0 = f_3 &= y_1 y_2+y_3(1-y_2)-t.
\end{aligned}
\end{equation}
\begin{align*}\tgnstretch
\Sigma = 
\begin{blockarray}{rccc ll}
& y_1 & y_2 & y_3 & \s{c_i} & \\
\begin{block}{r @{\hspace{10pt}}[ccc]ll}
f_1 & 1^\bullet  & -  & 0 & \s0 \\
f_2 & -  & 0^\bullet  & - & \s0 \\
f_3 & \lgo  & 0  & 0^\bullet & \s0  \\
\end{block}
 \s{d_j}& \s1 &\s0 &\s0 & \valSig{\Sigma}{1} \\
 \end{blockarray}
\Jac = \begin{blockarray}{rccc cc}
& y_1 & y_2 & y_3 \\
\begin{block}{r @{\hspace{10pt}}[ccc]cc}
f_1 & -1  &0    & 1  \\
f_2 & 0  & 1-2y_2    &0  \\
f_3 & 0 & y_1-y_3   & 1-y_2  \\
\end{block}
\mcdetJac{5}{\detJac{\Jac}{-(1-2y_2)(1-y_2)}}
\end{blockarray}
\end{align*}
SA gives $\nuS=1$, and $\det(\Jac)$ depends solely on $y_2$. From $f_2=0$, either $y_2=0$ or $y_2=1$. 
To examine if $\Jac$ is nonsingular, we consider each of the following two cases.
\begin{itemize}
\item If $y_2=0$, then $\det(\Jac)=-1$ and SA succeeds. In this case \rf{ascherpetzoldex} is of d-index 1.

\item If $y_2=1$, then $\det(\Jac)=0$ and SA fails. This failure comes as no surprise because \rf{ascherpetzoldex} is now of d-index 2 and SA underestimates its index; see the discussion in \CHref{background}.
\end{itemize}
\end{example}


\begin{remark}\label{rm:SIP}
For a structurally ill-posed (SIP) DAE, there does not exist a finite transversal in its $\Sigma$---every transversal in $\Sigma$ contains at least one $-\infty$. In this case, there exists no valid offsets $\~c,\~d$, not to mention a system Jacobian that depends on these offsets. In contrast, a structurally singular DAE has valid offsets and a system Jacobian that is identically singular. Hereby we distinguish the difference between a SIP DAE and a structurally singular DAE.
\end{remark}

%
%
%
%

Suppose $\Jac$ is generically nonsingular. If $\Jac$ is singular when evaluated at a point along a solution, then we say the DAE is {\em locally unsolvable} at this point, and we call it a {\em singularity point}. See \EXref{singularpoint}.


\begin{example}\label{ex:singularpoint}\cite{Lamour2015a}
Consider
\begin{equation}\label{eq:singularpoint} 
\begin{aligned}
0 = f_1 &= -x'+ y \\
0 = f_2 &= x + \cos(t)y.
\end{aligned}
\end{equation}
\begin{align*}\tgnstretch
\Sigma = 
\begin{blockarray}{rcc ll}
&  x  &   y  & \s{c_i} & \\
\begin{block}{r @{\hspace{10pt}}[cc]ll}
f_1 & 1^\bullet   & 0 & \s0 \\
f_2 & \lgo  & 0^\bullet  & \s0 \\
\end{block}
 \s{d_j} &\s1 &\s0 & \valSig{\Sigma}{1} \\
 \end{blockarray}
\Jac = \begin{blockarray}{rcc cc}
& x &  y\\
\begin{block}{r @{\hspace{10pt}}[cc]cc}
f_1 & -1 & 1  \\
f_2 & 0  & \cos(t)  \\
\end{block}
\mcdetJac{4}{\detJac{\Jac}{-\cos(t)}}
\end{blockarray}
\end{align*}

Since $\det(\~J)$ is generically nonzero, \rf{singularpoint} is structurally nonsingular. We can integrate this problem from $t=0$ with any consistent initial value $\left( x(0),y(0) \right) = (x_0,y_0)$, and the problem is index-1 (both differentiation and structural indices) as long as $\det(\~J)\neq 0$. However, $\~J$ is singular at $t=t_k=\left(k+1/2 \right)\pi$, $k=0,1,\ldots$. Hence, we say the DAE has a singularity point at $t_k$.
\end{example}

\section{Identifying structural analysis's failure}\label{sc:idfailure}
We give below a definition for the {\em true highest-order derivative} (HOD) of a variable $x_j$ in a function $u$.
\begin{definition}The {\em true HOD} of $x_j$ in $u$ is
\begin{align*}
\hoder{x_j}{u} &=
\casemod{ll}{ 
\text{the highest order derivatives of $x_j$ on which $u$ {\em truly} depends; or}\\
-\infty  \quad \text{ if $u$ does not depend on any derivative of $x_j$ (including $x_j$).}
} \label{eq:hoddef}
\end{align*}
\end{definition}

By ``truly'' we mean that, if $r=\hoder{x_j}{u}>-\infty$, then $u$ is not a constant with respect to $x_j^{(r)}$. For example, $u=x'+\cos^2 x''+\sin^2 x''=x'+1$ truly depends on $x'$ but not $x''$, resulting in $\hoder{x}{u}=1$.



In practice, however, we usually find the {\em formal} HOD of $x_j$ in $u$, denoted by $\comphoder{x_j}{u}$, instead of the {\em true} HOD. By ``formal'' we mean the dependence of an expression (or function) on a derivative {\em without symbolic simplifications}. For example, $u=x'+\cos^2 x''+\sin^2 x''$ formally depends on $x''$ and hence $\comphoder{x}{u}=2$, while $u=x'+1$ and $\hoder{x}{u}=1$.

We denote also $\compsij{i}{j}=\comphoder{x_j}{f_i}$ corresponding to $\sij{i}{j}$. The \daets and \daesa codes implement \cite[Algorithm 4.1 (Signature matrix)]{nedialkov2007solving} for finding formal $\compsij{i}{j}$.

Since the formal dependence is also used in \cite[\S 4]{nedialkov2007solving}, we can adopt the rules in \cite[Lemma 4.1]{nedialkov2007solving}, which indicate how to propagate the formal HOD in an expression. The most useful rules are:
\begin{itemize}
\item if a variable $v$ is a purely algebraic function of a set $U$ of variables $u$, then
\begin{equation}\label{eq:codelist}
\comphoder{x_j}{v} = \max_{u\in U} \comphoder{x_j}{u},
\end{equation}
and
\item if $v=\text{d}^p u/\text{d}t^p$, where $p>0$, then
\begin{equation}\label{eq:codeDif}
\comphoder{x_j}{v} = \comphoder{x_j}{u}+p.
\end{equation}
\end{itemize}
These rules are proved in \cite{nedialkov2007solving}, to which we refer for more details. We illustrate the rules in \EXref{formalhod}.


\begin{example}\label{ex:formalhod}
Let $u=(x_1x_2)'-x'_1x_2$. Applying \rf{codelist} and \rf{codeDif}, we derive the formal HOD of $x_1$ in $u$:
\begin{align*}
\comphoder{x_1}{u}
&= \max\setbg{\comphoder{x_1}{(x_1x_2)'},\, \comphoder{x_1}{x'_1x_2}} \\
&= \max\setbg{\comphoder{x_1}{x_1x_2}+1,\, \max\setbg{\comphoder{x_1}{x'_1}, \comphoder{x_1}{x_2}}} \\
&= \max\setbg{\max\setbg{\comphoder{x_1}{x_1},\, \comphoder{x_1}{x_2}}+1, \max\setbg{1,-\infty}} \\
&= \max\setbg{\max\setbg{0,-\infty}+1,\, 1} \\
&= \max\setbg{0+1,\, 1}\\
&= 1.
\end{align*}
Similarly $\comphoder{x_2}{u}=1$. Simplifying $u=(x_1x_2)'-x'_1x_2$ results in $u=x_1x'_2$. Hence, the true HOD of $x_1$ in $u$ is $\hoder{x_1}{u}=0$, and that of $x_2$ in $u$ is $\hoder{x_2}{u}=1$.
\end{example}

When such a {\em hidden symbolic cancellation} occurs, $\comphoder{x_j}{u}$ can overestimate the true $\hoder{x_j}{u}$. If $u$ is an equation $f_i$, then the formal HOD $\comphoder{x_j}{f_i}$ may not be the true $\sij{i}{j}$. We write $\compsij{i}{j}=\comphoder{x_j}{f_i}$ corresponding to $\sij{i}{j}=\hoder{x_j}{f_i}$. We call the matrix $\compSig=(\compsij{i}{j})$ the ``formal'' signature matrix. Also, let $\compc,\compd$ be any valid offsets for $\compSig$, and let $\compJ$ be the resulting Jacobian defined by \rf{sysjac} with $\compSig$ and $\compc,\compd$.

If $\compsij{i}{j}>\sij{i}{j}$, then $f_i$ does not depend truly on $x_j^{(\compsij{i}{j})}$. That is, $f_i$ is a constant with respect to $x_j^{(\compsij{i}{j})}$.
Then $\compJij{i}{j}=0$, and $(i,j)$ is a structural zero in $\compJ$. Due to such  cancellations, $\compJ$ has more structural zeros than $\Jac$ does, and hence $\compJ$ is more likely to be structurally singular. It is also possible that the DAE itself is structurally ill posed.

Since $\compsij{i}{j}\ge\sij{i}{j}$ for all $i,j=\oneton$, we can write $\compSig\ge\Sigma$ meaning ``elementwise greater or equal''. 

We define the {\em essential sparsity pattern} $\Sess$ of $\Sigma$ to be the union of the HVTs of $\Sigma$.  That is, the set of all $(i,j)$ positions that lie on any HVT. We give two theorems below, which are Theorems 5.1 and 5.2 in \cite{NedialkovPryce05a}. In the following, we use the term ``offset vector" to refer to the vector $(\~c,\~d) = (c_1,\ldots,c_n, d_1,\ldots,d_n)$.
\begin{theorem}\label{th:SigJac}
Suppose that a valid offset vector $(\~c,\~d)$ for $\Sigma$ gives a nonsingular $\Jac$ as defined by \rf{sysjac} at some consistent point. Then every valid offset vector gives a nonsingular $\newJ$ (not necessarily the same as $\Jac$) at this point.
All resulting $\newJ$, including $\Jac$, are equal on $\Sess$, and all have the same determinant $\det(\newJ)=\det(\Jac)$.
\end{theorem}

By ``equal on $\Sess$'' we mean $\newJ_{ij}=\Jij{i}{j}$ for all $(i,j)\in\Sess$.
 
\begin{theorem}\label{th:compSigma}
Assume that $\Jac$, resulting from $\Sigma$ and a valid offset vector $(\~c,\~d)$, is generically nonsingular. Let $(\compc,\;\compd)$ be a valid offset vector for the formal signature matrix $\compSig$, and let $\compJ$ be the Jacobian resulting from $\compSig$ and $(\compc,\;\compd)$.
In exact arithmetic, one of the following two alternatives must occur:
\begin{enumerate}[(i)]
\item $\val{\compSig}=\val{\Sigma}$. Then every HVT of $\Sigma$ is a HVT of $\compSig$, and $\compc,\compd$ are valid offsets for $\Sigma$. Consequently, $\compJ$ is also generically nonsingular.
\item $\val{\compSig}>\val{\Sigma}$. Then $\compJ$ is structurally singular.
\end{enumerate}
\end{theorem}

\THref{compSigma} shows that $\compJ$, resulting from $\compSig\ge\Sigma$ and a valid offset vector $(\compc,\;\compd)$, is {\em either} 
\begin{enumerate}[(1)]
\item nonsingular, and SA is using valid, but not necessarily canonical, offsets for the true $\Sigma$; {\em or} 
\item structurally singular, and SA fails due to symbolic cancellations, in a way that may be detected.
\end{enumerate} 

In the latter case, this failure may be avoided by performing symbolic simplification on some or all of the $f_i$'s. However, ``no clever symbolic manipulation can overcome the hidden cancellation problem, because the task of determining whether some expression is exactly zero is known to be undecidable in any algebra closed under the basic arithmetic operations together with the exponential function'' \cite{NedialkovPryce05a}.

\begin{example}
Consider
\begin{equation}\label{eq:algsys}
\begin{aligned}
0=f_1 &= (xy)'-x'y-xy' + 2x+y-3 \\
0=f_2 &= x+y-2.
\end{aligned}
\end{equation}
\begin{align*}\tgnstretch
\compSig=
\begin{blockarray}{rcc ll}
& \clsp x\clsp & \clsp y \clsp& \s{c_i} \\
\begin{block}{r @{\hspace{10pt}}[cc]ll}
f_1 & 1^\bullet & 1    &\s0  \\
f_2 & 0  & 0^\bullet   &\s1  \\
\end{block}
\s{d_j}& \s1 &\s1 & \valSig{\compSig}{1} \\
\end{blockarray}
\compJ = 
\begin{blockarray}{rcc cc}
& \clsp x\clsp & \clsp y \clsp& \\
\begin{block}{r @{\hspace{10pt}}[cc]cc}
f_1 & 0 & 0   \\
f_2 & 1& 1   \\
\end{block}
\mcdetJac{3}{\detJac{\compJ}{0}}
\end{blockarray}
\end{align*}
Here, the signature matrix and Jacobian are the formal ones. Since $\det(\compJ)=0$, SA fails.
Simplifying $f_1$ to $f_1 = 2x+y-3$ reveals that \rf{algsys} is a simple linear algebraic system:
\begin{equation*}
\begin{aligned}
0=f_1 &= 2x+y-3 \\
0=f_2 &= x+y-2.
\end{aligned}
\end{equation*}
\begin{align*}\tgnstretch
\Sigma = 
\begin{blockarray}{rcc ll}
& \clsp x\clsp & \clsp y \clsp& \s{c_i} \\
\begin{block}{r @{\hspace{10pt}}[cc]ll}
f_1 & 0^\bullet & 0    &\s0  \\
f_2 & 0  & 0^\bullet   &\s0  \\
\end{block}
\s{d_j}& \s0 &\s0 &\valSig{\Sigma}{0}\\
\end{blockarray}
\Jac = 
\begin{blockarray}{rcc cc}
& \clsp x\clsp & \clsp y \clsp& \\
\begin{block}{r @{\hspace{10pt}}[cc]cc}
f_1 & 2& 1     \\
f_2 & 1& 1     \\
\end{block}
\mcdetJac{3}{\detJac{\Jac}{1}}
\end{blockarray}
\end{align*}
\end{example}

Another kind of SA's failure occurs when $\Jac$ is not structurally singular, but is  identically singular. 
Examples~\exref{lenaintro} and \exref{correctindexbutfails} illustrate this case.

\begin{example}\label{ex:lenaintro}
Consider the coupled DAE from \cite{scholz2013combined}
\footnote{We consider this DAE with parameters $\beta=\epsilon=1$, $\alpha_1=\alpha_2=\delta=1$, and $\gamma=-1$. In \cite{scholz2013combined} superscripts are used as indices, while we use subscripts instead. We also change the (original) equation names $g_1,g_2$ to $f_3, f_4$, and the (original) variable names $y_1,y_2$ to $x_3, x_4$.}
:
\begin{equation}\label{eq:lenaDAE}
\begin{aligned}
0=f_1 &= -x'_1 + x_3 + b_1(t)\\
0=f_2 &= -x'_2 + x_4 + b_2(t)\\
0=f_3 &= \phantom{-}x_2 + x_3 + x_4 + c_1(t)\\
0=f_4 &= -x_1 + x_3 + x_4 + c_2(t).
\end{aligned}
\end{equation}
\begin{align*}\tgnstretch
\Sigma = \begin{blockarray}{rccccll}
 & x_{1} & x_{2} & x_{3} & x_{4} & \s{c_i} \\
\begin{block}{r @{\hspace{10pt}}[cccc]ll}
 f_{1}&1^\bullet&-&0&-&\s0\\ 
 f_{2}&-&1^\bullet&-&0&\s0\\ 
 f_{3}&-&\OK{0}&0^\bullet&0&\s0\\ 
 f_{4}&\OK{0}&-&0&0^\bullet&\s0\\ 
\end{block}
\s{d_j} &\s1&\s1&\s0&\s0 &\valSig{\Sigma}{2}\\
 \end{blockarray}
\Jac = \begin{blockarray}{rcccccc}
 & x_{1} & x_{2} & x_{3} & x_{4} \\
\begin{block}{r @{\hspace{10pt}}[cccc]cc}
 f_{1}&-1&0&1&0\\ 
 f_{2}&0&-1&0&1\\ 
 f_{3}&0&0&1&1\\ 
 f_{4}&0&0&1&1\\ 
\end{block}
& \mcdetJac{5}{\detJac{\Jac}{0}}  \\
\end{blockarray}
\end{align*}
This DAE is of d-index 3, while SA finds structural index 1 and singular $\Jac$. Hence SA fails.
\end{example}

\begin{example}\label{ex:correctindexbutfails}
In the following DAE, SA reports the correct d-index 2 but still fails.
\begin{equation}\label{eq:LinConst01_1}
\begin{aligned}
0=f_1&=-x'_{1}-x'_{3}+x_{1}+x_{2}+g_1(t)\\
0=f_2&=-x'_{2}-x'_{3}+x_{1}+x_{2}+x_{3}+x_{4}+g_2(t)\\
0=f_3&=x_{2}+x_{3}+g_3(t)\\
0=f_4&=x_{1}-x_{4}+g_4(t)
\end{aligned}
\end{equation}
\begin{align*}\tgnstretch
\Sigma = \begin{blockarray}{rccccll}
 & x_{1} & x_{2} & x_{3} & x_{4} & \s{c_i} \\
\begin{block}{r @{\hspace{10pt}}[cccc]ll}
 f_{1}&1^\bullet&\OK{0}&1&-&\s0\\ 
 f_{2}&\OK{0}&1^\bullet&1&0&\s0\\ 
 f_{3}&-&0&0^\bullet&-&\s1\\ 
 f_{4}&\OK{0}&-&-&0^\bullet&\s0\\ 
\end{block}
\s{d_j} &\s1&\s1&\s1&\s0 &\valSig{\Sigma}{2} \\
 \end{blockarray}
\Jac = \begin{blockarray}{rcccccc}
 & x_{1} & x_{2} & x_{3} & x_{4} \\
\begin{block}{r @{\hspace{10pt}}[cccc]cc}
 f_{1}&-1&0&-1&0\\ 
 f_{2}&0&-1&-1&1\\ 
 f_{3}&0&-1&-1&0\\ 
 f_{4}&0&0&0&1\\ 
\end{block}
& \mcdetJac{4}{\detJac{\Jac}{0}}
\end{blockarray}
\end{align*}
Using the solution scheme derived from the SA result, we would try to solve at stage $k=0$ the linear system $0=f_1,f_2,f'_3,f_4$ for $x'_1,x'_2,x'_3,x_4$, where the matrix is $\Jac$. Since it is singular, the solution scheme fails in solving \rf{LinConst01_1} at this stage; see \TBref{LinConst01_1}.

\begin{table}[th]
  \[
  \tgnstretch
  \begin{array}{ll@{\hspace{5mm}}l@{\hspace{5mm}}l@{\hspace{5mm}}l}
  \text{stage $k$}&  \text{solve}  &\text{for} &\text{using} &\text{comment}\\ \hline
   -1 & 0=f_3     & x_2,x_3    &-   &\text{initialize $x_1$}\\  
   \phantom{-}0 &0=f_1,f_2,f'_3,f_4    &  x'_1,x'_2,x'_3,x_4  & x_1,x_2,x_3  &\text{$\Jac$ is singular; solution scheme fails} \\
  \end{array}
  \]
  \caption{Solution scheme for \protect\rf{LinConst01_1}}
  \label{tb:LinConst01_1}
 \end{table}

Now we replace $f_2$ by $\newf_2=f_2+f'_3$ to obtain
\begin{equation}\label{eq:LinConst01_2}
\begin{aligned}
0=f_1&=-x'_{1}-x'_{3}+x_{1}+x_{2}+g_1(t)\\
0=f_2+f'_3 =\newf_2 &=x_{1}+x_{2}+x_{3}+x_{4}+g_2(t)+g'_3(t)\\
0=f_3&=x_{2}+x_{3}+g_3(t)\\
0=f_4&=x_{1}-x_{4}+g_4(t).
\end{aligned}
\end{equation}
\begin{align*}\tgnstretch
\newSig = \begin{blockarray}{rccccll}
 & x_{1} & x_{2} & x_{3} & x_{4} & \s{c_i} \\
\begin{block}{r @{\hspace{10pt}}[cccc]ll}
 f_{1}&1&\OK{0}&1^\bullet&-&\s0\\ 
 \newf_{2}&0^\bullet&0&0&0&\s1\\ 
 f_{3}&-&0^\bullet&0&-&\s1\\ 
 f_{4}&0&-&-&0^\bullet&\s1\\ 
\end{block}
\s{d_j} &\s1&\s1&\s1&\s1 &\valSig{\newSig}{1} \\
 \end{blockarray}
\newJ = \begin{blockarray}{rcccccc}
 & x_{1} & x_{2} & x_{3} & x_{4} \\
\begin{block}{r @{\hspace{10pt}}[cccc]cc}
 f_{1}&-1&0&-1&0\\ 
 \newf_{2}&1&1&1&1\\ 
 f_{3}&0&1&1&0\\ 
 f_{4}&1&0&0&-1\\ 
\end{block}
& \mcdetJac{5}{\detJac{\newJ}{2}}  \\
\end{blockarray}
\end{align*}
The solution scheme succeeds; see \TBref{LinConst01_2}. The resulting DAE \rf{LinConst01_2} is of structural index $\nu_S=1$, which equals the differentiation index. 

At stage $k=0,$ we solve $0=\,f_1,\newf'_2,f'_3,f'_4\,$ for $\, x'_1,x'_2,x'_3,x'_4\,$ using $x_1,x_2,x_3,x_4$. Since $\newf'_2=f'_2+f''_3$, we need $f''_3$ to find these first-order derivatives. Therefore, the original DAE \rf{LinConst01_1} is of differentiation index 2.

Note that by setting $f_2=\newf_2-f'_3$ we can immediately recover the original system. It can be easily verified that a vector function 
\[
\~x(t)=\left( x_1(t),\, x_2(t),\, x_3(t),\, x_4(t) \right)^T
\]
 that satisfies \rf{LinConst01_2} also satisfies \rf{LinConst01_1}, and vice versa. We explain in \CHref{LCmethod} how this conversion makes SA succeed.

\newcommand\mrow[3]{\multirow{#1}{#2}{\centering #3}}
\begin{table}[th]
  \[
  \tgnstretch
  \begin{array}{ll@{\hspace{5mm}}l@{\hspace{5mm}}l@{\hspace{5mm}}l}
  \text{stage $k$}&  \text{solve}  &\text{for} &\text{using} &\text{comment}\\ \hline
   -1 &0=\newf_2,f_3,f_4     & x_1,x_2,x_3,x_4    &-   & -\\
\phantom{-}0 & 0=f_1,\newf'_2,f'_3,f'_4    & x'_1,x'_2,x'_3,x'_4 & x_1,x_2,x_3,x_4  &\text{$\Jac$ is nonsingular;} \\[-1ex]
&&&& \text{solution scheme succeeds}\\
  \end{array}
  \]
  \caption{Solution scheme for \protect\rf{LinConst01_2}}
  \label{tb:LinConst01_2}
 \end{table}
\end{example}

In Examples~\exref{lenaintro} and~\exref{correctindexbutfails}, $\Jac$ is not structurally singular, but is still identically singular. No symbolic cancellation occurs in the equations therein. Therefore, this kind of failure is more difficult to detect and remedy, and we wish to find techniques to deal with such failures.

We call our techniques {\em conversion methods}, and describe them in the upcoming chapters. We wish to convert a structurally singular DAE into a structurally nonsingular problem, provided some conditions are satisfied and allow us to perform a conversion step. The original DAE and the converted one are {\em equivalent} in the sense that they have (at least locally) the same solution set. 
We shall also elaborate on this equivalence issue.


\chapter{The linear combination method}\label{ch:LCmethod}
In this chapter we introduce the linear combination method, or the LC method for short. We present in \SCref{LCprelim} some preliminary lemmas. Then we describe in \SCref{convLC} how to perform a conversion step. In \SCref{equivalent} we give definitions and results about equivalence of DAEs and address how equivalence is related to the LC method.

For simplicity, throughout this report, we consider only the second type of SA's failures described in \SCref{idfailure}: ``singular'' means identically singular but not structurally singular.
Based on this assumption, symbolic cancellations are not considered an issue that makes the \Sigmeth fail. 

\section{Preliminary lemmas}\label{sc:LCprelim}

\begin{lemma}\label{le:griewank}
{\bf\em (Griewank's Lemma)}\cite[Lemma 5.1]{nedialkov2007solving}
Let $v$ be a function of $t$, $x_j$'s and derivatives of them $(j=\oneton)$. Denote $v^{(p)}=\text{d}^pv/\text{d}t^p$, where $p>0$. If $\hoder{x_j}{v}\le q$,
then
\begin{equation*}
\pp{v}{x_j^{(q)}}=\pp{v'}{x_j^{(q+1)}}.
\end{equation*}
Hence
\begin{equation}\label{eq:griewank2}
\pp{v}{x_j^{(q)}} = \pp{v'}{x_j^{(q+1)}} \cdots = \pp{v^{(p)}}{x_j^{(q+p)}}.
\end{equation}
\end{lemma}


\begin{lemma}\label{le:sigred0}
Let $\Sigma$ and $\newSig$ be $n\times n$ signature matrices. 
Assume $\val{\Sigma}$ is finite,  $\~c,\~d$ are valid offsets for $\Sigma$, 
and $\newsij{i}{j} \le d_j-c_i$ for all $i,j=\oneton$. 
If a HVT in $\newSig$ contains a position $(i,j)$ where $\newsij{i}{j} < d_j-c_i$, then $\val{\newSig} < \val{\Sigma}$.
\end{lemma}
\begin{proof}
Let $T$ be a HVT in $\newSig$. Then
\begin{align}\label{eq:lcstrictless}
\val{\newSig} &= \sum_{(i,j)\in T} \newsij{i}{j} < \sum_{j=1}^n d_j - \sum_{i=1}^n c_i = \val{\Sigma}.\qed
\end{align}
\renewcommand{\qed}{}
\end{proof}

\begin{corollary}\label{co:sigred}
For a row index $\indxk$, let
\begin{align*}\tgnstretch
\Biggl\{
\begin{array}{ll}
\newsij{i}{j}=\sij{i}{j} &\qquad\text{for all $i\neq \indxk$ and all $j$, and}\\
\newsij{\indxk}{j}< d_j-c_\indxk &\qquad\text{for all $j$.}
\end{array}
\end{align*}
Then $\val{\newSig} < \val{\Sigma}$.
\end{corollary}

\begin{proof}
Since $\newsij{\indxk}{j} < d_j-c_\indxk$ for all $j$, the intersection of a HVT in $\newSig$ with positions in row $\indxk$ is a position $(\indxk,r)$ with $\newsij{\indxk}{r} < d_r-c_\indxk$. By \LEref{sigred0},  
$\val{\newSig} < \val{\Sigma}$.
\end{proof}

This lemma shows that, if we replace a row $\indxk$ in $\Sigma$ with a row with entries less than $d_j-c_\indxk$ for each column $j$, then the value of this signature matrix decreases.

\section{Conversion step}\label{sc:convLC}





Given a SWP DAE of the form \rf{maineq}, assume that we apply the $\Sigma$-method and obtain a singular system Jacobian $\Jac$. We seek a reformulation of this DAE so that the system Jacobian $\newJ$ of the new DAE may be generically nonsingular. We denote by $\Sigma$ and $\newSig$ the signature matrices of the original DAE and this new DAE, respectively. Denote by $\~c,\~d$ the valid offsets for $\Sigma$.

We describe below how to perform a conversion step using a linear combination (LC) of equations.  We call this conversion technique the {\em LC conversion method}, or simply the {\em LC method}.
The main result from this conversion is that, under certain conditions, we can obtain an equation in a row, say $\indxk$, such that $x_j$ occurs in this row of order $<d_j-c_\indxk$ for all $j$. Hence by \COref{sigred}, $\val{\newSig}< \val{\Sigma$}.

We assume $n\ge 2$. Let $u$ be a nonzero vector function from the null space of $\Jac^T$.
Here, $\Jac$ and $u$ are considered as functions of $t$, $x_j$'s and appropriate derivatives of them.

Denote by $\eqsetI(u)$ the set of indices for which the $i$th component of $u$ is not identically zero
\begin{align}\label{eq:idef}
\eqsetI(u) = \{ \, i \mid  u_i\neq 0 \,\},
\end{align}
and let
\begin{align}\label{eq:thdef}
\theta(u) = \min_{i \in \eqsetI(u)} c_i. 
\end{align}
Since $u$ is nonzero and $\Jac$ is identically singular, $\eqsetI(u)$ has at least two elements. Otherwise $\Jac$ has a row of identical zeros and is structurally singular.

\begin{remark}
We consider $u$ in its simplest form in the sense that its elements do not have a common factor comprising $t$, $x_j$'s, or/and derivatives of them. For instance, in \EXref{lenaintro}, we do not use $u=(0,0,x'_1,-x'_1)^T$ though $\Jac^Tu=\~0$, but use $u=(0,0,1,-1)^T$.

Also, we do not consider $u$ with any fractions. For example, we use $u=(0,0,x'_1, x_1x_2)^T$ instead of 
$(0,0,x_1^{-1},x_2(x'_1)^{-1})^T$. 
\end{remark}

The {\em sufficient condition} for applying the LC method is the following: for a nonzero $u\in \ker(\Jac^T)$,
\begin{equation}\label{eq:LCcond}
\boxed{\hoder{x_j}{u}< d_j-\theta(u) \qquad \text{for all $j=\oneton$}}
\end{equation}
 If this condition is satisfied, then we can perform a conversion step. We explain this ``sufficiency" in \RMref{suffLC}.

Denote by $\nzset(u)\subseteq \eqsetI(u)$ the set of indices $\indxk$ such that the $\indxk$th component of $u$ is not an identical zero and $c_\indxk=\theta(u)=\min_{i \in \eqsetI(u)} c_i$:
\begin{align}\label{eq:LCsetK}
\nzset(u) = \setbg{\,\indxk\in \eqsetI(u) \mid\, c_\indxk=\theta (u)}.
\end{align}
From \rf{thdef}, there exists at least one $\indxk\in \eqsetI(u)$ such that $c_\indxk=\theta(u)$, so $\nzset(u)\neq\emptyset$.

We choose an $\indxk\in \nzset(u)$ and replace $f_\indxk$ by 
\begin{align}\label{eq:replacing}
\newf_\indxk = \sum_{i\in \eqsetI(u)} u_i f_i^{\left(c_i-\theta(u)\right)}.
\end{align}
We refer to \rf{replacing} as a {\em conversion step} using the LC method
and to the resulting DAE as a {\em converted} DAE. Critical for the success of the LC method is the following lemma.
\begin{theorem}\label{le:LC}
For a SWP DAE with identically singular $\Jac$, let $u$ be a nonzero $n$-vector such that $\Jac^Tu=\~0$.
If 
\begin{equation*}
\hoder{x_j}{u}< d_j-\theta(u) \qquad \text{for all $j=\oneton$}
\end{equation*}
and we replace $f_\indxk$ by $\newf_\indxk$ in \rf{replacing}, 
then the converted DAE has $\newSig$ with $\val{\newSig} < \val{\Sigma}$.
\end{theorem}

First, we illustrate with an example how to perform a conversion step, and then we prove this lemma. Since $u$ is fixed during a conversion step, for brevity we write $\eqsetI(u)$, $\theta(u)$, and $\nzset(u)$ as $\eqsetI$, $\theta$, and $\nzset$, respectively\footnote{This set $L$ is not to be confused with the constant $L$ in the pendulum-related DAEs.}.

\begin{example}\label{ex:FGxy}
Consider
\begin{equation}\label{eq:FGxy}
\begin{aligned}
0= f_1 &= -x'_1 + x_3\\
0= f_2 &= -x'_2+ x_4\\
0= f_3 &= F(x_1,x_2) \\
0= f_4 &= x_3 F_{x_1}(x_1,x_2) + x_4 F_{x_2}(x_1,x_2) + G(x_1,x_2).
\end{aligned}
\end{equation}
Here $F_{x_1}(x_1,x_2)=\partial F(x_1,x_2)/\partial x_1$, and similarly we write $F_{x_2}(x_1,x_2)$, $G_{x_1}(x_1,x_2)$, and $G_{x_2}(x_1,x_2)$. 

\begin{align*}\renewcommand{\arraystretch}{1.5}
\Sigma=
\begin{blockarray}{rcccc ll}
&  x_1 &   x_2 &  x_3 &  x_4 & \s{c_i} \\
\begin{block}{r @{\hspace{10pt}}[cccc]ll}
f_1 & 1^\bullet  &-  & 0& -    &\s0  \\
f_2 & -  &1  & -& 0^\bullet    &\s0  \\
f_3 & 0  &0^\bullet  & -& -    &\s1  \\
f_4 & \lgo &\lgo  & 0^\bullet & 0    &\s0  \\
\end{block}
 \s{d_j}& \s1 &\s1&\s0  &\s0 & \valSig{\Sigma}{1}\\
 \end{blockarray}
\Jac = 
\begin{blockarray}{rcccc cc}
&  x_1 &   x_2 &  x_3 &  x_4 \\
\begin{block}{r @{\hspace{10pt}}[cccc]cc}
f_1 & -1  &0  & 1& 0      \\
f_2 & 0  &-1  & 0& 1      \\
f_3 & F_{x_1}  & F_{x_2}  &0 &0\\
f_4 & 0  &0  & F_{x_1} & F_{x_2}     \\
\end{block}
& \mcdetJac{5}{\detJac{\Jac}{0}}
\end{blockarray}
\end{align*}
Because of singular $\Jac$, the SA fails. It reports structural index 2, but the differentiation index is 3. If we take $u = \bigl( F_{x_1},\, F_{x_2},\, 1,\, -1\bigr)^T$, then $\Jac^T u=\~0$.

We illustrate \rrf{idef}{LCsetK}:
\begin{align*}
\eqsetI &= \setbg{\,i \mid\, u_i\neq 0\,} = \setbg{1,2,3,4}, \\
\theta &= \min_{i\in \eqsetI}c_i=0, \\
\nzset &= \setbg{\,\indxk\in \eqsetI \mid\, c_\indxk=\theta=0} = \setbg{1,2,4}.
\end{align*}
Then we check if condition \rf{LCcond} holds:
\begin{align*}
\hoder{x_1}{u} &= \;\;\, 0 \;\;< 1 = d_1-\theta, \\
\hoder{x_2}{u} &= \;\;\, 0 \;\;< 1 = d_2-\theta, \\
\hoder{x_3}{u} &= -\infty < 0 = d_3-\theta,\quad\text{and}\quad \\
\hoder{x_4}{u} &= -\infty < 0 = d_4-\theta. \qquad
\end{align*}
Hence $\hoder{x_j}{u}<d_j-\theta$ for all $j$.

Using \rf{replacing} gives
\begin{align*}
\newf &= \sum_{i\in \eqsetI} u_i f_i^{(c_i-\theta)} 
= \sum_{i\in \eqsetI} u_i f_i^{(c_i)} 
\\&= F_{x_1} f_1 + F_{x_2} f_2 + f'_3 - f_4 \\
&= F_{x_1} (-x'_1+x_3) + F_{x_2} (-x'_2+x_4) + (F(x_1,x_2))' - (x_3 F_{x_1}+ x_4 F_{x_2}+G) \\
&= - x'_1F_{x_1} + x_3F_{x_1} -  x'_2F_{x_2} + x_4F_{x_2} + x'_1F_{x_1} + x'_2F_{x_2} - x_3F_{x_1} - x_4F_{x_2} -G \\
&= -G.
\end{align*}
Now, with $\theta=0$,
\begin{align*}
\hoder{x_1}{\newf} &= \;\;\, 0 \;\; < 1 = d_1-\theta, \\
\hoder{x_2}{\newf} &= \;\;\, 0 \;\; < 1 = d_2-\theta, \\
\hoder{x_3}{\newf} &= -\infty < 0 = d_3-\theta, \quad\text{and}\quad\\
\hoder{x_4}{\newf} &= -\infty < 0 = d_4-\theta. \qquad
\end{align*}
That is, $\hoder{x_j}{\newf}<d_j-\theta$ for all $j$.

For each $\indxk\in \nzset=\setbg{1,2,4}$, assuming $u_\indxk\neq 0$, we can replace $f_\indxk$ by $\newf_\indxk=\newf$. We show in the following the three possible converted DAEs, each with $\val{\newSig}=0$ and generically nonsingular $\newJ$.

\begin{itemize}
\item $\indxk=1$:
\begin{equation}\label{eq:FGxyk=1}
\begin{aligned}
0= \newf_1 &= -G(x_1,x_2)\\
0= f_2 &= -x'_2+ x_4\\
0= f_3 &= F(x_1,x_2) \\
0= f_4 &= x_3 F_{x_1}(x_1,x_2) + x_4 F_{x_2}(x_1,x_2) + G(x_1,x_2)
\end{aligned}
\end{equation}
\begin{align*}\tgnstretch
\newSig=
\begin{blockarray}{rcccc ll}
&  x_1 &   x_2 &  x_3 &  x_4 & \s{c_i} \\
\begin{block}{r @{\hspace{10pt}}[cccc]ll}
\newf_1 & 0^\bullet  &0  & -& -    &\s1  \\
f_2 & -  &1  & -& 0^\bullet    &\s0  \\
f_3 & 0  &0^\bullet  & -& -    &\s1  \\
f_4 & \lgo  &\lgo  & 0^\bullet & 0    &\s0  \\
\end{block}
 \s{d_j}& \s1 &\s1&\s0  &\s0 &\valSig{\newSig}{0}\\
 \end{blockarray}
\newJ = 
\begin{blockarray}{rcccc cc}
&  x_1 &   x_2 &  x_3 &  x_4 \\
\begin{block}{r @{\hspace{10pt}}[cccc]cc}
\newf_1 & -G_{x_1}  & -G_{x_2}  & 0 & 0      \\
f_2 & 0  &-1  & 0& 1      \\
f_3 & F_{x_1}  & F_{x_2}  &0 &0\\
f_4 & 0  &0  & F_{x_1} & F_{x_2}     \\
\end{block}
\mcdetJac{6}{\detJac{\newJ}{F_{x_1}(F_{x_1}G_{x_2} - F_{x_2}G_{x_1})}}
\end{blockarray}
\end{align*}
When $u_1 = F_{x_1}\neq 0$ and $F_{x_1}G_{x_2} \neq F_{x_2}G_{x_1}$, the determinant is nonzero and the SA succeeds.

\item $\indxk=2$:
\begin{equation}\label{eq:FGxyk=2}
\begin{aligned}
0= f_1 &= -x'_1+ x_3\\
0= \newf_2 &= -G(x_1,x_2)\\
0= f_3 &= F(x_1,x_2)\\
0= f_4 &= x_3 F_{x_1}(x_1,x_2) + x_4 F_{x_2}(x_1,x_2) + G(x_1,x_2)
\end{aligned}
\end{equation}
\begin{align*}\tgnstretch
\newSig=
\begin{blockarray}{rcccc ll}
&  x_1 &   x_2 &  x_3 &  x_4 & \s{c_i} \\
\begin{block}{r @{\hspace{10pt}}[cccc]ll}
f_1 & 1  &-  & 0^\bullet& -    &\s0  \\
\newf_2 & 0^\bullet  &0  & -& -    &\s1  \\
f_3 & 0  &0^\bullet  & -& -    &\s1  \\
f_4 & \lgo  &\lgo  & 0 & 0^\bullet    &\s0  \\
\end{block}
 \s{d_j}& \s1 &\s1&\s0  &\s0 &\valSig{\newSig}{0}\\
 \end{blockarray}
\newJ = 
\begin{blockarray}{rcccc cc}
&  x_1 &   x_2 &  x_3 &  x_4  \\
\begin{block}{r @{\hspace{10pt}}[cccc]cc}
f_1 & -1 & 0 & 1 & 0 \\
\newf_2 & -G_{x_1}  & -G_{x_2}  & 0 & 0      \\
f_3 & F_{x_1}  & F_{x_2}  &0 &0\\
f_4 & 0  &0  & F_{x_1} & F_{x_2}     \\
\end{block}
\mcdetJac{6}{\detJac{\newJ}{F_{x_2}(F_{x_1}G_{x_2} - F_{x_2}G_{x_1})}}
\end{blockarray}
\end{align*}
Similarly, the SA succeeds when $u_2=F_{x_2}\neq 0$ and $F_{x_1} G_{x_2}\neq F_{x_2} G_{x_1}$.

\item $\indxk=4$:
\begin{equation}\label{eq:FGxyk=4}
\begin{aligned}
0= f_1 &= -x'_1+ x_3\\
0= f_2 &= -x'_2 + x_4\\
0= f_3 &= F(x_1,x_2) \\
0= \newf_4 &= -G(x_1,x_2)
\end{aligned}
\end{equation}
\begin{align*}\tgnstretch
\newSig=
\begin{blockarray}{rcccc ll}
&  x_1 &   x_2 &  x_3 &  x_4 & \s{c_i}  \\
\begin{block}{r @{\hspace{10pt}}[cccc]ll}
f_1 & 1  &-  & 0^\bullet& -    &\s0  \\
f_2 & -  &1  & -& 0^\bullet    &\s0  \\
f_3 & 0  &0^\bullet  & -& -    &\s1  \\
\newf_4 & 0^\bullet  &0  & -& -    &\s1  \\
\end{block}
\s{d_j}& \s1 &\s1&\s0  &\s0 &\valSig{\newSig}{0}\\
\end{blockarray}
\newJ = 
\begin{blockarray}{rcccc cc}
&  x_1 &   x_2 &  x_3 &  x_4  \\
\begin{block}{r @{\hspace{10pt}}[cccc]cc}
f_1 & -1  & 0 & 1& 0    \\
f_2 & 0  &-1  & 0& 1    \\
f_3 & F_{x_1} & F_{x_2} & 0& 0    \\
\newf_4 & -G_{x_1}  &-G_{x_2}  & 0& 0    \\
\end{block}
\mcdetJac{6}{\detJac{\newJ}{-F_{x_1}G_{x_2} + F_{x_2}G_{x_1}}}
\end{blockarray}
\end{align*}
In this case, SA's success requires only $F_{x_1}G_{x_2} \neq F_{x_2}G_{x_1}$.
\end{itemize}

\end{example}

Using the LC method, we obtain three converted DAEs from \rf{FGxy}: \rf{FGxyk=1}, \rf{FGxyk=2}, and \rf{FGxyk=4}. 
However, it is not guaranteed that all converted DAEs and the original DAE have exactly the same solution sets.
We will cover this equivalence issue in \SCref{equivalent}.

Now we prove \LEref{LC}.
\begin{proof}
We show first that 
\begin{align*}
\newsij{\indxk}{j} = \hoder{x_j}{\newf_\indxk} < d_j-c_\indxk \qquad \text{for all $j=\oneton$.}
\end{align*}

Since $\theta=\min_{i\in \eqsetI} c_i$, $c_i-\theta \ge 0$ for $i\in \eqsetI$. By \rf{cidj}, $\hoder{x_j}{f_i}=\sij{i}{j}\le d_j-c_i$. Applying Griewank's Lemma \rf{griewank2} to \rf{sysjac}, with $q=c_i-\theta$, gives
\begin{align}\label{eq:gl}
\Jac_{ij} 
= \frac{\partial f_i}{\partial x_j^{(d_j-c_i)}} 
= \frac{\partial f_i^{(c_i-\theta)}}{\partial x_j^{(d_j-c_i+c_i-\theta)}} 
= \frac{\partial f_i^{(c_i-\theta)}}{\partial x_j^{(d_j-\theta)}} \qquad \text{for $i\in \eqsetI$}.
\end{align}
Then for all $j=\oneton$,
\begin{alignat}{3}
0 &= (\Jac^Tu)_j = \sum_{i=1}^n u_i(\Jac^T)_{ji}= 
\sum_{i \in \eqsetI}^n u_i \Jac_{ij}  &\hspace{10mm}&\text{using $\Jac^Tu=\~0$} \nonumber\\
&=\sum_{i \in \eqsetI} u_i \frac{\partial f_i}{\partial x_j^{(d_j-c_i)}} 
=\sum_{i \in \eqsetI} u_i \frac{\partial f_i^{(c_i-\theta)}}{\partial x_j^{(d_j-\theta)}} 
&&\text{using \rf{gl}}
\nonumber\\
&= \frac{\partial\left(
\sum_{i \in \eqsetI} u_i  f_i^{(c_i-\theta)}\right)}
{\partial x_j^{(d_j-\theta)}} 
 &&\text{using $\hoder{x_j}{u}<d_j-\theta$ for all $j$}\hspace{-5mm}
\label{eq:notdepend}\\
&= \frac{\partial \newf_\indxk}{\partial x_j^{(d_j-\theta)}}
 &&\text{using \rf{replacing}} \nonumber.
\end{alignat}
This shows that $\newf_\indxk$ does not truly depend on $x_j^{(d_j-\theta)}$, that is,
\begin{align*}
\newsij{\indxk}{j}=\hoder{x_j}{\newf_\indxk} < d_j-\theta = d_j-c_\indxk\qquad\text{for all $j=\oneton$}.\label{newseq}
\end{align*}
By \COref{sigred}, $\val{\newSig}< \val{\Sigma$}.
\end{proof}

\begin{remark}\label{rm:nameLC}
We name this method ``LC'' because of the following considerations. The vector $u$ from the null space of $\Jac^T$ that satisfies \rf{LCcond} does not comprise the leading derivatives $x_j^{(d_j-\theta)}$ for all $j$ in equation $f_i^{\left(c_i-\theta\right)}$ for all $i\in \eqsetI$. We consider here each $u_i$ as a ``constant'' in 
\[
\newf_\indxk = \sum_{i\in \eqsetI} u_i f_i^{\left(c_i-\theta\right)},
\]
and $\newf_\indxk$ as a ``linear combination'' of equations $f_i^{\left(c_i-\theta\right)}$.
\end{remark}

\begin{remark}\label{rm:suffLC}
If $\hoder{x_j}{u}= d_j-\theta$ for some $j$, then $\val{\newSig}$ is not guaranteed $< \val{\Sigma}$. In this case, we cannot swap the sum and the differentiation operator in \rf{notdepend}. Therefore, we cannot prove $\partial\newf_\indxk / \partial x_j^{(d_j-\theta)}=0$. Then, in $\newSig$, it can happen that
\[
\newsij{\indxk}{j} = \hoder{x_j}{\newf_\indxk}=d_j-\theta = d_j-c_\indxk \qquad \text{for some $j$},
\]
giving $\le$ instead of strictly $<$ in \rf{lcstrictless}.

However, if $\hoder{x_j}{u}<d_j-\theta$ holds for $j$ from a particular set, we can still achieve $\val{\newSig}<\val{\Sigma}$. We leave this investigation for future work, consider the condition \rf{LCcond} sufficient for now, and require it to be satisfied for the LC method.
\end{remark}

If $u$ is a constant vector, then $\hoder{x_j}{u}=-\infty$ for every $x_j$, and the condition \rf{LCcond} is automatically satisfied. In this case we do not need to check it. We illustrate this in the next example.
\begin{example}\label{ex:xyzt}
Consider
\begin{equation*}
\begin{aligned}
0 = f_1 &= x_1 + tx_2 + t^2x_3 + g_1(t)\\
0 = f_2 &= x'_1 + tx'_2 + t^2x'_3 + g_2(t)\\
0 = f_3 &= x''_1 + tx''_2 + 2t^2x''_3 + g_3(t).
\end{aligned}
\end{equation*}
\begin{align*}\tgnstretch
\Sigma=
\begin{blockarray}{rccc ll}
& x_1 &  x_2 &  x_3 &  \s{c_i} \\
\begin{block}{r @{\hspace{10pt}}[ccc]ll}
f_1 & 0^\bullet  & 0 &0   &   \s2 \\
f_2 & 1 &1^\bullet  &1 &   \s1  \\ 
f_3 & 2  &2 & 2^\bullet & \s0  \\
\end{block}
 \s{d_j}& \s2 &\s2 &\s2 &\valSig{\Sigma}{3}
 \end{blockarray}
\Jac = \begin{blockarray}{rccc cc}
& x_1 & x_2 & x_3 \\
\begin{block}{r @{\hspace{10pt}}[ccc]cc}
f_1 & 1  &t    &t^2      \\
f_2 & 1  &t    &t^2      \\
f_3 & 1  &t    &2t^2    \\  
\end{block}
&\mcdetJac{3}{\detJac{\Jac}{0}}
\end{blockarray}
\end{align*}
For $u=(-1,1,0)^T$, $\Jac^Tu=0$. Using \rrf{idef}{LCsetK} gives
\[
\eqsetI=\setbg{1,2},\quad \theta=c_2=1,\quad \text{and}\quad \nzset=\setbg{2}.
\]
Since $u$ is a constant vector, condition \rf{LCcond} is satisfied. We replace $f_2$ by
\begin{align*}
\newf_2 &= u_1 f^{(2-1)}_1 + u_2 f^{(1-1)}_2 \\
&= -f'_1 + f_2 \\
&= -\bigl( x_1 + tx_2 + t^2x_3 - g_1 \bigr)'+ (x'_1 + tx'_2 + t^2x'_3 + g_2) \\
&= - x_2 - 2tx_3 - g'_1+g_2.
\end{align*}

The converted DAE is
\begin{alignat*}{3}
0 &= f_1 &&= x_1 + tx_2 + t^2x_3 + g_1\\
0 &= \newf_2 &&= -x_2 - 2tx_3 - g'_1+g_2\\
0 &= f_3 &&= x''_1 + tx''_2 + 2t^2x''_3 + g_3.
\end{alignat*}
\begin{align*}\tgnstretch
\newSig=
\begin{blockarray}{rccc ll}
& x_1 &  x_2 &  x_3 &  \s{c_i} \\
\begin{block}{r @{\hspace{10pt}}[ccc]ll}
f_1 & 0^\bullet  & 0 &0   &   \s2 \\
\newf_2 &  - &0^\bullet  &0 &   \s2  \\ 
f_3 & 2  &2 & 2^\bullet & \s0  \\
\end{block}
 \s{d_j}& \s2 &\s2 &\s2  &\valSig{\newSig}{2} \\
 \end{blockarray}
\newJ = \begin{blockarray}{rccc cc}
& x_1 &  x_2 & x_3 \\
\begin{block}{r @{\hspace{10pt}}[ccc]cc}
f_1 & 1  &t    &t^2      \\
\newf_2 & 0  &-1  &-2t     \\
f_3 & 1  &t    &2t^2      \\
\end{block}
&\mcdetJac{3}{\detJac{\Jac}{-t^2}}
\end{blockarray}
\end{align*}
If $t>\sqrt{\epsilon}$ for a suitable $\epsilon$ depending on the machine precision, then $\Jac$ is computably nonsingular.
\end{example}

\section{Equivalent DAEs}\label{sc:equivalent}
First, we give a definition for equivalent DAEs.
\begin{definition}\label{df:equivalent}
Let $\daeF$ and $\newdaeF$ denote two DAEs. They are {\em equivalent} (on some interval for $t$)
if a solution of $\daeF$ is a solution to $\newdaeF$
and vice versa.
\end{definition}

In the following context, we denote by $\daeF$ the original DAE with equations $f_i$, $i=\oneton$, and singular Jacobian $\Jac$. After a conversion step using the LC method, we obtain a (converted) DAE, denoted by $\newdaeF$, with equations $\newf_i$, $i=\oneton$, and Jacobian $\newJ$, which may be singular.

\begin{theorem}\label{th:equivLC}
After a conversion step using the LC method, DAEs $\daeF$ and $\newdaeF$ are equivalent on some real time interval $\intvI$ for $t$, if $u_\indxk\neq 0$ for all $t\in\intvI$.
\end{theorem}

\begin{proof}
Let a solution of $\daeF$, over some interval $\intvI\subset \bbR$, be a vector-valued function 
\[
\~x(t)=\bigl(x_1(t),\ldots,x_n(t)\bigr)^T
\]
that satisfies \rf{maineq} for all $t\in \intvI$.

We denote the vector used in the LC method by $u = (u_1,\ldots, u_n)$, where its $i$th component is of the form
\begin{equation*}
u_i = u_i
\left(t;\, x_1,x'_1,\ldots,x_1^{(d_1-\theta-1)};\, \ldots ;\,x_n,x'_n,\ldots,\,x_n^{(d_n-\theta-1)}
\right).
\end{equation*}

If $u$ is defined at $(t,\~x(t))$ for all $t\in \intvI$, then
\[
\newf_\indxk  =  \sum_{i\in \eqsetI} u_i f_i^{(c_i-\theta)} \quad\text{and}\quad \newf_i = f_i \,\text{ for }\, i\neq l
\]
 vanish at $ (t,\~x(t))$, and thus  $\~x(t)$ is a solution to $\newdaeF$.

Conversely, assume that $\newx(t)$ is a solution of $\newdaeF$ on $\intvI$. If $u$ is defined at $(t,\newx(t))$ for all $t\in \intvI$ and
$u_l\neq 0$, then 
\begin{equation}\label{eq:LCsolution2}
f_\indxk =\frac{1}{u_\indxk}\biggl( \newf_\indxk- \sum_{i\in \eqsetI\backslash \{\indxk\}} u_i \newf_i^{(c_i-\theta)}\biggr)
 \quad\text{and}\quad f_i=\newf_i \,\text{ for }\, i\neq l
\end{equation}
vanish at $(t,\newx(t))$, and thus $\newx(t)$ is a solution to $\daeF$.

By \DFref{equivalent}, $\daeF$ and $\newdaeF$ are equivalent.
\end{proof}

\begin{remark}\label{rm:goodchoiceLC}
We can see from \rf{LCsolution2} that, if we have a choice for $l$, it is desirable to choose it such that $u_l$ is identically nonzero, e.g., a nonzero constant,  $x_1^2+1$, or $2+\cos^2 x_3$. In this case, $\daeF$ and $\newdaeF$ are {\em always} equivalent---we do not need to check the {\em equivalence condition} $u_l\neq 0$ when we solve $\newdaeF$.
\end{remark}

\begin{example}
In \EXref{FGxy}, case $\indxk=1$ [resp. $\indxk=2$]  requires $F_{x_1}\neq 0$ [resp. $F_{x_2}\neq 0$] to recover the original DAE \rf{FGxy} from \rf{FGxyk=1} [resp. from \rf{FGxyk=2}].  However, for case $\indxk=4$,  $u_4=1$ is a nonzero constant for any $t$. Therefore this choice is more desirable than the other two.
\end{example}

Below we define an ill-posed DAE using the structural posedness defined in the \daesa papers \cite{NedialkovPryce2012a,NedialkovPryce2012b}.
\begin{definition}\label{df:illposed}
A DAE is ill posed if it has an equivalent DAE that is structurally ill-posed (SIP); otherwise it is well posed.
\end{definition}

\begin{example}
Consider problem \rf{pendmess1}. Using $0=f_3=x^2+y^2-\pendL^2$, we reduce $f_1$ to the trivial $0=\newf_1=0$. This is just performing a simple substitution, and is not applying the LC method.
The signature matrix
\begin{align}\label{eq:SigIllposed}
\tgnstretch
\newSig = 
\begin{blockarray}{rccc cc}
&  x &   y & \lam  \\
\begin{block}{r @{\hspace{10pt}}[ccc]cc}
\newf_1 & -  & -  & -   \\
f_2 & -  & 2  & 0  \\
f_3 & 0  & 0  & -   \\
\end{block}
 \end{blockarray}
 \end{align}
 does not have a finite HVT, so the resulting DAE is SIP. Hence, by \DFref{illposed} , the original SWP DAE \rf{pendmess1}  is ill posed.
\end{example}

\begin{corollary}\label{co:illposed}
If a structurally well-posed DAE can be converted, by the LC method, to an equivalent DAE that is structurally ill-posed, then the original DAE is ill posed.
\end{corollary}

\begin{proof}
This follows from \THref{equivLC} and \DFref{illposed}.
\end{proof}

\begin{example}\label{ex:pendmess2}
Consider the following SWP DAE
\begin{equation}\label{eq:pendmess2}
\begin{aligned}
0 = f_1 &= y'''+y'\lam+y\lam' \\
0 = f_2 &= y''+y\lam -\pendG\\
0 = f_3 &= x^2+y^2-\pendL^2.
\end{aligned}
\end{equation}
\begin{align*}\tgnstretch
\Sigma = 
\begin{blockarray}{rccc ll}
&  x &   y & \lam  & \s{c_i} \\
\begin{block}{r @{\hspace{10pt}}[ccc]ll}
f_1 & -  & 3  & 1^\bullet & \s0 \\
f_2 & -  & 2^\bullet  & 0 & \s1 \\
f_3 & 0^\bullet  & 0  & - & \s0  \\
\end{block}
 \s{d_j}& \s0 &\s3 &\s1  &\valSig{\Sigma}{3} \\
 \end{blockarray}
\Jac = \begin{blockarray}{rccc cc}
& x &  y & \lam \\
\begin{block}{r @{\hspace{10pt}}[ccc]cc}
f_1 & 0  &1    &y  \\
f_2 & 0  &1    &y  \\
f_3 & 2x  &0    &0  \\
\end{block}
&\mcdetJac{3}{\detJac{\Jac}{0}}
\end{blockarray}
\end{align*}

For $u=(1,-1,0)^T$, $\Jac^Tu=0$. Using \rrf{idef}{LCsetK} gives
\[
\eqsetI=\setbg{1,2},\quad \theta=c_1=0,\quad \text{and}\quad \nzset=\setbg{1}.
\]
Since $u$ is a constant vector, condition \rf{LCcond} is satisfied. We replace $f_1$ by
\begin{align*}
\newf_1 &= f_1-f'_2 = (y'''+y'\lam+y\lam') - (y''+y\lam-g)' = 0.
\end{align*}
The signature matrix of the resulting problem is exactly \rf{SigIllposed}. Hence, by \COref{illposed}, \rf{pendmess2} is ill posed.
\end{example}

If the Jacobian of the converted DAE is still singular, we may be able to apply the LC method iteratively, provided condition \rf{LCcond} is satisfied on each iteration. Since after each conversion step we reduce the value of the signature matrix by at least 1, the number of iterations does not exceed $\val{\Sigma}$, where $\Sigma$ is for the original DAE. We use \EXref{messpend} to show how we can iterate with the LC method.
\begin{example}\label{ex:messpend}
We construct the following (artificial) \modpenda DAE from \pend \rf{pend}:
\begin{equation}\label{eq:messpendLC}
\begin{aligned}
0 = A &= f_3+f_1' = x^2+y^2-\pendL^2 + (x''+x\lam)'\\
0 = B &= f_1+A'' = x''+x\lam + \bigl(x^2+y^2-\pendL^2 + (x''+x\lam)'\bigr)''\\
0 = C &= f_2+A''' = y''+y\lam-\pendG + \bigl(x^2+y^2-\pendL^2 + (x''+x\lam)'\bigr)'''.
\end{aligned}
\end{equation}
\begin{align*}\tgnstretch
\Siter{0} = 
\begin{blockarray}{rccc ll}
&  x &   y & \lam  & \s{c_i} \\
\begin{block}{r @{\hspace{10pt}}[ccc]ll}
A & 3^\bullet  & 0  & 1 & \s3 \\
B & 5  & 2^\bullet  & 3 & \s1 \\
C & 6  & 3  & 4^\bullet & \s0  \\
\end{block}
 \s{d_j}& \s6 &\s3 &\s4  &\valSig{\Siter{0}}{9} \\
 \end{blockarray}
\Jiter{0} = \begin{blockarray}{rccc cc}
& x &  y & \lam \\
\begin{block}{r @{\hspace{10pt}}[ccc]cc}
A & 1  &2y    &x  \\
B & 1  &2y    &x  \\
C & 1  &2y    &x  \\
\end{block}
&\mcdetJac{3}{\detJac{\Jiter{0}}{0}}
\end{blockarray}
\end{align*}
Here, a superscript denotes an iteration number, not a power. We show how to recover the simple pendulum problem.

We find a vector in $\ker(\JiterT{0})$: $\uiter{0}=(-1,1,0)^T$. Then 
\[
\Iiter{0}=\setbg{1,2},\quad \thiter{0}=1,\quad \text{and} \quad \Kiter{0}=\setbg{2}.
\]
We replace the second equation $B$ by
\begin{align*}
-A^{(3-1)}+B = -A''+(A''+f_1) = f_1 = x''+x\lam.
\end{align*}
The converted DAE is
\begin{alignat*}{3}
0 &= A &&= x^2+y^2-\pendL^2 + (x''+x\lam)'\\
0 &= f_1 &&= x''+x\lam\\
0 &= C &&=  y''+y\lam-\pendG + \bigl(x^2+y^2-\pendL^2 + (x''+x\lam)'\bigr)'''.
\end{alignat*}
\begin{align*}\tgnstretch
\Siter{1} = 
\begin{blockarray}{rccc ll}
&  x &   y & \lam  & \s{c_i}  \\
\begin{block}{r @{\hspace{10pt}}[ccc]ll}
A & 3^\bullet  & 0  & 1 & \s3 \\
f_1 & 2  & -  & 0^\bullet & \s4 \\
C & 6  & 3^\bullet  & 4 & \s0  \\
\end{block}
 \s{d_j}& \s6 &\s3 &\s4  &\valSig{\Siter{1}}{6} \\
 \end{blockarray}
\Jiter{1} = \begin{blockarray}{rccc cc}
& x &  y & \lam \\
\begin{block}{r @{\hspace{10pt}}[ccc]cc}
A & 1  &2y    &x  \\
f_1 & 1  &0    &x  \\
C & 1  &2y    &x  \\
\end{block}
&\mcdetJac{3}{\detJac{\Jiter{1}}{0}}
\end{blockarray}
\end{align*}

Although $\val{\Siter{1}}=6<9=\val{\Siter{0}$}, matrix $\Jiter{1}$ is still singular. If $\uiter{1}=(-1,0,1)^T$, then $\JiterT{1} \uiter{1}=\~0$. This gives 
\[
\Iiter{1}=\setbg{1,3},\quad \thiter{1}=0,\quad \text{and} \quad \Kiter{1}=\setbg{3}. 
\]
We replace the third equation $C$ by
\begin{align*}
-A^{(3-0)}+C = -A'''+(f_2+A''') = f_2 = y''+y\lam-\pendG.
\end{align*}
The converted DAE is
\begin{alignat*}{3}
0 &= A &&= x^2+y^2-\pendL^2 + (x''+x\lam)'\\
0 &= f_1 &&= x''+x\lam\\
0 &= f_2 &&= y''+y\lam -\pendG.
\end{alignat*}
\begin{align*}\tgnstretch
\Siter{2} = 
\begin{blockarray}{rccc ll}
&  x &   y & \lam  & \s{c_i} \\
\begin{block}{r @{\hspace{10pt}}[ccc]ll}
A & 3^\bullet  & \lgo  & 1 & \s0  \\
f_1 & 2  & -  & 0^\bullet    & \s1  \\
f_2 & -  & 2^\bullet  & \lgo & \s0 \\
\end{block}
\s{d_j}& \s3 &\s2 &\s1 &\valSig{\Siter{2}}{5} \\
\end{blockarray}
\Jiter{2} = \begin{blockarray}{rccc cc}
& x &  y & \lam \\
\begin{block}{r @{\hspace{10pt}}[ccc]cc}
A & 1     &0   &x    \\
f_1 & 1     &0   &x    \\ 
f_2 & 0     &1   &0    \\
\end{block}
&\mcdetJac{3}{\detJac{\Jiter{2}}{0}}
\end{blockarray}.
\end{align*}

We have $\val{\Siter{2}}=5<6=\val{\Siter{1}}$, but $\Jiter{2}$ is still singular. We find $u=(1,-1,0)^T$ such that $\JiterT{2}\uiter{2}=\~0$. Then 
\[
\Iiter{2}=\{1,2\},\quad \thiter{2}=0,\quad \text{and}\quad \Kiter{2}=\setbg{1}. 
\]
Replacing the first equation $A$ by
\begin{align*}
A-f_1' = (f_3+f_1')-f_1' = f_3 = x^2+y^2-\pendL^2,
\end{align*}
we recover $f_1,f_2,f_3$ from \rf{messpendLC}. This is exactly the DAE \pend\rf{pend}, with $\val{\Sigma}=2$ and $\det(\Jac)=-2\pendL^2$; cf. \EXref{simplepend}.

Since each $u$ in every conversion iteration is a constant vector, each $u_\indxk$ we pick is a nonzero constant. By \RMref{goodchoiceLC}, the original DAE \rf{messpendLC} and \pend are {\em always} equivalent. Hence, we can solve \rf{messpendLC} by simply solving \pend.
\end{example}

\chapter{The expression substitution method}\label{ch:ESmethod}
We develop in this chapter the expression substitution conversion method. In \SCref{condES}, we introduce some notation. We describe in \SCref{convES} how to perform a conversion step using this method and address in \SCref{equivalentES} the equivalence issue.

\section{Preliminaries}\label{sc:condES}

A conversion using the LC method seeks a row in $\Sigma$, replaces the corresponding equation by a linear combination of existing equations, and constructs a new DAE with a signature matrix of a smaller value. Inspired by the LC method, our goal is to develop a conversion method that seeks a column in $\Sigma$, performs a change of certain variables, and constructs a new DAE with $\newSig$ such that  $\val{\newSig}<\val{\Sigma}$. We refer to this approach as the {\em  expression substitution (ES) conversion method}, or the {\em ES method}.

Again, we start from a SWP DAE with a signature matrix $\Sigma$, offsets $\~c,\,\~d$, and identically singular Jacobian $\Jac$. To start our analysis, we give some notation below.

Let $u$ be a vector function from the null space of $\Jac$, that is, $\Jac u=\~0$. Denote by $\nzset(u)$ the set of indices $j$ for which the $j$th component of $u$ is not identically zero
\begin{align}\label{eq:nzset}
\nzset(u) =\setbg{\, j\,\mid\, u_j\neq 0\,},
\end{align}
and denote $s(u)$ by the number of elements in $\nzset(u)$:
\begin{equation}\label{eq:cardK}
s(u) = |\nzset(u)|.
\end{equation}
Note that $s \ge 2$. Otherwise $\Jac$ has a column that is identically the zero vector, and hence $\Jac$ is structurally singular.

Let
\begin{align}\label{eq:eqsetI}
\eqsetI(u) = \setbg{\, i\,\mid\, d_j-c_i=\sij{i}{j}\;\;\text{for some $j\in \nzset(u)$} }.
\end{align}
Denote also 
\begin{align}\label{eq:maxci}
\cm{u} = \max_{i\in \eqsetI(u)} c_i.
\end{align}

Now we illustrate (\ref{eq:nzset}-\ref{eq:maxci}).
\begin{example}\label{ex:ESexam1}
Consider
\begin{equation}\label{eq:ESexam1}
\begin{aligned}
0 = f_1 &= x_1 + e^{-x_1'-x_2x_2''} + g_1(t) \\
0 = f_2 &= x_1 + x_2x_2' + x_2^2+g_2(t).
\end{aligned}
\end{equation}
\begin{align*}\tgnstretch
\Sigma=
\begin{blockarray}{rcc ll}
&  x_1 &  x_2 & \s{c_i} \\
\begin{block}{r @{\hspace{10pt}}[cc]ll}
f_1 & 1^\bullet & 2   &\s0  \\
f_2 & 0 & 1^\bullet   &\s1  \\
\end{block}
\s{d_j}& \s1 &\s2 &\valSig{\Sigma}{2} \\
\end{blockarray}
\Jac = 
\begin{blockarray}{rcc cc}
&  x_1  &  x_2   \\
\begin{block}{r @{\hspace{10pt}}[cc]cc}
f_1 & -\mu &  -\mu x_2     \\
f_2 & 1  &  x_2     \\
\end{block}
&\mcdetJac{3}{\detJac{\Jac}{0}} \\
\end{blockarray}
\end{align*}
In $\Jac$, $\mu=e^{-x_1'-x_2x_2''}$. Using $u=(x_2,-1)^T$ for which $\Jac u=0$, \rrf{nzset}{maxci} become
\begin{equation}\label{eq:setESexam1}
\begin{aligned}
\nzset(u) &=\setbg{1,2},\quad s(u)=|\nzset(u)|=2,\\
\eqsetI(u)&=\setbg{1,2}, \\
\text{and}\quad C(u)&=\max_{i\in \eqsetI(u)} c_i= c_2 = 1.
\end{aligned}
\end{equation}
We show later how the ES method works on this problem.
\end{example}

\begin{remark}\label{rm:ESexam1}
Assume that we apply the LC method to \rf{ESexam1}. 
First, we find $u=\left( 1,\mu \right)^T$ from $\ker(\Jac^T)$.
Using the notation in the LC method, we find $I=\setbg{1,2}$, $\theta=0$, and $k=1$. Since
\[
\hoder{x_1}{u} = \hoder{x_1}{\mu} = 1 =1-0= d_1-\theta,
\]
the condition \rf{LCcond} is not satisfied. After a conversion step, the resulting DAE is still structurally singular with $\val{\newSig}=\val{\Sigma}=2$ and identically singular $\newJ$. See \SCref{ESexam1LC} for more details.
\end{remark}

\section{A conversion step using expression substitution}\label{sc:convES}

We can perform a conversion step using the ES method, if the following conditions hold for some nonzero $u$ such that $\Jac u=\~0$:
\begin{equation}
\boxed{
\hoder{x_j}{u} \le \casemod{ll}{
d_j-\cm{u}-1 & \text{if } j\in \nzset(u) \\[1ex]
d_j-\cm{u} & \text{otherwise}} \label{eq:hodxju}
}
\end{equation}
and
\begin{equation} \label{eq:ESsuffcond}
\boxed{d_j-C(u) \ge 0 \qquad \text{for all $j\in\nzset(u)$}}
\end{equation}
We call \rf{hodxju} and \rf{ESsuffcond} the {\em sufficient conditions for applying the ES method}.

\renewcommand{\cm}[1]{C}

Picking an $\indxk\in\nzset$, we show below how to perform a conversion step. Since we use the same $u$ throughout the following analysis, we omit the argument $u$ and simply write $\nzset,\; s,\; \eqsetI$, and $C$.

Without loss of generality, assume that the nonzero entries of $u$ are in its first $s$ positions:
\begin{align*}
u = (u_1,\ldots, u_s,0,\ldots ,0)^T.
\end{align*}
Then $\nzset=\setbg{j\mid u_j\neq 0} = \setrnge{1}{s}$, where $s=|\nzset|$.

We introduce $s$ variables $y_1,\ldots,y_s$ and let
\begin{equation}\label{eq:gjgk}
\left\{
\begin{aligned}
y_j &= 
x_j^{(d_j-\cm{u})}-\frac{u_j}{u_\indxk} x_\indxk^{(d_\indxk-\cm{u})} \qquad\text{for $j\in\nzset\backslash\setbg{\indxk}$,}
\\
y_\indxk &= x_\indxk^{(d_\indxk-\cm{u})}
\phantom{-\frac{u_j}{u_\indxk} x_\indxk^{(d_\indxk-\cm{u})}}
 \qquad\,\, \text{for $j=\indxk$.}
\end{aligned}
\right.
\end{equation}
(The condition \rf{ESsuffcond} guarantees that the order of $x_j$, $j\in \nzset$, in \rf{gjgk} is nonnegative.)

Written in matrix form, \rf{gjgk} is
\[\renewcommand{\arraystretch}{0.7}
\left[\begin{array}{c}
y_1\\ \vdots \\ y_\indxk \\ \vdots \\ y_s\\
\end{array}\right]
=
\left[\begin{array}{ccccc}
1 & & -u_1/u_\indxk\\ 
&\ddots &\vdots  \\ 
&& 1 \\ 
& &\vdots &\ddots \\ 
&& -u_s/u_\indxk & & 1\\
\end{array}\right]
\left[\begin{array}{c}
x_1^{(d_1-\cm{u})}\\ \vdots \\ x_\indxk^{(d_\indxk-\cm{u})} \\ \vdots \\ x_s^{(d_s-\cm{u})}\\
\end{array}\right].
\]
This $s\times s$ square matrix is nonsingular with determinant 1.

We write the first part of \rf{gjgk} as
\begin{align} \label{eq:x1C}
x_j^{(d_j-\cm{u})} &= y_j+\frac{u_j}{u_\indxk} x_\indxk^{(d_\indxk-\cm{u})}
\qquad \text{for $j\in \nzset\backslash\{\indxk\}$}. 
\end{align}
By \rf{maxci}, $c_i\le \cm{u}$ for all $i\in\eqsetI$. Differentiating  \rf{x1C} $\cm{u}-c_i\ge 0$ times yields
\begin{align*}
\left( x_j^{(d_j-\cm{u})}\right) ^{(\cm{u} - c_i)} 
=x_j^{(d_j-c_i)} 
= \biggl( y_j+\frac{u_j}{u_\indxk} x_\indxk^{(d_\indxk-\cm{u})} \biggr) ^{(\cm{u} - c_i)}.
\end{align*}

In each $f_i$ with $i\in \eqsetI$, we replace every $x_j^{(\sij{i}{j})}$ with $\sij{i}{j}=d_j-c_i$ and  $j\in\nzset\setminus\setbg{\indxk}$ by
\begin{align*}
\Big( y_j+\frac{u_j}{u_\indxk} x_\indxk^{(d_\indxk-\cm{u})} \Big)^{(\cm{u}-c_i)}.
\end{align*}
Denote by $\newf_i$ the equations resulting from these substitutions. For $i\notin \eqsetI$, we set $\newf_i=f_i$.

From \rf{gjgk}, we introduce the equations that prescribe the substitutions:
\begin{equation}\label{eq:gj}
\left\{
\begin{aligned}
0 = g_j &= -y_j + x_j^{(d_j-\cm{u})} - \frac{u_j}{u_\indxk}x_\indxk^{(d_\indxk-\cm{u})} \qquad \text{for $j\in \nzset\backslash\{\indxk\}$} \\
0 = g_\indxk &= -y_\indxk + x_\indxk^{(d_\indxk-\cm{u})} 
\phantom{+ \frac{u_j}{u_\indxk}x_\indxk^{(d_\indxk-\cm{u})}}
\qquad\,\,\,\, \text{for $j=\indxk$.}
\end{aligned}
\right.
\end{equation}
We append these equations to the $\newf_i$'s and construct an augmented DAE that comprises 
\[
\text{equations}\quad \newf_1,\ldots \newf_n;\ g_1, \ldots, g_s \quad\text{in variables}\quad x_1,\ldots, x_n;\ y_1, \ldots y_s.
\]
 We write the signature matrix and system Jacobian of this converted problem as $\newSig$ and $\newJ$, respectively.

\begin{example}\label{ex:expxyk=2}
For \rf{ESexam1}, assume we pick $\indxk=2$. Since
\[
\nzset\setminus \setbg{\indxk} = \setbg{1,2}\setminus\setbg{2}=\setbg{1},
\]
we introduce $y_j=y_1$.
Since
\[
d_1=1,\quad d_2=2,\quad c_1=0,\quad C= c_2 = 1,\quad \text{and}\quad u=(x_2,-1)^T,
\]
\rf{x1C} becomes
\begin{align*}
x_1^{(d_1-\cm{2})}  
=  x_1
=  y_1 + \frac{u_1}{u_2} x_2^{(d_2-\cm{2})} 
= y_1 - x_2x'_2.
\end{align*}
In $f_1$, we replace $x_1^{(d_1-c_1)} = x_1^{(1-0)}= x'_1$ by 
\[
(y_1 - x_2x'_2)^{(C-c_1)} = (y_1 - x_2x'_2)^{(1-0)} = (y_1 - x_2x'_2)'= y'_1-x'^2_2-x_2x''_2.
\]
Similarly, in $f_2$ we replace $x_1$ by $y_1 - x_2x'_2$.
Taking these substitutions into account and appending $g_1$ and $g_2$, we obtain
\begin{equation}\label{eq:ESexamk=2}
\begin{aligned}
0 = \newf_1 &= x_1 + e^{-(y_1 - x_2x'_2)'-x_2x_2''} + g_1(t) \\
&= x_1 + e^{-y'_1 +x'^2_2} + g_1(t) \\
0 = \newf_2 &= (y_1 - x_2x'_2) + x_2x'_2 + x_2^2+g_2(t) \\
&= y_1+x_2^2+g_2(t) \\
0 = g_1 &= -y_1 + x_1 + x_2x'_2 \\
0 = g_2 &= -y_2 + x_2.
\end{aligned}
\end{equation}
\begin{align*}\tgnstretch
\newSig =
\begin{blockarray}{rcccc ll}
&  x_1 &  x_2  &  y_1  &  y_2  & \s{c_i} \\
\begin{block}{r @{\hspace{10pt}}[cccc]ll}
\newf_1 & 0 & 1^\bullet  & 1    &- &\s0  \\
\newf_2 & - & 0  & 0^\bullet    &- &\s1  \\
g_1 & 0^\bullet & 1  & \lgo &- &\s0  \\
g_2 & - &\lgo & -    &0^\bullet &\s0  \\
\end{block}
\s{d_j} & \s0 & \s1 &\s1 &\s0 &\valSig{\newSig}{1}
\end{blockarray}
\~{\newJ} = 
\begin{blockarray}{rcccc cc}
&  x_1 &  x_2  &  y_1  &  y_2   \\
\begin{block}{r @{\hspace{10pt}}[cccc]cc}
\newf_1 & 1 &\clsp 2x'_2\alpha \clsp & \clsp -\alpha \clsp & 0 \\
\newf_2 & 0 & 2x_2  & 1 & 0 \\
g_1 & 1 & x_2  & 0 & 0 \\
g_2 & 0 & 0  & 0 & -1 \\
\end{block}
\mcdetJac{5}{\detJac{\newJ}{x_2-2\alpha(x_2+x'_2)}}
\end{blockarray}
\end{align*}
Here $\alpha = e^{-y'_1 +x'^2_2}$. If $\det(\newJ)\neq 0$, then SA succeeds on \rf{ESexamk=2}.
\end{example}

Our aim is to show that $\val{\newSig}<\val{\Sigma}$ after a conversion step, provided that the sufficient conditions \rf{hodxju} and \rf{ESsuffcond} hold. Before proving this inequality, we 
state two lemmas related to the structure of $\newSig$.

%
%

\begin{lemma}\label{le:ESnewSig}
Let $\~c=(c_1,\ldots,c_n)$ and $\~d=(d_1,\ldots,d_n)$ be the valid offsets for $\Sigma$. Let $\~{\newc}$ and $\~{\newd}$ be the two $(n+s)$-vectors defined as
\begin{align}\tgnstretch
\newd_j &= \casemod{ll}
{
d_j &\qquad\text{if $j=\oneton$}\\[1ex]
\cm{u}  &\qquad\text{if $j=n+\oneton+s,$\quad and}
} \label{eq:newdES}\\ 
\newc_i &=\casemod{ll}
{
c_i &\qquad\text{if $i=\oneton$}\\[1ex]
\cm{u} &\qquad\text{if $i=n+\oneton+s$.}
} \label{eq:newcES}
\end{align}
Then $\newSig$ is of the form in \FGref{ESnewSig}.
\begin{figure}[ht]
\begin{align*}
\renewcommand{\arraystretch}{2}
\begin{blockarray}{rccccccrc}
& x_1\,\,\,\cdots\,\,\, x_{\indxk-1}  & x_{\indxk}  & x_{\indxk+1}\,\,\,\cdots\,\,\, x_s  & x_{s+1}\,\,\,\cdots\,\,\,x_n & y_1\,\,\,\cdots\,\,\, y_{\indxk-1}   & y_\indxk   & y_{\indxk+1}\,\,\,\cdots\,\,\, y_s & \;\newc_i \\
\begin{block}{r @{\hspace{10pt}}[lcl|c|ccc@{\hspace{-1pt}}]c}
\newf_1 &&&&&&-\infty && \;\;\;\; c_1     \\
\vdots &   & \HugeLess &  & \HugeLe  & \HugeLe &\vdots & \HugeLe & \hspace{4mm}\vdots\\
\newf_n &&&&&&-\infty & & \;\;\;\; c_n \\  
\cline{2-8}
g_1 & \clsp = \qquad \multirow{2}{0.5cm}{\parbox{12pt}{\HugeLess}}   &  =  &\qquad\multirow{3}{0.5cm}{\LargeLess} & \multirow{2}{0.5cm}{\parbox{12pt}{\HugeLe}}& \mc{1}{l}{\phantom{=}$0$} &&&  C\\
\vdots  & \clsp\phantom{===}\ddots  & \vdots &&& \phantom{==}\ddots & \mc{2}{c}{\HugeInf} & \;\;\vdots \\
g_{\indxk} &   & = &  & -\infty \cdots -\infty && 0 && \hspace{4mm}C\\
\vdots & \qquad\multirow{1}{0.5cm}{\LargeLess} & \vdots  & \clsp\phantom{==}\ddots  & \multirow{2}{0.5cm}{\parbox{12pt}{\HugeLe}} & \mc{2}{c}{\HugeInf}  & \ddots\phantom{==} & \hspace{2mm}\vdots\\
g_s &  & = & \multirow{-3}{0.5cm}{\LargeLess}  \clsp\phantom{===}=&  &   & & \mc{1}{r}{$0$\phantom{==}} &C\\ 
\end{block}
\newd_j & d_1\,\,\,\cdots\,\,\, d_{\indxk-1}  & d_{\indxk}  & d_{\indxk+1}\,\,\,\cdots\,\,\, d_s  & d_{s+1}\,\,\,\cdots\,\,\,d_n & C\,\,\,\cdots\,\,\, C   & C   & C\,\,\,\cdots\,\,\, C
\end{blockarray}
\end{align*}\caption{\label{fg:ESnewSig} The form of $\newSig$ for the converted DAE using the ES method. The $<,\;\le,\;=$ mean the relations between $\newsig_{ij}$ and $\newd_j-\newc_i$, respectively. For example, an $(i,j)$ position with $\le$ has $\newsig_{i,j}\le\newd_j-\newc_i$.}
\end{figure}
\end{lemma}

The proof is rather technical, and we present it in Appendix~\chref{ProofES}.

\begin{lemma}\label{le:ES}
After a conversion step using the ES method, $\val{\newSig}<\val{\Sigma}$.
\end{lemma}

\begin{proof}
If $\newSig$ does not have a finite HVT, then $\val{\newSig}=-\infty$, while the original DAE is SWP with a finite $\val\Sig$. Hence $\val{\newSig}<\val{\Sigma}$ holds.

On the other hand, if $\newSig$ has a finite HVT $\newT$ so that $\val{\newSig}>-\infty$, then
\begin{alignat*}{3}
\val{\newSig} &= \sum_{(i,j)\in \newT} \newsij{i}{j} 
\\ &\le \sum_{(i,j)\in \newT}(\newd_j-\newc_i) &\hspace{10mm}& \text{since $\newd_j-\newc_i\ge \newsij{i}{j}$ for all $i,j=\rnge{1}{n+s}$} 
\\&=\sum_{j=1}^{n+s} \newd_j - \sum_{i=1}^{n+s} \newc_i
\\&= \left( \sum_{j=1}^n d_j + s\cm{u} \right) - \left( \sum_{i=1}^n c_i + s\cm{u} \right) &&\text{using \rf{newdES} and \rf{newcES},}
\\&= \sum_{j=1}^{n} d_j - \sum_{i=1}^{n} c_i = \val{\Sigma}.
\end{alignat*}

We prove in the following that $\val{\newSig}=\val{\Sigma}$ leads to a contradiction.
 Assume this equality holds. Then there exists a transversal $\newT$ of $(n+s)$ positions in $\newSig$ such that 
\begin{align}\label{eq:ass2}
\newd_j-\newc_i = \newsig_{ij}>-\infty \qquad\qquad \text{for all $(i,j)\in \newT$}.
\end{align}

The column corresponding to $y_\indxk$ has only one finite entry $\newsig_{n+\indxk,n+\indxk}=0$, and therefore\\
$(n+\indxk,n+\indxk)\in \newT$.
Consider 
$(i_1,1), \ldots,  (i_s,s) \in \newT$. Since $(n+\indxk,n+\indxk)\in \newT$, row numbers 
$i_1,\ldots,i_s$ take values among 
\begin{equation}\label{eq:noidxk}
1,\,2,\,\ldots ,\,n,\, n+1,\,\ldots,\, n+\indxk-1,\, n+\indxk+1, \,\ldots ,\,n+s.
\end{equation}
In \rf{noidxk} only $s-1$ numbers are greater than $n$. Hence at least one of these row numbers is among $\rnge{1}{n}$.
That is, there exists $(i_r,r)\in \newT$ with $1\le i_r\le n$ and $1\le r\le s$. This entry is in $\newSig(1:n,1:s)$\footnote{Using \matlab notation.} in \FGref{ESnewSig}, so
$\newd_r - \newc_{i_r} < \newsig_{i_r,r}$,
which contradicts to our assumption in \rf{ass2}. Therefore $\val{\newSig}<\val{\Sigma}$.
\end{proof}

\medskip

\begin{remark}\label{rm:ES}
We give several remarks about the ES method.
\begin{itemize}
\item After a conversion step, we perform symbolic simplifications
on $\newf_i$, for $i\in\eqsetI$. By doing this we ensure that the $x_j^{(d_j-c_i)}$'s for $j\in\nzset=\setrnge{1}{s}$ disappear from these equations. That is, $\hoder{x_j}{\newf_i}<d_j-c_i$ for $j=\rnge{1}{s}$ and $i\in\eqsetI$.
\item Clearly, $y_\indxk$ appears only in $g_\indxk$. We mark down the positions in $\newT$ on $\newSig$, and then remove row $n+\indxk$ (corresponding to $g_\indxk$) and column $n+\indxk$ (corresponding to $y_\indxk$). Because $(n+\indxk,n+\indxk)\in \newT$, the remaining marked positions still form a HVT $\widetilde T$ in the resulting $(n+s-1)\times (n+s-1)$ signature matrix $\compSig$. 

Since $\newsicj{n+\indxk}{n+\indxk}=0$, $\val{\compSig}=\val{\newSig}$. 
The purpose to use $g_\indxk$ and $y_\indxk$ in the above proof and analysis is for our convenience. In practice, we can exclude $g_\indxk$ and $y_\indxk$ in the resulting DAE. For consistency, after removing $g_\indxk$ and $y_\indxk$, we still use $\newSig$ and $\newT$ to denote the signature matrix and system Jacobian, respectively, for the resulting DAE. See \EXref{canremove}.

\item If some derivative $x_j^{(d_j-c_i)}$, for $i=\oneton$ and $j\in\nzset\setminus\setbg{\indxk}$, appears implicitly in an expression in $f_i$, then we need to write this expression into a form in which $x_j^{(d_j-c_i)}$ appears explicitly. See \EXref{makeexplicit}.
\end{itemize}
\end{remark}


\begin{example}\label{ex:canremove}
Removing $g_2$ and $y_2$ in \rf{ESexamk=2} results in a DAE with the following signature matrix and Jacobian
\begin{align*}\tgnstretch
\newSig =
\begin{blockarray}{rccc ll}
&  x_1 &  x_2  &  y_1  & \s{c_i} \\
\begin{block}{r @{\hspace{10pt}}[ccc]ll}
\newf_1 & 0 & 1  & 1^\bullet    &\s0  \\
\newf_2 & - & 0^\bullet  & 0    &\s1  \\
g_1 & 0^\bullet & 1  & \lgo     &\s0  \\
\end{block}
\s{d_j} & \s0 & \s1 &\s1 &\valSig{\newSig}{1}
\end{blockarray}
\~{\newJ} = 
\begin{blockarray}{rccc cc}
&  x_1 &  x_2  &  y_1   \\
\begin{block}{r @{\hspace{10pt}}[ccc]cc}
\newf_1 & 1 & 2x'_2\alpha  &  -\alpha   \\
\newf_2 & 0 & 2x_2  & 1\\
g_1 & 1 & x_2  & 0 \\
\end{block}
\mcdetJac{4}{\detJac{\newJ}{-x_2+2\alpha(x_2+x'_2)}}
\end{blockarray}
\end{align*}
Here $\alpha = e^{-y'_1 +x'^2_2}$.
\end{example}

\begin{example}\label{ex:makeexplicit}
Suppose that $x'''_1$ appears implicitly in $(\sin 2x_1')''$ in $f_1$, and that  the ES method finds
\[
 \nzset=\setbg{1,2},\quad \indxk=2, \quad d_1=d_2=3,\quad \text{and} \quad \cm{u}=c_1=0.
\]
We want to replace $x_1^{(d_1-c_1)} = x'''_1$ by
\[
\Big( y_1+\frac{u_1}{u_2} x_2^{(d_2-\cm{u})} \Big)^{(\cm{u}-c_1)}
= y_1  +\frac{u_1}{u_2} x'''_2.
\]
To make $x'''_1$ appear explicitly in $f_1$, we expand $(\sin 2x_1')''$ and write $2x'''_1\cos x'_1 - 4(x''_1)^2\sin x'_1$ instead. Now we can perform the substitution for $x'''_1$.
\end{example}



\section{Equivalence for the ES method}\label{sc:equivalentES}

We discuss here the equivalence for the ES method. Our approach below is similar to that for deriving the equivalence for the LC method.

We denote by $\daeF$ the original DAE with equations $f_i$, $i=\oneton$, and a singular Jacobian $\Jac$. After a conversion step using the ES method, we obtain a (converted) DAE $\newdaeF$ with equations $\newf_i$, $i=\rnge{1}{n+s}$, and a Jacobian $\newJ$, which may be singular. Here $\newf_{n+j}=g_j$, $j=\onetos$. 

%

Assume that 
\[
\~x(t)=\bigl(x_1(t),\ldots,x_n(t)\bigr)^T
\]
is a solution of $\daeF$ on some real time interval $\intvI\subset \bbR$. That is, every $f_i$, $i=\oneton$, vanishes at $(t,\~x(t))$.
Assume also $u$ is defined at $(t,\~x(t))$ for all $t\in \intvI$. 
We can substitute $\~x(t)$ in \rf{gjgk} to find 
\[
\~y(t) = (y_1(t), \ldots, y_s(t))^T
\]
such that every $\newf_{n+j}=g_j$, $j=\onetos$, in \rf{gj} vanishes at $(t,\~x(t),\~y(t))$.
Using \rf{gjgk}, we perform substitutions in $f_i$, $i\in\eqsetI$, to obtain $\newf_i$. We let $\newf_i=f_i$ for $i\notin\eqsetI$.
Since these substitutions do not change the value of each equation, each $\newf_i$ also vanishes at $(t,\~x(t),\~y(t))$. Therefore $(\~x(t),\~y(t))$ is a solution to $\newdaeF$.

Conversely, assume that $(\newx(t),\newy(t))$ is a solution of $\newdaeF$ on $\intvI\subset\bbR$. Assume also that $u$ is defined at $(t,\newx(t),\newy(t))$ for all $t\in \intvI$. Note here $u$ depends merely on $t$ and $\newx(t)$.
Since $u_\indxk$ is a denominator in each $g_j$ in \rf{gj}, this solution requires $u_\indxk(t)\neq 0$ on $\intvI$. Given that each $g_j$  vanishes at $(t,\newx(t),\newy(t))$, from \rf{gjgk} we have
\begin{equation*}
y_j^{(q)} =
\left\{
\begin{aligned} 
\left( x_j^{(d_j-\cm{u})} - \frac{u_j}{u_\indxk} x_\indxk^{(d_\indxk-\cm{u})} \right)^{(q)}
&\qquad j\in \nzset\setminus\{\indxk\}\\
\left( x_\indxk^{(d_\indxk-\cm{u})} \right)^{(q)}
\phantom{-\,\,\,\,\frac{u_j}{u_\indxk} x_\indxk^{(d_\indxk-\cm{u})}}
&\qquad j=\indxk,
\end{aligned}
\right.
\end{equation*}
where $q\ge 0$.
Substituting the expressions on the right-hand side for the derivatives of $y_j$ in each $\newf_i$ recovers $f_i$ and does not change its value. Therefore, each $f_i$ also vanishes at $(t,\newx(t),\newy(t))$, or simply $(t,\newx(t))$ since $\newy(t)$ does not appear in $f_i$. Then $\newx(t)$ is a solution to $\daeF$.
\medskip

The above discussion gives
\begin{lemma}\label{le:equivES}
After a conversion step using the ES method, DAEs $\daeF$ and $\newdaeF$ are equivalent if $u_\indxk\neq 0$ for all $t\in\intvI$.
\end{lemma}

Again, if we have a choice for $l$, it is desirable to choose one (whenever possible) such that $u_l$ is identically nonzero. In this case, $\daeF$ and $\newdaeF$ are {\em always} equivalent and we do not need to check $u_l\neq 0$ when we solve $\newdaeF$.

\begin{example}
In \rf{ESexam1}, assume we pick $\indxk=1$. Then \rf{x1C} becomes
\begin{align*}
x'_2 =  x_2^{(d_2-\cm{1})} =  y_2 + \frac{u_2}{u_1} x_1^{(d_1-\cm{1})} 
= y_2 - x_1/x_2.
\end{align*}
Here we use
\[
d_1=1,\quad d_2=2,\quad C=1,\quad \text{and}\quad u=(x_2,-1)^T.
\]
%
Then we 
\[
  \tgnstretch
  \begin{array}{lll}
 \text{substitute} & \text{for} &\text{in}  \\ \hline
(y_2 - x_1/x_2)' & x''_2 & f_1\\
y_2 - x_1/x_2 & x'_2 & f_2\\
  \end{array}
\]

The equations $g_j$ derived from \rf{gj} are
\begin{align*}
0 = g_1 &= - y_1 + x_1 \\
0 = g_2 &= - y_2 + x'_2 + x_1/x_2.
\end{align*}
As $\indxk=1$, we can remove $g_1$ and $y_1$, append equation $g=g_2$, and obtain the resulting DAE
\begin{equation*}
\begin{aligned}
0 = \newf_1 &= x_1 + e^{-x'_1-x_2\cdot (y_2 - x_1/x_2)'} + g_1(t) \\
&= x_1 + e^{-x'_1-x_2y'_2 - x'_2x_1/x_2 + x'_1} + g_1(t)\\
&= x_1 + e^{-x_2y'_2 - x'_2x_1/x_2} + g_1(t)\\
0 = \newf_2 &= x_1 + x_2(y_2 - x_1/x_2) + x_2^2+g_2(t) \\
&= x_2y_2 + x_2^2+g_2(t) \\
0 = g &= - y_2 + x'_2 + x_1/x_2.
\end{aligned}
\end{equation*}
\begin{align*}\tgnstretch
\newSig =
\begin{blockarray}{rccc ll}
& \clsp x_1\clsp & \clsp x_2 \clsp & \clsp y_2 \clsp & \s{c_i} \\
\begin{block}{r @{\hspace{10pt}}[ccc]ll}
\newf_1 & 0 & 1  & 1^\bullet    &\s0  \\
\newf_2 & - & 0^\bullet  & 0    &\s1  \\
g & 0^\bullet & 1  & \lgo     &\s0  \\
\end{block}
\s{d_j} & \s0 & \s1 &\s1 &\valSig{\newSig}{1}
\end{blockarray}
\~{\newJ} = 
\begin{blockarray}{rccc cc}
&  x_1 &  x_2  &  y_2   \\
\begin{block}{r @{\hspace{10pt}}[ccc]cc}
\newf_1 & 1-x'_2\beta/x_2 & -x_1\beta/x_2  &  -x_2\beta   \\
\newf_2 & 0 & 2x_2+y_2  & x_2\\
g & 1/x_2 & 1  & 0\\
\end{block}
\mcdetJac{4}{\detJac{\newJ}{-x_2 + \beta(2x_2+y_2+x'_2-x_1/x_2)}}
\end{blockarray}
\end{align*}
In $\newJ$, $\beta = \exp (-x_2y'_2 - x'_2x_1/x_2)$. If $\det(\newJ)\neq 0$, then SA succeeds and gives structural index $\nu_S =2$.
 Here $\val{\newSig}=1<2=\val{\Sigma}$.

However, the original DAE and the resulting one are equivalent only if  $u_1=x_2\neq 0$ on some time interval $\intvI$. In practice, it is more desirable to choose $\indxk=2$ since $u_\indxk=-1$ is identically nonzero; see also \EXref{expxyk=2}.
\end{example}

%
%


\chapter{Examples}\label{ch:example}

In this chapter, we illustrate how to apply the LC method and the ES method to several structurally singular DAEs. When a conversion method {\em succeeds}, we obtain an equivalent structurally regular DAE with a nonsingular system Jacobian.

In \SCref{lenaDAE}, we apply both conversion methods to the $4\times 4$ linear constant coefficient (coupled) DAE \rf{lenaDAE}. The LC method succeeds in converting this problem to a structurally regular DAE in two iterations, reducing the value of the the signature matrix by 2. In contrast, the ES method reduces the value of the signature matrix by 1 in the first iteration. In the second iteration, the condition for applying the ES method is not satisfied, and hence it cannot be applied further.

In \SCref{pendabc}, we illustrate both methods on an artificially complicated problem \modpendb derived from the simple pendulum DAE \pend\rf{pend}. We show in \SSCref{pendabc-ES} how the ES method succeeds in converting this problem to a structurally regular DAE, which has a relatively simple structure. In \SSCref{pendabc-LC}, the LC method is applied, but yields a considerably more complicated result.

In \SCref{ESexam1LC}, we address \RMref{ESexam1} in more detail: the condition for applying the LC method is not satisfied for \rf{ESexam1}. If we perform a conversion step, then the value of the signature matrix is not guaranteed to decrease.


\section{A simple coupled DAE}\label{sc:lenaDAE}

Recall the $4\times 4$ linear constant coefficient (coupled) DAE \rf{lenaDAE}:
\begin{equation}\label{eq:LenaDAEb}
\daeiter{0}:\left\{
\begin{alignedat}{5}
0=f_1 &= -&&x'_1 + x_3 + b_1(t)\\
0=f_2 &= -&&x'_2 + x_4 + b_2(t)\\
0=f_3 &= &&x_2 + x_3 + x_4 + c_1(t)\\
0=f_4 &= -&&x_1 + x_3 + x_4 + c_2(t).
\end{alignedat}
\right.
\end{equation}
\begin{align*}\tgnstretch
\Siter{0} = \begin{blockarray}{rccccll}
 & x_{1} & x_{2} & x_{3} & x_{4} & \s{c_i} \\
\begin{block}{r @{\hspace{10pt}}[cccc]ll}
 f_{1}&1^\bullet&-&0&-&\s0\\ 
 f_{2}&-&1^\bullet&-&0&\s0\\ 
 f_{3}&-&\OK{0}&0^\bullet&0&\s0\\ 
 f_{4}&\OK{0}&-&0&0^\bullet&\s0\\ 
\end{block}
\s{d_j} &\s1&\s1&\s0&\s0 &\valSig{\Siter{0}}{2}\\
 \end{blockarray}
\Jiter{0} = \begin{blockarray}{rcccccc}
 & x_{1} & x_{2} & x_{3} & x_{4} \\
\begin{block}{r @{\hspace{10pt}}[cccc]cc}
 f_{1}&-1&0&1&0\\ 
 f_{2}&0&-1&0&1\\ 
 f_{3}&0&0&1&1\\ 
 f_{4}&0&0&1&1\\ 
\end{block}
\mcdetJac{6}{\detJac{\Jiter{0}}{0}}
\end{blockarray}
\end{align*}
We use\footnote{We use a superscript to mean an iteration number, not a power.}
$\daeiter{0}$ to denote the original problem. We let
$\Siter{0}$ and $\Jiter{0}$ denote the signature matrix and Jacobian of the original problem, respectively.

\subsection{LC method}\label{ssc:lenaDAE-LC} 
We show how to apply the LC method to this problem, and finally obtain an equivalent structurally regular DAE on which SA succeeds.

Let $\uiter{0}=(0,0,-1,1)^T$. Then $\JiterT{0} \uiter{0}=\~0$. Using \rrf{idef}{LCsetK} gives
\[
\Iiter{0}=\setbg{3,4},\quad \thiter{0}=0,\quad\text{and}\quad \Kiter{0}=\setbg{3,4}. 
\]
We choose $\kiter{0}=3\in\Iiter{0}$ and replace $f_3$ by
\begin{align*}
\newf_3 = \uiter{0}_3 f_3+\uiter{0}_4 f_4 =  -f_3+f_4 = -x_1 -x_2 - c_1(t) + c_2(t).
\end{align*}
The converted DAE is
\begin{equation*}
\daeiter{1}:\left\{
\begin{aligned}
0= f_1&= -x'_1 + x_3 + b_1(t)\\
0= f_2&= -x'_2 + x_4 + b_2(t)\\
0= \newf_3&= -x_1 -x_2 - c_1(t) + c_2(t)\\
0= f_4 &= -x_1 + x_3+x_4 + c_2(t).
\end{aligned}
\right.
\end{equation*}
\begin{align*}\tgnstretch
\Siter{1}=
\begin{blockarray}{rcccc ll}
&\clsp x_1\clsp & \clsp x_2\clsp & \clsp x_3&\clsp x_4\clsp & \clsp \s{c_i} \\
\begin{block}{r @{\hspace{10pt}}[cccc]ll}
f_1 & 1^\bullet  & - &0  & - &   \s0 \\ 
f_2 &  - &1  & - &0^\bullet &   \s0  \\ 
\newf_3 & 0  &0^\bullet & - &-& \s1  \\
f_4 & \lgo & - & 0^\bullet &0& \s0  \\
\end{block}
 \s{d_j}& \s1 &\s1 &\s0 &\s0 &\valSig{\Siter{1}}{1} \\
 \end{blockarray}
\Jiter{1} = \begin{blockarray}{rcccc cc}
&\clsp x_1\clsp & \clsp x_2\clsp & \clsp x_3&\clsp x_4\clsp \\
\begin{block}{r @{\hspace{10pt}}[cccc]cc}
f_1 & -1  & 0 &1  & 0 &    \\ 
f_2 & 0  & -1  & 0 &1 &    \\ 
\newf_3 &-1   &-1 & 0 & 0 &   \\
f_4 & 0 & 0 & 1 & 1 &   \\
\end{block}
\mcdetJac{6}{\detJac{\Jiter{1}}{0}}
\end{blockarray}
\end{align*}

Since $\Jiter{1}$ is still singular, we apply again the LC method. Let $\uiter{1}=(-1,-1,1,1)^T$. Then $\JiterT{1} u=0$. Now
\[
\Iiter{1}=\setbg{1,2,3,4},\quad \thiter{1}=0,\quad\text{and}\quad \Kiter{1}=\setbg{1,2,4}.
\]
We choose $\kiter{1}=1\in\Iiter{1}$ and replace $f_1$ by
\begin{align*}
\newf_1  
&= \uiter{1}_1 f_1 + \uiter{1}_2 f_2+ \uiter{1}_3 \newf'_3 + \uiter{1}_4 f_4 \\
&= -f_1-f_2+\newf'_3 +f_4 \\
&= -\left[-x'_1+x_3+b_1(t)\right] 
- \left[-x'_2+x_4+b_2(t)\right]
+ \left[-x_1-x_2-c_1(t)+c_2(t)\right]' \\
&\phantom{= -[-x'_1\,\,\,\,}+ \left[-x_1+x_3+x_4+c_2(t)\right] \\
& =-x_1-b_1(t)-b_2(t)-c'_1(t)+c'_2(t)+c_2(t).
\end{align*}
The converted DAE is
\begin{equation*}
\daeiter{2}:\left\{
\begin{aligned}
0= \newf_1&= -x_1-b_1(t)-b_2(t)-c'_1(t)+c'_2(t)+c_2(t)\\
0= f_2&= -x'_2 + x_4 + b_2(t)\\
0= \newf_3&= -x_1 -x_2 - c_1(t) + c_2(t)\\
0= f_4&= -x_1 + x_3+x_4 + c_2(t).
\end{aligned}
\right.
\end{equation*}

\begin{align*}\tgnstretch
\Siter{2} =
\begin{blockarray}{rccccll}
&\clsp x_1\clsp & \clsp x_2\clsp & \clsp x_3&\clsp x_4\clsp & \clsp \s{\newc_i} \\
\begin{block}{r @{\hspace{10pt}}[cccc]ll}
\newf_1 & 0^\bullet & - & - &-& \s1\\
f_2 & -  &1  & - &0^\bullet &   \s0  \\
\newf_3 &  0 & 0^\bullet & - &-& \s1 \\
f_4 & \lgo  & - &0^\bullet  & 0 &   \s0\\ 
\end{block}
\s{\newd_j}&\s1 &\s1 & \s0 &\s0  &\valSig{\Siter{2}}{0} \\ 
\end{blockarray}
\Jiter{2} = \begin{blockarray}{rccccrr}
&\clsp x_1\clsp & \clsp x_2\clsp & \clsp x_3&\clsp x_4\clsp \\
\begin{block}{r @{\hspace{10pt}}[cccc]rr}
\newf_1 &-1  &0  & 0 & 0   \\
f_2 & 0  & -1  & 0 & 1    \\ 
\newf_3 & -1  & -1 & 0 & 0  \\
f_4 & 0  & 0 & 1 & 1     \\ 
\end{block}
\mcdetJac{6}{\detJac{\Jiter{2}}{1}}
\end{blockarray}
\end{align*}
The SA succeeds on this converted DAE and gives structural index $\nu_S=2$. Since $\uiter{0}_{\kiter{0}}$ and $\uiter{1}_{\kiter{1}}$ are nonzero constants, $\daeiter{2}$, $\daeiter{1}$, and $\daeiter{0}$ are always equivalent.

\subsection{ES method}\label{ssc:lenaDAE-ES}

For $\daeiter{0}$ in  \rf{LenaDAEb}, the ES method cannot convert it  to a structurally regular DAE.
We illustrate this argument with one particular choice of $\indxk\in\nzset$ in each iteration of the ES method, and do not explore all possible combinations of such choices. To handle the limitation of the ES method, further development is required, which is left as future work. 

Let $u=(1,-1,1,-1)^T$. Then $\Jiter{0} u=\~0$. Using \rrf{nzset}{maxci} finds
\begin{align}\label{eq:setESlena}
\nzset &=\setbg{1,2,3,4},\quad s=|\nzset|=4,\quad \eqsetI=\setbg{1,2,3,4},\quad\text{and}\quad C=\max_{i\in \eqsetI} c_i=0.
\end{align}

Assume we pick $\indxk=3$. Using \rf{gjgk}, we introduce $y_j$ for each $j\in \nzset\setminus\setbg{\indxk}=\setbg{1,2,4}$:
\begin{equation}\label{eq:lenaDAEsubs}
\begin{aligned}
y_1 &= x_1^{(d_1-C)} -\inlinefrac{u_1}{u_3} x_3^{(d_3-C)} = x'_1-x_3\\
y_2 &= x_2^{(d_2-C)} -\inlinefrac{u_2}{u_3} x_3^{(d_3-C)} = x'_2+x_3\\
y_4 &= x_4^{(d_4-C)} -\inlinefrac{u_4}{u_3} x_3^{(d_3-C)} = x_4+x_3.
\end{aligned}
\end{equation}
From \rf{lenaDAEsubs}, we construct $g_j$'s in \rf{gj}. 
Since $\indxk=3$, we can exclude $g_3$ and $y_3$ in the converted DAE; see \RMref{ES}.

By \rf{x1C} and from \rf{lenaDAEsubs}, we write
\begin{align*}
x'_1 = y_1+x_3, \quad
x'_2 = y_2-x_3, \quad\text{and}\quad 
x_4  = y_4-x_3.
\end{align*}
In \rf{lenaDAE}, we
\[
  \tgnstretch
  \begin{array}{lll}
 \text{substitute} & \text{for} &\text{in}  \\ \hline
y_1+x_3 & x'_1 & f_1\\
y_2-x_3 & x'_2 & f_2\\
y_4-x_3 & x_4  & f_2,\,f_3,\,f_4
  \end{array}
\]

The converted DAE is
\begin{equation}\label{eq:lenaDAE_ES1}
\begin{alignedat}{5}
0 = f_1 &= -&&y_1 + b_1(t)\\
0 = f_2 &= &&y_4-y_2 + b_2(t)\\
0 = f_3 &= &&x_2 + y_4 + c_1(t)\\
0 = f_4 &= - &&x_1 + y_4 + c_2(t) \\
0 = g_1 &= -&&y_1+x'_1-x_3 \\
0 = g_2 &= -&&y_2+x'_2+x_3 \\
0 = g_4 &= -&&y_4+x_4+x_3.
\end{alignedat}
\end{equation}

\begin{align*}\tgnstretch
\newSig =
\begin{blockarray}{rccccccc ll}
&\clsp x_1\clsp & \clsp x_2\clsp & \clsp x_3 &\clsp x_4\clsp &\clsp y_1\clsp &\clsp y_2\clsp &\clsp y_4\clsp & \s{c_i} \\
\begin{block}{r @{\hspace{10pt}}[ccccccc]ll}
f_1 & -  & - & -  & - & 0^\bullet & - & - & \s0 \\ 
f_2 & -  & - & -  & - & - & 0^\bullet & \OK{0} & \s0  \\
f_3 & -  & 0\bullet & -  & - & - & - & 0 & \s1  \\
f_4 & 0  & - & -  & - & - & - & 0^\bullet & \s1  \\
g_1 & 1^\bullet  & - & 0  & - & 0 & - & - & \s0  \\
g_2 & -  & 1 & 0^\bullet  & - & - & 0 & - & \s0  \\
g_4 & -  & - & 0  & 0^\bullet & - & - & \OK{0} & \s0  \\
\end{block}
 \s{d_j}& \s1 &\s1 &\s0 &\s0 &\s0 &\s0 &\s1 &\valSig{\newSig}{1}  \\
 \end{blockarray}
\end{align*}
\begin{align*}\tgnstretch
\newJ = 
\begin{blockarray}{rccccccc cc}
&\clsp x_1\clsp & \clsp x_2\clsp & \clsp x_3 &\clsp x_4\clsp &\clsp y_1\clsp &\clsp y_2\clsp &\clsp y_4\clsp  \\
\begin{block}{r @{\hspace{10pt}}[ccccccc]cc}
f_1 & 0  & 0 & 0  & 0 & -1 & 0 & 0  \\ 
f_2 & 0  & 0 & 0  & 0 & 0 & -1 & 0  \\
f_3 & 0  & 1 & 0  & 0 & 0 & 0 & 1   \\
f_4 & -1  & 0 & 0  & 0 & 0 & 0 & 1  \\
g_1 & 1  & 0 & -1  & 0 & -1 & 0 & 0  \\
g_2 & 0  & 1 & 1  & 0 & 0 & -1 & 0  \\
g_4 & 0  & 0 & 1  & 1 & 0 & 0 & 0  \\
\end{block}
\mcdetJac{9}{\detJac{\newJ}{0}}
\end{blockarray}
\end{align*}
If $u=(1,-1,1,-1,0,0,1)^T$, then $\newJ u=\~0$. We use \rrf{nzset}{maxci} again to find
\begin{align*}
\nzset &=\setbg{1,2,3,4,7},\quad s=|\nzset|=5,\quad \eqsetI=\setbg{3,4,5,6,7},\quad\text{and}\quad C=\max_{i\in \eqsetI} c_i=1.
\end{align*}
Since $j=3,4\in\nzset$ and 
\[
d_3-C = d_4-C = 0-1 = -1<0, 
\]
condition \rf{ESsuffcond} is not satisfied, and thus applying the ES method to \rf{lenaDAE_ES1} does not guarantee a strict decrease in the value of $\newSig$.

\section{Artificially modified pendulum \modpendb}\label{sc:pendabc}
From \pend\rf{pend}, we construct a problem \modpendb by performing a linear transformation on the state variables:
\begin{align*}\tgnstretch
\left[
\begin{array}{c}
x\\y\\ \lam
\end{array}
\right] = \left[
\begin{array}{ccc}
1 & 1 & 0\\ 0 & 1 & 1\\ 1 & 0 & 1
\end{array}
\right] \left[
\begin{array}{c}
z_1\\ z_2\\ z_3
\end{array}
\right]
.
\end{align*} 
The resulting DAE is
\begin{equation}\label{eq:pendabc}
\begin{aligned}
0 = f_1 &= (z_1+z_2)''+(z_1+z_2)(z_3+z_1)\\
0 = f_2 &= (z_2+z_3)''+(z_2+z_3)(z_3+z_1) -\pendG\\
0 = f_3 &= (z_1+z_2)^2+(z_2+z_3)^2-\pendL^2.
\end{aligned}
\end{equation}

\begin{align*}\tgnstretch
\Siter{0}=
\begin{blockarray}{rccc ll}
&\clsp z_1\clsp & \clsp z_2\clsp & \clsp z_3\clsp & \clsp \s{c_i} \\
\begin{block}{r @{\hspace{10pt}}[ccc]ll}
f_1 & 2  & 2 & \OK{0}   &   \s0 \\
f_2 & \OK{0} &2  &2 &   \s0  \\ 
f_3 & 0  &0 & 0 & \s2  \\
\end{block}
 \s{d_j}& \s2 &\s2 &\s2 &\valSig{\Siter{0}}{2}
 \end{blockarray}
\Jiter{0} = \begin{blockarray}{rccc @{\hspace{3pt}}cc}
&\clsp z_1\clsp & \clsp z_2\clsp &\clsp z_3\clsp \\
\begin{block}{r @{\hspace{10pt}}[ccc@{\hspace{6pt}}]cc}
f_1 & 1    &1    & 0     \\
f_2 & 0  &1    &1      \\
f_3 & 2\alpha  &2(\alpha+\beta)    &2\beta    \\  
\end{block}
\mcdetJac{4}{\detJac{\Jiter{0}}{0}}
\end{blockarray}
\end{align*}
Here $\alpha=z_1+z_2$ and $\beta=z_2+z_3$.

\subsection{ES method}\label{ssc:pendabc-ES}

If $u=(1,-1,1)^T$, then $\Jiter{0} u=\~0$. 
We apply the ES method, and \rrf{nzset}{maxci} give
\begin{align*}
\nzset &=\setbg{1,2,3},\quad s=|\nzset|=3,\quad I=\setbg{1,2,3},\quad \text{and} \quad C = \max_{i\in \eqsetI} c_i= c_3 = 2.
\end{align*}
Since $u$ is a constant vector, picking any $\indxk\in \nzset=\setbg{1,2,3}$ gives an equivalent converted DAE.  We show below the conversion for case $\indxk=1$.

Since $\nzset\setminus \setbg{\indxk} =\setbg{2,3}$, we introduce variables $w_2$ and $w_3$ corresponding to $z_2$ and $z_3$, respectively. Using \rf{x1C} gives
\begin{equation}\label{eq:pendabcsubs}
\begin{aligned}
z_2 &=  z_2^{(d_2-\cm{2})} 
=  w_2 + \frac{u_2}{u_1} z_1^{(d_1-\cm{1})} 
= w_2-z_1 \\
z_3 &=  z_3^{(d_3-\cm{2})} 
=  w_3 + \frac{u_3}{u_1} z_1^{(d_1-\cm{1})} 
= w_3+z_1.
\end{aligned}
\end{equation}
To perform substitutions, we first write the derivatives $z''_1,\ z''_2$ and $z''_3$ in $f_1$ and $f_2$ explicitly:
\begin{align*}
0 = f_1 &= z''_1+z''_2 + (z_1+z_2)(z_3+z_1)\\
0 = f_2 &= z''_2+z''_3 + (z_2+z_3)(z_3+z_1) -\pendG
\end{align*}
Then we
\[
  \tgnstretch
  \begin{array}{lll}
 \text{substitute} & \text{for} &\text{in}  \\ \hline
w''_2-z''_1 & z''_2 & f_1,\; f_2\\
w''_3+z''_1 & z''_3 & f_2\\
w_2-z_1 & z_2 & f_3\\
w_3+z_1 & z_3 & f_3
  \end{array}
\]
Taking \rf{pendabcsubs} into consideration, we find the resulting DAE (with $g_1$ and $y_1$ removed as $\indxk=1$)
\begin{equation}\label{eq:pendabcES}
\begin{alignedat}{3}
0 = \newf_1 &= &&w_2''+w_2(2z_1+w_3)\\
0 = \newf_2 &= &&(w_2+w_3)''+(w_2+w_3)(2z_1+w_3) -\pendG\\
0 = \newf_3 &= &&w_2^2+(w_2+w_3)^2-\pendL^2 \\
0 = g_2 &= -&&z_2+w_2-z_1 \\
0 = g_3 &= -&&z_3+w_3+z_1.
\end{alignedat}
\end{equation}
\begin{align*}\tgnstretch
\newSig = \begin{blockarray}{rcccccll}
 & z_{1} & z_{2} & z_{3} & w_{2} & w_{3} & \s{c_i} \\
\begin{block}{r @{\hspace{10pt}}[ccccc]ll}
 \newf_{1}&0&-&-&2^\bullet&\OK{0}&\s0\\ 
 \newf_{2}&0^\bullet&-&-&2&2&\s0\\ 
 \newf_{3}&-&-&-&0&0^\bullet&\s2\\ 
 g_{2}&0&0^\bullet&-&\OK{0}&-&\s0\\ 
 g_{3}&0&-&0^\bullet&-&\OK{0}&\s0\\ 
\end{block}
\s{d_j} &\s0&\s0&\s0&\s2&\s2 &\valSig{\newSig}{2} \\
 \end{blockarray}
\newJ = \begin{blockarray}{rccccccc}
 & z_{1} & z_{2} & z_{3} & w_{2} & w_{3} \\
\begin{block}{r @{\hspace{10pt}}[ccccc]cc}
 \newf_{1}&2w_2&0&0&1&0\\ 
 \newf_{2}&2\mu &0&0&1&1\\ 
 \newf_{3}&0&0&0&2(w_2+\mu) &2\mu \\ 
 g_{2}&-1&-1&0&0&0\\ 
 g_{3}&1&0&-1&0&0\\ 
\end{block}
\mcdetJac{7}{\detJac{\newJ}{-4\pendL^2}}
\end{blockarray}
\end{align*}
Here $\mu=w_2+w_3$. We use equation $\newf_3=0$ to obtain $\det(\newJ)$:
\[\det(\newJ)=-4(2w_2^2+2w_2w_3+w_3^2)=-4\pendL^2\neq 0.\]
Hence SA succeeds on \rf{pendabcES}. Because $u_1=1$ is a nonzero constant, \rf{pendabcES} and \rf{pendabc} are always equivalent.


\subsection{LC method}\label{ssc:pendabc-LC}
We show below how to apply the LC method to \rf{pendabc}. The resulting DAE is relatively complicated, and its equivalence to the original problem requires two conditions to be satisfied.


 Let $\uiter{0}=\bigl( \alpha,\beta,-1/2 \bigr)^T$. Then $\JiterT{0} \uiter{0}=\~0$. Using \rrf{idef}{LCsetK} gives 
\[
\Iiter{0}=\setbg{1,2,3},\quad \thiter{0}=0,\quad \text{and}\quad \Kiter{0}=\setbg{1,2}.
\]
 Since $\uiter{0}_1=\alpha$ and $\uiter{0}_2=\beta$ are not identically nonzero, the converted DAE is equivalent to \rf{pendabc} only if $\uiter{0}_\indxk\neq 0$ for the $\indxk$ we pick.

Assume that $\uiter{0}_1=\alpha=z_1+z_2\neq 0$. We pick $\indxk=1$ and replace $f_1$ by 
\begin{align*}
\newf_1 &= \uiter{0}_1 f_1 + \uiter{0}_2 f_2 +  \uiter{0}_3 f''_3 \\
&= (z_1+z_2)f_1 + (z_2+z_3)f_2 - f''_3/2 \\
&= \uline{(z_1+z_2)(z_1+z_2)''} + (z_1+z_2)^2(z_3+z_1) \uuline{+ (z_2+z_3)(z_2+z_3)''} + (z_2+z_3)^2(z_3+z_1) \\
&\phantom{= (z_1\,\,}- \pendG(z_2+z_3) \uline{- (z_1+z_2)(z_1+z_2)''} - (z_1'+z_2')^2 \uuline{- (z_2+z_3)(z_2+z_3)''} - (z_2'+z_3')^2 \\
&= \left[(z_1+z_2)^2+ (z_2+z_3)^2\right] (z_3+z_1) -\pendG(z_2+z_3) - (z_1'+z_2')^2 - (z_2'+z_3')^2 \\
&= \pendL^2(z_3+z_1) -\pendG(z_2+z_3) - (z_1'+z_2')^2 - (z_2'+z_3')^2 .
\end{align*}

The resulting DAE is
\begin{equation*}
\daeiter{1}:\left\{
\begin{aligned}
0 = \newf_1 &= \pendL^2 (z_3+z_1) -\pendG(z_2+z_3) - (z_1'+z_2')^2 - (z_2'+z_3')^2\\
0 = f_2 &= (z_2+z_3)''+(z_2+z_3)(z_3+z_1) -\pendG\\
0 = f_3 &= (z_1+z_2)^2+(z_2+z_3)^2-\pendL^2.
\end{aligned}
\right.
\end{equation*}
\begin{align*}\tgnstretch
\Siter{1}=
\begin{blockarray}{rccc ll}
&\clsp z_1 \clsp & \clsp z_2 \clsp & \clsp z_3 \clsp & \clsp \s{c_i} \\
\begin{block}{r @{\hspace{10pt}}[ccc]ll}
\newf_1 & 1  & 1 &1   &   \s1 \\
f_2 &  \OK{0} &2  &2 &   \s0  \\ 
f_3 & 0  &0 & 0 & \s2  \\
\end{block}
 \s{d_j}& \s2 &\s2 &\s2 &\valSig{\Siter{1}}{3}
 \end{blockarray}
\Jiter{1} = \begin{blockarray}{rccc @{\hspace{3pt}}cc}
&\clsp z_1 \clsp & \clsp z_2 \clsp &\clsp z_3 \clsp \\
\begin{block}{r @{\hspace{10pt}}[ccc@{\hspace{6pt}}]cc}
\newf_1 & -2\alpha'    &-2(\alpha+\beta)'    & -2\beta'     \\
f_2 & 0  &1    &1      \\
f_3 & 2\alpha  &2(\alpha+\beta)    &2\beta    \\  
\end{block}
\mcdetJac{5}{\detJac{\Jiter{1}}{0}}
\end{blockarray}
\end{align*}
We use $\alpha$ and $\beta$ to denote $z_1+z_2$ and $z_2+z_3$, respectively. Let also $\gamma$ denote $z_3+z_1$. Note that $(\alpha,\beta,\gamma)$ are actually $(x,y,\lam)$ in \rf{pend}---this notation is for simplicity only, and we shall not substitute $\alpha,\beta,\gamma$ for $z_1+z_2,\ z_2+z_3$, and $z_3+z_1$, respectively. That is, we do not use the ES method here.

Matrix $\Jiter{1}$ is still identically singular. We let $\uiter{1}=\bigl( \alpha,\, 2\alpha\beta'-2\beta\alpha',\, \alpha' \bigr)^T$. Then $\JiterT{1} \uiter{1}=0$. Using \rrf{idef}{LCsetK} again gives
\[
\Iiter{1}=\setbg{1,2,3},\quad \thiter{1}=0,\quad\text{and}\quad \Kiter{1}=\setbg{2}.
 \]
Suppose 
\[
\frac{\uiter{1}_2}{2} = \alpha\beta'-\beta\alpha' = (z_1+z_2)(z'_2+z'_3)-(z_2+z_3)(z'_1+z'_2) \neq 0.
\]
Then the converted DAE is equivalent to $\daeiter{1}$.
We replace $f_2$ by 
\begin{equation*}
\begin{aligned}
\newf_2 &= \uiter{1}_1 f'_1 + \uiter{1}_2 f_2 +  \uiter{1}_3 f''_3 \\
& = \alpha f'_1 + 2(\alpha\beta'-\alpha'\beta) f_2 + \alpha' f''_3 \\
&= \alpha (
\pendL^2\gamma' 
-\pendG\beta' 
-2\alpha' \underline{\alpha''} -2\beta' \underline{\beta''} 
)
+ 2(\alpha\beta' -\alpha'\beta) (\underline{\beta''} +\beta\gamma-\pendG) \\
&\phantom{= \alpha (\pendL^2\gamma'\,\,}
+2\alpha'(\alpha'^2+ \alpha \underline{\alpha''} +\beta'^2+\beta \underline{\beta''})\\
&= \alpha(\pendL^2 \gamma'-\pendG\beta')
 +2(\alpha\beta' -\alpha'\beta) (\beta\gamma-\pendG)
+2\alpha'(\alpha'^2+\beta'^2)  \\
&= (z_1+z_2) \Bigl[ \pendL^2 (z'_3+z'_1)-\pendG(z'_1+z'_2) \Bigr] \\
&\phantom{= (z_1\,\,} +  2\Bigl[ (z_1+z_2)(z'_2+z'_3) -(z'_1+z'_2)(z_2+z_3) \Bigr]
\Bigl[ (z_2+z_3)(z_3+z_1)-\pendG \Bigr]  \\
&\phantom{= (z_1\,\,} + 2(z'_1+z'_2) \Bigl[(z'_1+z'_2)^2+(z'_2+z'_3)^2\Bigr] . 
\end{aligned}
\end{equation*}

The resulting DAE is
\begin{equation*}
\daeiter{2}:\left\{
\begin{aligned}
0 = \newf_1 &= \pendL^2 (z_3+z_1) -\pendG(z_2+z_3) - (z_1'+z_2')^2 - (z_2'+z_3')^2\\
0 = \newf_2 &= (z_1+z_2) \Bigl[ \pendL^2 (z'_3+z'_1)-\pendG(z'_1+z'_2) \Bigr] \\
&\phantom{= (z_1\,\,} +  2\Bigl[ (z_1+z_2)(z'_2+z'_3) -(z'_1+z'_2)(z_2+z_3) \Bigr]
\Bigl[ (z_2+z_3)(z_3+z_1)-\pendG \Bigr]  \\
&\phantom{= (z_1\,\,} + 2(z'_1+z'_2) \Bigl[(z'_1+z'_2)^2+(z'_2+z'_3)^2\Bigr]\\
0 = f_3 &= (z_1+z_2)^2+(z_2+z_3)^2-\pendL^2.
\end{aligned}
\right.
\end{equation*}
\begin{align*}\tgnstretch
\Siter{2} = \begin{blockarray}{rcccll}
 & z_{1} & z_{2} & z_{3} & \s{c_i} \\
\begin{block}{r @{\hspace{10pt}}[ccc]ll}
 \newf_{1}&1&1&1^\bullet&\s0\\ 
 \newf_{2}&1^\bullet&1&1&\s0\\ 
 f_{3}&0&0^\bullet&0&\s1\\ 
\end{block}
\s{d_j} &\s1&\s1&\s1 &\valSig{\Siter{2}}{2}\\
 \end{blockarray}
\end{align*}
Jacobian $\Jiter{2}$ is complicated, and we do not show it here. Its determinant is
\begin{align*}
\det(\Jiter{2}) &= -4\pendL^2 \left({z_{1}} + {z_{2}}\right) 
\left[(z_1+z_2)(z'_2+z'_3)-(z_2+z_3)(z'_1+z'_2)\right] \\
&= -4\alpha \pendL^2(\alpha\beta' -\beta\alpha' ) \neq 0,
\end{align*}
since we already assume
\[
\uiter{0}_1=\alpha=z_1+z_2\neq 0 \quad \text{and} \quad 
\uiter{1}_2/2 = \alpha\beta'-\beta\alpha'  \neq 0.
\]
Therofore SA succeeds and gives structural index $\nu_S=1$.

\medskip

Now we consider the case $\alpha\beta'-\beta\alpha'=0$. Since 
\[
0=h' = 2\alpha\alpha' +2\beta\beta' \qquad \text{and}\qquad \alpha\neq 0,
\]
we have
\[
0=\alpha\beta'-\beta\alpha' = \alpha\beta' +\beta\cdot (\beta\beta')/\alpha = \beta' (\alpha^2+\beta^2)/\alpha = \beta'\pendL^2/\alpha.
\]
So $\beta'=\alpha'=0$. 
Since $\uiter{1}=\bigl( \alpha, 2\alpha\beta'-2\beta\alpha', \alpha' \bigr)^T = (\alpha,0,0)^T$, the first row in $\Jiter{1}$ is identically zero and the LC method is not applicable here.

Hence, the DAEs $\daeiter{2}$ and $\daeiter{0}$ are equivalent under the conditions
\[
\uiter{0}_1=\alpha=z_1+z_2\neq 0 \quad \text{and} \quad \uiter{1}_2=\beta'=z'_2+z'_3 \neq 0.
\]

\section{DAE \protect\rf{ESexam1} and LC method}\label{sc:ESexam1LC}

We show below that applying the LC method to \rf{ESexam1} does not convert it into a structurally nonsingular DAE, because the condition for the LC method is not satisfied. Recall \rf{ESexam1} and its SA result.
\begin{equation*}
\begin{aligned}
0 = f_1 &= x_1 + e^{-x_1'-x_2x_2''} + g_1(t) \\
0 = f_2 &= x_1 + x_2x_2' + x_2^2+g_2(t).
\end{aligned}
\end{equation*}
\begin{align*}\tgnstretch
\Sigma=
\begin{blockarray}{rcc ll}
& \clsp x_1\clsp & \clsp x_2 \clsp& \s{c_i} \\
\begin{block}{r @{\hspace{10pt}}[cc]ll}
f_1 & 1^\bullet & 2   &\s0  \\
f_2 & 0 & 1^\bullet   &\s1  \\
\end{block}
\s{d_j}& \s1 &\s2  &\valSig{\Sigma}{2}\\
\end{blockarray}
\Jac = 
\begin{blockarray}{rcc @{\hspace{3pt}}cc}
&  x_1  &  x_2   \\
\begin{block}{r @{\hspace{10pt}}[cc@{\hspace{6pt}}]cc}
f_1 & -\mu &  -\mu x_2     \\
f_2 & 1  &  x_2     \\
\end{block}
\mcdetJac{3}{\detJac{\Jac}{0}}
\end{blockarray}
\end{align*}
Here $\mu=e^{-x_1'-x_2x_2''}$. If $u=\left( 1,\mu \right)^T$, then $\Jac^T u=\~0$.
Using \rrf{idef}{LCsetK} gives 
\[
\eqsetI=\setbg{1,2}, \quad\theta=0, \quad\text{and}\quad \nzset=\{1\}.
\]
Let $\indxk=1$ and replace $f_1$ by
\begin{align*}
\newf_1 &= u_1 f_1 + u_2 f'_2 = f_1+ \mu f'_2 \\
&= x_1 + \mu + g_1(t)  + \mu (x_1 + x_2x_2' + x_2^2+g_2(t))' \\ 
&= x_1 + \mu + g_1(t) 
+ \mu \left( x'_1+x_2x''_2+(x'_2)^2+2x_2 x'_2 +g'_2(t) \right) \\
&= x_1 + g_1(t)
+ \mu\left(1+x'_1+x_2x''_2+(x'_2)^2 +2x_2 x'_2+g'_2(t)\right) .
\end{align*}

The resulting DAE is
\begin{equation}
\begin{aligned}
0 = \newf_1 &= x_1 + g_1(t)
+ e^{-x_1'-x_2x_2''}\left(1+x'_1+x_2x''_2+(x'_2)^2 +2x_2 x'_2+g'_2(t)\right) \\
0 = f_2 &= x_1 + x_2x_2' + x_2^2+g_2(t).
\end{aligned}
\end{equation}
\begin{align*}\tgnstretch
\newSig=
\begin{blockarray}{rcc ll}
& \clsp x_1\clsp & \clsp x_2 \clsp& \s{c_i} \\
\begin{block}{r @{\hspace{10pt}}[cc]ll}
\newf_1 & 1^\bullet & 2   &\s0  \\
f_2 & 0 & 1^\bullet   &\s1  \\
\end{block}
\s{d_j}& \s1 &\s2  &\valSig{\newSig}{2}  \\
\end{blockarray}
\newJ = 
\begin{blockarray}{rcc @{\hspace{3pt}}cc}
& \clsp x_1 \clsp & \clsp x_2 \clsp  \\
\begin{block}{r @{\hspace{10pt}}[cc@{\hspace{6pt}}]cc}
\newf_1 & -\alpha\mu  & -\alpha\mu x_2    \\
f_2 & 1 & x_2   \\
\end{block}
\mcdetJac{4}{\detJac{\newJ}{0}}
\end{blockarray}
\end{align*}
Here $\alpha=x'_1+x_2x''_2+(x'_2)^2 +2x_2 x'_2+g'_2(t)$ and $\mu=e^{-x_1'-x_2x_2''}$.

The conversion step does not reduce the value of signature matrix nor produce a nonsingular Jacobian, because \rf{LCcond} is not satisfied:
\begin{align*}
\hoder{x_1}{u} = \hoder{x_1}{\mu} &= 1 =1-0= d_1-\theta.
\end{align*}
\chapter{Conclusion and future work}\label{ch:conclu}

We identified two types of structural analysis's failure. For the first type, the system Jacobian is structurally singular, and the failure is likely due to hidden symbolic cancellations. One way to handle this is to perform symbolic simplifications before applying SA. 

We focused on dealing with the second type, where SA fails in a less obvious way. In this case, the Jacobian is not structurally singular but is still identically singular. We proposed two symbolic-numeric methods for converting a DAE with such singular Jacobian to an equivalent DAE on which SA succeeds with nonsingular 
 Jacobian, provided some conditions are satisfied. Such conditions can be checked automatically. These conversion methods provably succeed and thus allow SA to handle more DAE types. 
 Our methods enable SA to better recognize the true structure of a DAE, and thus SA is more likely to succeed and obtain correct structural information. Moreover, our methods provide insights into reasons for SA's failures, which were not well understood before.


We summarize the two conversion methods here.
The LC method is more straightforward: it keeps the size of the system and replaces only one equation within a conversion step. The ES method requires more conditions to apply. It augments the system and changes several equations within a conversion step, which generally takes more symbolic operations. The common goal of both methods is to reduce the value of the signature matrix; this value is also the number of degrees of freedom reported by the SA. We also need to ensure that the converted DAE is equivalent to the original one: on some time interval, a solution of the original DAE should be a solution to the converted DAE, and vice versa. 
Moreover, it is desirable to choose a conversion (if possible) such that we do not need to monitor the equivalence condition ($u_\indxk\neq 0$) when we solve the converted problem.

A practical question worth considering is how to choose the appropriate conversion method between the two for a given structurally singular DAE. For many of the examples we have studied, it is fairly common that conditions of only one method are satisfied; see 
\SCref{lenaDAE} and \SCref{ESexam1LC}.
For some other DAEs, provided conditions of both methods are satisfied, applying one method usually requires fewer symbolic manipulations than applying the other; see \modpendb in \SCref{pendabc}. 
In the latter case, we examine if a conversion has an identically nonzero $u_l$, such that the converted DAE and the original one are {\em always} equivalent.
If applying either method guarantees such equivalence, we prefer the LC method because it changes only one equation and maintains the problem size.


Our next goal is to combine these conversion methods with block triangular forms (BTFs) of a DAE \cite{NedialkovPryce2012a,Pryce2014a}. If the Jacobian is identically singular and the DAE has a non-trivial BTF, that is, it has two or more diagonal blocks, then we can locate the block that leads to the singularity, and perform a conversion step on this singular block. Using this approach, we have already made some progress and managed to convert several problems, including the Campbell-Griepentrog Robot Arm and the Ring Modulator, into SA success cases. However, more careful studies are required and details need to be worked out.

Future work also includes rigorous implementation of these conversion methods in \matlab. We currently have a prototype code, which builds upon our structural analyzer package \daesa \cite{NedialkovPryce2012b} and takes advantage of \matlab's symbolic toolbox. We can call \daesa functions to display a converted DAE's structure, perform quasilinearity analysis, and print out a solution scheme \cite{nedialkov2014a}. Our goal is to eventually build a complete tool for performing DAE conversions.


Another interesting direction for research is combining the dummy derivative index reduction method \cite{Matt93a} with the conversion techniques. McKenzie et al. \cite{DDandSA2015} show that dummy derivative method and the \Sigmeth share some similarities. The former method converts a high-index DAE into a low-index one by introducing dummy algebraic variables and augmenting the system. Suppose we adopt this methodology. When a condition for applying a conversion method is violated, applying some dummy-derivative-like strategy may help make a conversion possible.

Our conversion methods require a symbolic toolbox that is capable of performing nontrivial symbolic arithmetic operations, differentiation, and simplifications. However, we only focus on a linear combination of the equations (in the LC method) and a linear combination of derivatives of highest order (in the ES method). How to further exploit symbolic tools to develop other conversion methods should deserve some investigation. 

We conclude with a conjecture here. In all our experiments, when we transform a DAE with identically singular Jacobian to an equivalent solvable DAE with generically nonsingular Jacobian, the value of the signature matrix always decreases. As Pryce points out in \cite{Pryce98}, the solvability of a DAE lies within its inherent nature, not the way it is formulated nor the method that analyzes it. We conjecture that, if a reformulation of a structurally singular DAE results in an equivalent solvable DAE, then the value of the signature matrix 
always decreases. However, based on our current knowledge, it seems difficult to prove this conjecture.

 \appendix

\begin{appendices}

\chapter{Proofs for the ES method}\label{ch:ProofES}
For readers' convenience, we recall the notation from Sections \ref{sc:condES} and \ref{sc:convES}:
\begin{alignat*}{3}
\nzset &=\setbg{\, j\,\mid\, u_j\neq 0\,} = \rngeset{1}{s}, \qquad &&s= |\nzset|, \\
\eqsetI &= \setbg{\, i\,\mid\, d_j-c_i=\sij{i}{j}\;\;\;\text{for some $j\in \nzset$ }}, \qquad 
&&\cm{u} = \max_{i\in \eqsetI} c_i.
\end{alignat*}
We also assume the conditions for applying the ES method are satisfied:
\begin{align}
\hoder{x_j}{u} &\le 
\casemod{ll}{
d_j-\cm{u}-1 & \text{if } j\in \nzset \\
d_j-\cm{u} & \text{otherwise},} \label{eq:ESsuffcond2}
\end{align}
and
\begin{align*}
d_j-C&\ge 0 \qquad \text{for all $j\in\nzset$.} 
\end{align*}

Denote
\begin{alignat*}{3}
\neqsetI &= \rngeset{1}{n}\setminus \eqsetI &&= \setbg{\, i\,\mid\, d_j-c_i>\sij{i}{j} \;\;\;\text{for all $j\in \nzset$}\,},\quad\text{and} 
%
\\\zset &= \rngeset{1}{n}\setminus \nzset &&= \setbg{\, j\,\mid\, u_j=0} = \rngeset{s+1}{n}.
\end{alignat*}

In the following, we assume we pick a column index $\indxk\in\nzset$ in a conversion step using the ES method. We also assume $u_\indxk\neq 0$ for all $t$ in some $\intvI\subset \bbR$.

\section{Preliminary results for the proof of Lemma~\ref{le:ESnewSig}}\label{ap:proofhodinexp}

\begin{lemma}\label{le:hodinexp}
Let $\indxk\in\nzset$. If \rf{ESsuffcond2} holds, then for an $r \in \nzset\setminus\{\indxk\}$,
\begin{align}\label{eq:hodxjyr1}
\hoder{x_\jone}{y_\jtwo+\frac{u_\jtwo}{u_\indxk} x_\indxk^{(d_\indxk-\cm{\indxk})} } &\le
\casemod{ll}{
d_j-\cm{\indxk}-1& \text{if }j\in \nzset \setminus \{ \indxk\} =   
\setbg{1,\ldots,l-1,l+1,\ldots,s}
\\ [1ex]
d_j-\cm{\indxk} &  
\text{if }j\in \zset \cup \{ \indxk\} =   \setbg{l, s+1, s+2, \ldots, n}.
}   
\end{align}
\end{lemma}

\begin{proof}
According to our assumptions at the beginning of \CHref{LCmethod}, no symbolic cancellation occurs in a structurally singular DAE. Hence the formal HOD and the true HOD of a variable in a function are the same. Using \rf{codelist} gives
\begin{align}
\hoder{x_\jone}{y_\jtwo+\frac{u_\jtwo}{u_\indxk} x_\indxk^{(d_\indxk-\cm{\indxk})}} 
&\le \max \left\{
\hoder{x_\jone}{u},\,
\hoder{x_\jone}{x_\indxk^{(d_\indxk-\cm{\indxk})}}
\right\} 
%
.\label{eq:hodxjyr}
\end{align}

\begin{enumerate}[(a)]
\item Consider the case $j=\indxk\in\nzset$. Using \rf{ESsuffcond2} gives $\hoder{x_\indxk}{u}\le d_\indxk-C-1$, so
\begin{align} \label{eq:hodxjyr11}
\text{RHS of \rf{hodxjyr}} 
&= \max \left\{
\hoder{x_\indxk}{u}
,\hoder{x_\indxk}{x_\indxk^{(d_\indxk-\cm{\indxk})}}
\right\} \nonumber\\
&= \hoder{x_\indxk}{x_\indxk^{(d_\indxk-\cm{\indxk})}} 
= d_\indxk-\cm{\indxk}.
\end{align}
\item Consider the case $j\neq \indxk$, that is, $j\in\setbg{1,\ldots,l-1,l+1,\ldots,n}$.
 Using \rf{ESsuffcond2} again, we have
\begin{align}\label{eq:hodxjyr12}
\text{RHS of \rf{hodxjyr}} &= \max \left\{
\hoder{x_\jone}{u}
,\hoder{x_\jone}{x_\indxk^{(d_\indxk-\cm{\indxk})}}
\right\} \nonumber\\
&= \hoder{x_j}{u}
\le \casemod{ll}{
d_j-\cm{\indxk}-1  \qquad &\text{if $j\in\nzset\setminus\{\indxk\}=\setbg{1,\ldots,l-1,l+1,\ldots,s}$} \\[1ex]
d_j-\cm{\indxk}     \qquad &\text{if $j\notin \nzset=\setbg{l,s+1,\ldots,n}$.}
}
\end{align}
\end{enumerate}

Combining \rf{hodxjyr11} and \rf{hodxjyr12} results in \rf{hodxjyr1} and completes this proof.
\end{proof}

\begin{corollary}\label{co:hodinexp}
For an $i\in \eqsetI$,
\begin{align}
\hoder{x_\jone}{\Big( y_\jtwo+\frac{u_\jtwo}{u_\indxk} x_\indxk^{(d_\indxk-\cm{\indxk})} \Big)^{(\cm{\indxk}-c_i)}} &\le
\casemod{ll}{
d_j-c_i-1& \text{if }j\in \nzset \setminus \{ \indxk\}=\setbg{1,\ldots,l-1,l+1,\ldots,s}   \\ [1ex]
d_j-c_i &  \text{otherwise}.
}  \label{eq:hodxjyr2}
\end{align}
\end{corollary}

\begin{proof}
Since $C=\max_{i\in \eqsetI} c_i$, the order $C-c_i\ge 0$ for $i\in \eqsetI$. We have
\begin{alignat*}{3}
\text{LHS of \rf{hodxjyr2}} 
%
&= \hoder{x_\jone}{ y_\jtwo+\frac{u_\jtwo}{u_\indxk} x_\indxk^{(d_\indxk-\cm{\indxk})} } +(\cm{\indxk}-c_i)  && 
\\ &\le (\cm{\indxk}-c_i)+\casemod{ll}{
d_j-\cm{\indxk}-1 & \text{if }j\in \nzset \setminus \{ \indxk\}   \\ 
d_j-\cm{\indxk} &  \text{otherwise}
} &\qquad\qquad\qquad& \text{using \rf{hodxjyr1}}
\\ &= \casemod{ll}{
d_j-c_i-1 & \text{if }j\in \nzset \setminus \{ \indxk\}   \\ [1ex]
d_j-c_i &  \text{otherwise}
}
\\ &=\text{RHS of \rf{hodxjyr2}}. &&&\qquad\ \ \qed
\end{alignat*}
\renewcommand{\qed}{}
\end{proof}

\section{Proof of Lemma~{\protect\ref{le:ESnewSig}}}\label{ap:proofESnewSig}

\begin{proof}
We write $\newSig$ from \FGref{ESnewSig} into the following $2\times 3$ block form
\begin{align*}
\tgnstretch
\newSig = \left[
\ifbool{ACM}{}{\renewcommand{\arraystretch}{1.5}}
\begin{array}{c|c|c}
\newSig_{1,1} & \,\newSig_{1,2} & \,\newSig_{1,3} \\ \cline{1-3}
\newSig_{2,1} & \,\newSig_{2,2} & \,\newSig_{2,3}
\end{array}
\right].
\end{align*}
We aim to prove the relations between $\newsij{i}{j}$ and $\newd_j-\newc_i$ in each block. 

Recall \rf{newdES} and \rf{newcES}:
\begin{align*}\tgnstretch
\casemod{ll}
{\newd_j =d_j,\,\, \newc_i =c_i &\qquad\text{if $i,j=\oneton$} \\[1ex]
\newd_j =\newc_i =C &\qquad\text{if $i,j=n+\oneton+s$.}
}  
\end{align*}
We prove for $\left[\newSig_{1,1} |\ \newSig_{1,2}\right]$, $\newSig_{1,3}$, 
$\left[\newSig_{2,1} |\ \newSig_{2,2}\right]$, and $\newSig_{2,3}$ in order.

\begin{enumerate}
\item Consider
\begin{align*}
\tgnstretch
\left[
\ifbool{ACM}{}{\renewcommand{\arraystretch}{1.5}}
\begin{array}{c|c}
\newSig_{1,1} & \,\newSig_{1,2} 
\end{array}
\right] = 
\renewcommand{\arraystretch}{2}
\begin{blockarray}{rcccccc}
& x_1\,\,\,\cdots\,\,\, x_{\indxk-1}  & x_{\indxk}  & x_{\indxk+1}\,\,\,\cdots\,\,\, x_s  & x_{s+1}\,\,\,\cdots\,\,\,x_n & \newc_i \\
\begin{block}{r @{\hspace{10pt}}[ccc|c]cc}
\newf_1 &&&&&  c_1     \\
\vdots &   & \HugeLess &  & \HugeLe  & \vdots\\
\newf_n &&&&& c_n \\  
\end{block}
\newd_j & d_1\,\,\,\cdots\,\,\, d_{\indxk-1}  & d_{\indxk}  & d_{\indxk+1}\,\,\,\cdots\,\,\, d_s  & d_{s+1}\,\,\,\cdots\,\,\,d_n 
\end{blockarray}.
\end{align*}

We need to show that, for $i=\rnge{1}{n}$, 
\begin{align}\label{eq:newSig11}
\newsij{i}{j}=\hoder{x_j}{\newf_i}\le 
\casemod{ll}{
\newd_j - \newc_i - 1 \qquad &\text{if $j\in\nzset=\rngeset{1}{s}$,} \\[1ex]
\newd_j - \newc_i \qquad &\text{if $j\in\zset=\rngeset{s+1}{n}$.}
}
\end{align}

We consider the cases 
\begin{enumerate}[(a)]
\item $j\in\nzset$ and  $i\in\eqsetI$,
\item $j\in\zset$ and  $i\in\eqsetI$, and 
\item $j\in  \nzset \cup \zset$ and $i\in\neqsetI$.
\end{enumerate}

Let $\jtwo \in \nzset\backslash\setbg{\indxk} = \setbg{1,\ldots,l-1,l+1,\ldots,s}$. 

\begin{enumerate}[(a)]
\item Consider $j\in \nzset\setminus\{\indxk\}$. By \COref{hodinexp}, the HOD of $x_j$ is $\le d_j-c_i-1$ in every 
\[
\Big( y_r+ \frac{u_r}{u_\indxk} x_\indxk^{(d_\indxk-\cm{u})} \Big)^{(\cm{u}-c_i)}
\]
that replaces $x_r^{(d_r-c_i)}$ in an $f_i$---here $\jtwo \in \nzset\backslash\setbg{\indxk}$, which includes $j$. Therefore, 
\begin{align}
\hoder{x_{j}}{\newf_i} 
&\le d_{j}-c_i-1 \qquad \text{for $j\in\nzset\backslash\setbg{\indxk}, \ i\in \eqsetI$}.  \label{eq:hodxjneqk1}
\end{align}

Now consider $j=\indxk \in\nzset$. We show below that  
  $\partial \newf_i/\partial x_\indxk^{(d_\indxk-c_i)} = 0$, which implies $x_\indxk^{(d_\indxk-c_i)}$ does not appear in $\newf_i$, $i\in\eqsetI$. 
That is,  
\begin{equation}\label{eq:hodxjeqk}
\hoder{x_{\indxk}}{\newf_i}  \le d_{\indxk}-c_i-1 \qquad\text{for $i\in \eqsetI$}.
\end{equation}

Using \rf{ESsuffcond2} gives
\begin{align*}
\hoder{x_\indxk}{\frac{u_r}{u_\indxk}} \le  \hoder{x_\indxk}{u} \le d_\indxk-\cm{\indxk}-1.
\end{align*}
Also
\begin{align*}
\hoder{x_\indxk}{y_{\jtwo}+\frac{u_r}{u_\indxk} x_\indxk^{(d_\indxk-\cm{\indxk})}} 
= \max \left\{\hoder{x_\indxk}{u}
,\hoder{x_\indxk}{x_\indxk^{(d_\indxk-\cm{\indxk})}}\right\} 
= d_\indxk-\cm{\indxk}.
\end{align*}
 
Since $C-c_i\ge 0$ for $i\in \eqsetI$, we apply Griewank's lemma \rf{griewank2}, with $q=C-c_i$, to
\begin{equation}\label{eq:ur/ul}
x_r^{(d_r-\cm{\indxk})}=y_r+ \frac{u_r}{u_\indxk} x_\indxk^{(d_\indxk-\cm{\indxk})}
\end{equation}
from \rf{x1C}. Differentiating both sides of \rf{ur/ul} with respect to $x_\indxk^{(d_\indxk-C)}$ gives
\begin{align}
\frac{u_{\jtwo}}{u_\indxk}
= \pp{x_{\jtwo}^{(d_{\jtwo}-\cm{\indxk})}}{x_\indxk^{(d_\indxk-\cm{\indxk})}}
= \pp{x_{\jtwo}^{(d_{\jtwo}-\cm{\indxk}+\cm{\indxk}-c_i) }}{x_\indxk^{(d_\indxk-\cm{\indxk}+\cm{\indxk}-c_i)}}
= \pp{x_{\jtwo}^{(d_{\jtwo}-c_i)}}{x_\indxk^{(d_\indxk-c_i)}}.
\label{eq:ujuk}
\end{align}

Then
\begin{alignat*}{3}
\pp{\newf_i}{x_\indxk^{(d_\indxk-c_i)}} 
&=  \pp{f_i}{x_\indxk^{(d_\indxk-c_i)}} + \sum_{r\in \nzset\setminus\{\indxk\}}
\pp{f_i}{x_r^{(d_r-c_i)}} \cdot 
\pp{x_r^{(d_r-c_i)}}{x_\indxk^{(d_\indxk-c_i)}}
&\hspace{20mm}&\text{by the chain rule}  \nonumber\\
&= \Jij{i}{\indxk} + \sum_{r\in \nzset\setminus\{\indxk\}}
\Jij{i}{r} \cdot \frac{u_r}{u_\indxk} 
\nonumber &&\text{by \rf{ujuk}} \nonumber\\
&= \frac{1}{u_\indxk} \sum_{r\in \nzset}  \Jij{i}{r} u_r= \frac{1}{u_\indxk}(\~J u)_i = 0 &&\text{because $\~J u=\~0$}.
\end{alignat*}

\item Now  $j\in \zset =\rngeset{s+1}{n}$ and $i\in \eqsetI$. None of the $x_j^{(d_j-c_i)}$ with $j\in\zset$ and $i\in\eqsetI$ is replaced.
By Corollary~\ref{co:hodinexp}, the HOD of $x_j$ is $\le d_j-c_i$ in every 
\[
\Big( y_r+ \frac{u_r}{u_\indxk} x_\indxk^{(d_\indxk-\cm{u})} \Big)^{(\cm{u}-c_i)}
\]
that replaces $x_r^{(d_r-c_i)}$.
Hence
\begin{align}
\hoder{x_{j}}{\newf_i} 
&\le d_{j}-c_i \qquad \text{for $j\notin\nzset, \ i\in \eqsetI$}.  \label{eq:hodxjneqk2}
\end{align}

\item Consider $i\in \neqsetI$. Since $d_j-c_i> \sij{i}{j}$ for every $j\in\nzset$, we do not replace any derivative. Hence,
\begin{equation} \label{eq:hodxjfineqI}
\hoder{x_j}{\newf_i}=\hoder{x_j}{f_i}
\le \casemod{ll}{
 d_j-c_i-1 \qquad &\text{for $i\in\neqsetI$, $j\in\nzset=\rngeset{1}{s}$} \\[1ex]
d_j-c_i \qquad &\text{for $i\in\neqsetI$, $j\in\zset=\rngeset{s+1}{n}$.}
}
\end{equation}
\end{enumerate}

Since $\newd_j = d_j$ and $\newc_i = c_i$ for $i,j=\oneton$, combining \rf{hodxjneqk1,hodxjeqk,hodxjneqk2,hodxjfineqI} proves \rf{newSig11}.

\item For
\begin{align*}
\newSig_{1,3} = 
\renewcommand{\arraystretch}{2}
\begin{blockarray}{rccccc}
& y_1\,\,\,\cdots\,\,\, y_{\indxk-1}  & y_{\indxk}  & y_{\indxk+1}\,\,\,\cdots\,\,\, y_s  &  \newc_i \\
\begin{block}{r @{\hspace{10pt}}[ccc]cc}
\newf_1 &&-\infty &&  c_1     \\
\vdots &  \HugeLe & -\vdots   & \HugeLe  & \vdots\\
\newf_n &&-\infty && c_n \\  
\end{block}
\newd_j & C\,\,\,\cdots\,\,\, C  & C  & \,\,\,\,\,C\,\,\,\,\,\cdots\,\,\,\,\,\,\,\, C  
\end{blockarray},
\end{align*}
we need to show that, for $i=\rnge{1}{n}$,
\begin{align*}
\newsicj{i}{n+j}=\hoder{y_j}{\newf_i} 
\casemod{ll}{
\le \newd_j - \newc_i \qquad &\text{if $j\in\nzset\setminus\{\indxk\}$,} \\[1ex]
= -\infty \qquad &\text{if $j=\indxk$.}
}
\end{align*}
Here, the $(n+j)$th column corresponds to $y_j$.

\begin{enumerate}[(a)]
\item Consider $j\in \nzset\setminus\{\indxk\}$. 
For all $i\in \eqsetI$,
\begin{align*}
\hoder{y_{j}}{\Bigl(y_{j}+\frac{u_{j}}{u_\indxk} x_\indxk^{(d_\indxk-\cm{\indxk})}\Bigr)^{(\cm{\indxk}-c_i)}} 
&=  \hoder{y_{j}}{y_{j}+\frac{u_{j}}{u_\indxk} x_\indxk^{(d_\indxk-\cm{\indxk})}} + (\cm{\indxk}-c_i)\\
&= 0 +(\cm{\indxk}-c_i) = \cm{\indxk}-c_i.
\end{align*}
Then for all $i$,
\[
\newsig_{i,n+j} = \hoder{y_j}{\newf_i} =
\left\{
\begin{aligned}
&C-c_i && \text{if $i\in\eqsetI$ and   } \\
&  & &\ \text{ $x_j^{(d_j-c_i)}$ is replaced by $\Big( y_j+ \frac{u_j}{u_\indxk} x_\indxk^{(d_\indxk-\cm{u})} \Big)^{(\cm{u}-c_i)}$}\\
&-\infty && \text{otherwise.}
\end{aligned}
\right.
\]

To combine these two cases, we can write
\begin{align*}
\newsig_{i,n+j} \le \cm{k}-c_i = \newd_{n+j} - \newc_i
\qquad \text{for $j\in\nzset\backslash\setbg{\indxk}$ and all $i=\rnge{1}{n}$.}
\end{align*}

\item Now consider $j=\indxk$. Since $y_\indxk$ does not appear in any $\newf_i$,
\[
\newsig_{i,n+\indxk}=\hoder{y_\indxk}{\newf_i} = -\infty\qquad\text{for all $i=\oneton$.}
\]
\end{enumerate}

\renewcommand{\jtwo}{i}

\item Consider
\begin{align*}
\tgnstretch
\left[
\ifbool{ACM}{}{\renewcommand{\arraystretch}{1.5}}
\begin{array}{c|c}
\newSig_{2,1} & \,\newSig_{2,2} 
\end{array}
\right] = 
\renewcommand{\arraystretch}{2}
\begin{blockarray}{ccc cccc crc}
&& x_1\,\,\,\cdots\,\,\, x_{\indxk-1}  & x_{\indxk}  & x_{\indxk+1}\,\,\,\cdots\,\,\, x_s  & x_{s+1}\,\,\,\cdots\,\,\,x_n  & \newc_i\\
\begin{block}{cc @{\hspace{10pt}}[lcl|c]cccc}
\newf_{n+1}  &g_1 & \clsp = \qquad \multirow{2}{0.5cm}{\parbox{12pt}{\HugeLess}}   &  =  &\qquad\multirow{3}{0.5cm}{\LargeLess} &  \multirow{2}{0.5cm}{\parbox{12pt}{\HugeLe}} & \cm{\indxk}\\
\vdots & \vdots  & \clsp\phantom{===}\ddots  & \vdots &&& \vdots \\
\newf_{n+\indxk} & g_{\indxk} &   & = &  & -\infty \cdots -\infty & \cm{\indxk} \\
\vdots & \vdots & \qquad\multirow{1}{0.5cm}{\LargeLess} & \vdots  & \clsp\phantom{==}\ddots  & \multirow{2}{0.5cm}{\parbox{12pt}{\HugeLe}} & \vdots \\
\newf_{n+s} & g_s &  & = & \multirow{-3}{0.5cm}{\LargeLess}  \clsp\phantom{===}= & &\cm{\indxk}  \\ 
\end{block}
&\newd_j & d_1\,\,\,\cdots\,\,\, d_{\indxk-1}  & d_{\indxk}  & d_{\indxk+1}\,\,\,\cdots\,\,\, d_s  & d_{s+1}\,\,\,\cdots\,\,\,d_n 
\end{blockarray}.
\end{align*}

Recall $l\in \nzset$. Let row number $\jtwo \in\rngeset{1}{s}$.
We consider the following cases.
\begin{enumerate}[(a)]
\item $\jone=\indxk$ or $\jone=\jtwo$. That is, the entries in the $\indxk$th column in $\newSig_{2,1}$ or those on the (main) diagonal of $\newSig_{2,1}$.
\item $\jone\neq\indxk$ and $\jtwo=\indxk$. That is, the $\indxk$th row in $\left[\newSig_{2,1},\,\newSig_{2,2}\right]$ without the $l$th column.
\item[(c1)] $j=\rnge{1}{s}$ and $j,l,i$ are distinct. This case covers all the entries with `$<$' in $\newSig_{2,1}$.
\item[(c2)] $j=\rnge{s+1}{n}$ and $j,l,i$ are distinct. This case covers all the entries with `$\le$' in $\newSig_{2,2}$.
\end{enumerate}

 We shall prove
\begin{align}\label{eq:newSig2122}
\newsig_{n+\jtwo,\jone}=\hoder{x_\jone}{g_\jtwo} 
\casemod{ll}{
=  \newd_\jone - \newc_{n+\jtwo}   &\qquad\text{if $\jone=\indxk$ or $\jone=\jtwo$} 
\\[1ex]
=  -\infty   &\qquad\text{if $\jone\neq\indxk$ and $\jtwo=\indxk$} 
\\[1ex]
\le \newd_\jone - \newc_{n+\jtwo}-1  &\qquad\text{if $\jone=\rnge{1}{s}$, and $\jone,\,\indxk,\,\jtwo$ are distinct}
\\[1ex]
\le \newd_\jone - \newc_{n+\jtwo}  &\qquad\text{if $\jone=\rnge{s+1}{n}$ and $\jone,\,\indxk,\,\jtwo$ are distinct.}
}
\end{align}

Recall
\begin{equation*}
0 = g_\jtwo =
\left\{
\begin{aligned}
&-y_\jtwo + x_\jtwo^{(d_\jtwo-\cm{u})} - \frac{u_\jtwo}{u_\indxk}x_\indxk^{(d_\indxk-\cm{u})} &&\text{for $\jtwo\in \nzset\backslash\{\indxk\}$} \\
& -y_\indxk + x_\indxk^{(d_\indxk-\cm{u})} 
&& \text{for $\jtwo=\indxk$.}
\end{aligned}
\right.
\end{equation*}
\begin{enumerate}[(a)]
\item Since $x_\indxk^{(d_\indxk-\cm{\indxk})}$ and $x_\jtwo^{(d_\jtwo-\cm{\indxk})}$ (if $\indxk=\jtwo$ then both are the same) occur in $g_\jtwo$,
\begin{equation}\label{eq:newSig21}
\begin{aligned}
\hoder{x_\indxk}{g_\jtwo} &= d_\indxk-\cm{\indxk}= \newd_{\indxk}-\newc_{n+\jtwo}\quad\text{and} \\
\hoder{x_\jtwo}{g_\jtwo} &= d_\jtwo-\cm{\indxk}= \newd_{\jtwo}-\newc_{n+\jtwo}.
\end{aligned}
\end{equation}
\item Now 
\begin{equation}
\hoder{x_{\jone}}{g_{\indxk}} = \hoder{x_{\jone}}{ y_{\indxk}-x_{\indxk}^{(d_{\indxk}-\cm{\indxk})}} = -\infty \le d_\jone - \cm{\indxk}-1 = \newd_\jone - \newc_{n+\indxk}-1.
\end{equation}
\item[]\hspace{-7mm}(c1, c2) Consider the last two cases together: $\jone$, $\indxk$, and $\jtwo$ are distinct. We have
\begin{align*}
\hoder{x_{\jone}}{g_{\jtwo}} 
&=\hoder{x_{\jone}}{ y_{\jtwo}-x_{\jtwo}^{(d_{\jtwo}-\cm{\indxk})}+\frac{u_{\jtwo}}{u_\indxk} x_\indxk^{(d_\indxk-\cm{\indxk})}} \le \hoder{x_{\jone}}{u}.
\end{align*}
Using \rf{ESsuffcond2} gives
\begin{align}
\hoder{x_{\jone}}{g_{\jtwo}} \le  \hoder{x_{\jone}}{u} \le
\casemod{lll}{
d_{\jone}-C-1 &= \newd_{\jone}-\newc_{n+\jtwo}-1  &\qquad\text{if $j\in\nzset$}\\[1ex]
d_{\jone}-C &= \newd_{\jone}-\newc_{n+\jtwo} &\qquad\text{if $j\in\zset$.}
}\label{eq:newSig22}
\end{align}
\end{enumerate}
Combining \rrf{newSig21}{newSig22} gives \rf{newSig2122}.

\medskip

\item For
\begin{align*}
\newSig_{2,3}=
\renewcommand{\arraystretch}{2}
\begin{blockarray}{cccccc}
&&  y_1\,\,\,\cdots\,\,\, y_{\indxk-1}   & y_\indxk   & y_{\indxk+1}\,\,\,\cdots\,\,\, y_s \\
\begin{block}{cc @{\hspace{10pt}}[ccc@{\hspace{1pt}}] @{\hspace{-2pt}}c}
\newf_{n+1} & g_1 &  \mc{1}{l}{\phantom{=}$0$} &&& \,\cm{\indxk}  \\
\vdots &\vdots  &  \phantom{===}\ddots & \mc{2}{c}{\HugeInf} &\,\,\,\,\vdots \\
\newf_{n+\indxk} & g_{\indxk} && 0  && \quad \cm{\indxk}\\
\vdots &\vdots & \mc{2}{c}{\HugeInf}  &\ddots &\,\,\,\,\vdots \\
\newf_{n+s} & g_s &  &   &  \mc{1}{r}{$0$\phantom{=}} &\cm{\indxk}\\ 
\end{block}
&\newd_j & \cm{\indxk}\,\,\,\cdots\,\,\, \cm{\indxk}  & \cm{\indxk}  & \,\,\,\,\,\cm{\indxk}\,\,\,\,\,\cdots\,\,\,\,\,\,\,\, \cm{\indxk}  
\end{blockarray},
\end{align*}
we consider $\jtwo, \jone=\rnge{1}{s}$.
Then 
\begin{align*}
\newsig_{n+\jtwo,n+\jone} = \hoder{y_{\jone}}{g_{\jtwo}} &=\hoder{y_{\jone}}{ y_{\jtwo}-x_{\jtwo}^{(d_{\jtwo}-\cm{\indxk})}+\frac{u_{\jtwo}}{u_\indxk} x_\indxk^{(d_\indxk-\cm{\indxk})}} \\
&= \hoder{y_{\jone}}{y_{\jtwo}} \\
&= 
\casemod{ll}{
0=\cm{\indxk}-\cm{\indxk}=\newd_{n+\jone}-\newc_{n+\jtwo} \qquad\qquad &\text{if $\jtwo=\jone$} \\[1ex]
-\infty \qquad &\text{otherwise}.\qed
}
\end{align*}
\end{enumerate}
\renewcommand{\qed}{}
\end{proof}

\chapter{More  examples}\label{ch:Fixes}
We review how to remedy several structurally singular DAEs from the literature. 
We discuss the index-5 Campbell-Griepentrog Robot Arm DAE in \SCref{robotarm}, the transistor amplifier DAE in \SCref{transamp}, and the ring modulator problem in \SCref{ringmod}. We also show in \SCref{arbitrary} how to ``fix'' the index overestimation problem on Rei\ss ig's family of linear DAEs, for which SA produces a nonsingular system Jacobian but overestimates the index. After we apply a technique similar to the linear combination method, SA reports the correct index $\nu_S=1$ on this family of DAEs.

\section{Robot Arm}\label{sc:robotarm}
We slightly simplify the two-link robot arm problem in \cite{Campbell:1995:SGD:203046.203047} by writing the derivatives of $x_1, x_2$, and $x_3$ implicitly in the equations:
\begin{equation}\label{eq:robotarm}
\begin{aligned}
0 = f_1 =\ & x''_1 - \Bigl[2c(x_3)(x'_1+x'_3)^2 + x'^2_1 d(x_3) \\
&\phantom{x''_1} + (2x_3-x_2)\bigl(a(x_3)+2b(x_3)\bigr) + a(x_3)(u_1-u_2)\Bigr] \\
0 = f_2 =\ & x''_2 -\Bigl[ -2c(x_3)(x'_1+x'_3)^2-x'^2_1 d(x_3)\\
&\phantom{x''_2} +(2x_3-x_2)\bigl(1-3a(x_3)-2b(x_3) \bigr) - a(x_3)u_1+ \bigl( a(x_3)+1\bigr) u_2 \Bigr]\\
0 = f_3 =\ & x''_3 -\Bigl[- 2c(x_3)(x'_1+x'_3)^2-x'^2_1 d(x_3)
 +(2x_3-x_2)\bigl( a(x_3)-9b(x_3)\bigr)\\ 
&\phantom{x''_3} -2x'^2_1 c(x_3) - d(x_3)\bigl( x'_1+x'_3\bigr)^2 -\bigl( a(x_3)+b(x_3) \bigr) (u_1-u_2)\Bigr]\\
0 = f_4 =\ & \cos x_1 + \cos(x_1+x_3) - p_1(t)\\
0 = f_5 =\ & \sin x_1 + \sin(x_1+x_3) - p_2(t),
\end{aligned}
\end{equation}
where
\[
\begin{alignedat}{9}\tgnstretch
p_1(t) &= \cos(1-e^t)+\cos(1-t)  &\qquad\qquad&  &p_2(t)&= \sin(1-e^t)+\sin(1-t) \\
a(s) &= 2/(2-\cos^2 s) &&  &b(s)&= \cos s/(2-\cos^2 s)\\
c(s) &=\sin s/(2-\cos^2 s) && &d(s)&= \sin s\cos s/(2-\cos^2 s).\\
\end{alignedat}
\]

\begin{align*}
\tgnstretch
\hspace{-8mm}
\Sigma = \begin{blockarray}{rcccccll}
 &x_{1} &x_{2} &x_{3} &u_{1} &u_{2} & \s{c_i} \\
\begin{block}{r @{\hspace{10pt}}[ccccc]lc}
 f_{1}&2&\OK{0}&\OK{1}&0^\bullet&0&\s{0}\\ 
 f_{2}&\OK{1}&2^\bullet&\OK{1}&0&0&\s{0}\\ 
 f_{3}&\OK{1}&\OK{0}&2&0&0^\bullet&\s{0}\\ 
 f_{4}&0&&0^\bullet&&&\s{2}\\ 
 f_{5}&0^\bullet&&0&&&\s{2}\\ 
\end{block}
\s{d_j} &\s{2}&\s{2}&\s{2}&\s{0}&\s{0} &\valSig{\Sigma}{2} \\
 \end{blockarray}
\hspace{-3mm}
\Jac = \begin{blockarray}{rccccccc}
 & x_{1} & x_{2} & x_{3} & u_{1} & u_{2} \\
\begin{block}{r @{\hspace{10pt}}[ccccc]cc}
 f_{1}&1&&&-a_3&a_3\\ 
 f_{2}&&1&&a_3&-1-a_3\\ 
 f_{3}&&&1&a_3+b_3&-a_3-b_3\\ 
 f_{4}& \pp{f_4}{x_1}&& \pp{f_4}{x_3} &&\\ 
 f_{5}&\pp{f_5}{x_1}&&\pp{f_5}{x_3}&&\\ 
\end{block}
&\mcdetJac{5}{\detJac{\Jac}{0}}
\end{blockarray}
\end{align*}

Here
\begin{align}\label{eq:robotarmformulae}
\begin{alignedat}{5}
\ppin{f_4}{x_1} &= - \sin x_{1} - \sin(x_{1} + x_{3})     &\qquad\qquad& &a_3&=a(x_3)=2/(2-\cos^2 x_3)
\\ \ppin{f_4}{x_3} &= - \sin(x_{1} + x_{3})                      && &b_3&=b(x_3)=\cos x_3/(2-\cos^2 x_3)
\\ \ppin{f_5}{x_1} &= \cos x_{1} + \cos(x_{1} + x_{3})
\\ \ppin{f_5}{x_3} &= \cos(x_{1} + x_{3}).
\end{alignedat}
\end{align}

SA reports structural index 3, while the d-index is 5. 
\subsection{ES method}
Pryce\cite{Pryce98} fixes this failure by introducing a new variable $w$ and substituting it for $u_1-u_2$ in $f_1$ and $f_3$. These two equations become $\newf_1$ and $\newf_3$. We append $g=w-(u_1-u_2)$ that prescribes these substitutions and obtain the converted DAE:
\begin{equation}\label{eq:robotarmES}
\begin{aligned}
0 = \newf_1 =& x''_1 - \Bigl[2c(x_3)(x'_1+x'_3)^2 + x'^2_1 d(x_3) \\
&\phantom{x''_1} + (2x_3-x_2)\bigl(a(x_3)+2b(x_3)\bigr) + a(x_3)w\Bigr] \\
0 = f_2 =& x''_2 -\Bigl[ -2c(x_3)(x'_1+x'_3)^2-x'^2_1 d(x_3)\\
&\phantom{x''_2} +(2x_3-x_2)\bigl(1-3a(x_3)-2b(x_3) \bigr) - a(x_3)u_1+ \bigl( a(x_3)+1\bigr) u_2 \Bigr]\\
0 = \newf_3 =& x''_3 -\Bigl[- 2c(x_3)(x'_1+x'_3)^2-x'^2_1 d(x_3)
 +(2x_3-x_2)\bigl( a(x_3)-9b(x_3)\bigr)\\ 
&\phantom{x''_3} -2x'^2_1 c(x_3) - d(x_3)\bigl( x'_1+x'_3\bigr)^2 -\bigl( a(x_3)+b(x_3) \bigr) w\Bigr]\\
0 = f_4 =& \cos x_1 + \cos(x_1+x_3) - p_1(t)\\
0 = f_5 =& \sin x_1 + \sin(x_1+x_3) - p_2(t) \\
0 = g =& w - (u_1-u_2).
\end{aligned}
\end{equation}

\begin{align*}
\hspace{20mm}\newSig &= 
\begin{alignedat}{3}
\tgnstretch
\begin{blockarray}{rccccccll}
 &x_{1} &x_{2} &x_{3} &u_{1} &u_{2} &w & \s{c_i} \\
\begin{block}{r @{\hspace{10pt}}[cccccc]ll}
 \newf_{1}&2&0^\bullet&\OK{1}&&&0&\s{2}\\ 
 f_{2}&\OK{1}&2&\OK{1}&0^\bullet&0&&\s{0}\\ 
 \newf_{3}&\OK{1}&0&2&&&0^\bullet&\s{2}\\ 
 f_{4}&0&&0^\bullet&&&&\s{4}\\ 
 f_{5}&0^\bullet&&0&&&&\s{4}\\ 
g&&&&0&0^\bullet&\OK{0}&\s{0}\\ 
\end{block}
\s{d_j} &\s{4}&\s{2}&\s{4}&\s{0}&\s{0}&\s{2} &\valSig{\newSig}{0} \\
 \end{blockarray}
 \end{alignedat}\\
 \newJ &= 
\begin{alignedat}{3}
\tgnstretch
\begin{blockarray}{rccccccll}
 & x_{1} & x_{2} & x_{3} & u_1 & u_2 & w\\
\begin{block}{r @{\hspace{10pt}}[cccccc]ll}
 \newf_{1}&1&a_3+2b_3&&&&-a_3\\ 
 f_{2}&&1&&a_3&-a_3-1&\\ 
 \newf_{3}&&a_3-9b_3&1&&&a_3+b_3\\ 
 f_{4}& \pp{f_4}{x_1} &&\pp{f_4}{x_3}&&&\\ 
 f_{5}&\pp{f_5}{x_1}&&\pp{f_5}{x_3}&&&\\ 
g&&&&-1&1&\\ 
\end{block}
&\mcdetJac{6}{\detJac{\newJ}{-2\sin x_3(a_3^2-3a_3b_3+b_3^2)}}
\end{blockarray}
\end{alignedat}
\end{align*}
$\newJ$ is not identically singular; refer to \rf{robotarmformulae} for the entries in it. SA reports the correct index 5, and succeeds if $\det(\newJ)\neq 0$.

\subsection{LC method}
We replace $f_3$ by
\begin{align*}
\newf_3 &= f_1+ \frac{a_3}{a_3+b_3}f_3
\\ &= x''_1 - \Bigl[2c(x_3)(x'_1+x'_3)^2 + x'^2_1 d(x_3)  + (2x_3-x_2)\bigl(a_3+2b_3\bigr) + \uuline{a_3(u_1-u_2)}\Bigr]
\\ &\phantom{=x''_1\ }  
+ \frac{a_3}{a_3+b_3} x''_3 - \uuline{\frac{a_3}{a_3+b_3}}\Bigl[- 2c(x_3)(x'_1+x'_3)^2-x'^2_1 d(x_3)
 +(2x_3-x_2)\bigl( a_3-9b_3\bigr)
\\ &\phantom{=x''_1\ }   
-2x'^2_1 c(x_3) - d(x_3)\bigl( x'_1+x'_3\bigr)^2 \uuline{-\bigl( a_3+b_3 \bigr) (u_1-u_2)}\Bigr]
\\ &= x''_1 - \Bigl[2c(x_3)(x'_1+x'_3)^2 + x'^2_1 d(x_3)  + (2x_3-x_2)\bigl(a_3+2b_3\bigr) \Bigr] + \frac{a_3}{a_3+b_3} x''_3 
\\ &\phantom{=x''_1\ }   
 -\frac{a_3}{a_3+b_3}\Bigl[- 2c(x_3)(x'_1+x'_3)^2-x'^2_1 d(x_3)
 +(2x_3-x_2)\bigl( a_3-9b_3\bigr)
\\ &\phantom{=x''_1\ }   
-2x'^2_1 c(x_3) - d(x_3)\bigl( x'_1+x'_3\bigr)^2 \Bigr]
\end{align*}
(the underlined terms cancel out).
\begin{align*}
\tgnstretch
\hspace{-8mm}
\newSig = \begin{blockarray}{rcccccll}
 &x_{1} &x_{2} &x_{3} &u_{1} &u_{2} & \s{c_i} \\
\begin{block}{r @{\hspace{10pt}}[ccccc]lc}
 f_{1}&\OK{2}&\OK{0}&\OK{1}&0^\bullet&0&\s{0}\\ 
 f_{2}&\OK{1}&2^\bullet&\OK{1}&0&0^\bullet&\s{0}\\ 
 \newf_{3}&2&0^\bullet&2&&&\s{2}\\ 
 f_{4}&0^\bullet&&0&&&\s{4}\\ 
 f_{5}&0&&0^\bullet&&&\s{4}\\ 
\end{block}
\s{d_j} &\s{4}&\s{2}&\s{4}&\s{0}&\s{0} &\valSig{\newSig}{0} \\
 \end{blockarray}
\hspace{-3mm}
\newJ = \begin{blockarray}{rccccccc}
 & x_{1} & x_{2} & x_{3} & u_{1} & u_{2} \\
\begin{block}{r @{\hspace{10pt}}[ccccc]cc}
 f_{1}&&&&-a_3&a_3\\ 
 f_{2}&&1&&a_3&-1-a_3\\ 
 \newf_{3}&1&\pp{\newf_3}{x_2}&\frac{a_3}{a_3+b_3}&&\\ 
 f_{4}& \pp{f_4}{x_1}&& \pp{f_4}{x_3} &&\\ 
 f_{5}&\pp{f_5}{x_1}&&\pp{f_5}{x_3}&&\\ 
\end{block}
\phantom{d_j}
\end{blockarray}
\end{align*}
Here
\begin{align*}
\pp{\newf_3}{x_2} &= a_3+2b_3+\frac{a_3}{a_3+b_3}(a_3-9b_3), \\[1ex]
\det(\newJ) &= -2\sin x_3(a_3^2-3a_3b_3+b_3^2)a_3/(a_3+b_3).
\end{align*}
Refer to \rf{robotarmformulae} for the other entries in $\newJ$.
Since
\[
\frac{a_3}{a_3+b_3} = \frac{2}{2+\cos x_3} \neq 0 \quad \text{for all $x_3\in\bbR$,}
\]
the converted DAE is always equivalent to \rf{robotarm}. $\newJ$ is not identically singular. SA reports index 5, and succeeds if $\det(\newJ)\neq 0$.

\section{Transistor amplifier}\label{sc:transamp}

Below is a transistor amplifier problem originated from electrical circuit analysis\cite{TestSetIVP}. It is classified in \cite{TestSetIVP} as a stiff index-1 DAE consisting of 8 equations.
\begin{equation}\label{eq:transamp}
\begin{alignedat}{7}
0&=f_1&&=&&C_1(x'_1-x'_2) + \frac{x_1-U_e(t)}{R_0} \\
0&=f_2&&=-&&C_1(x'_1-x'_2) - \frac{U_b}{R_2}+x_2\left(\frac{1}{R_1}+\frac{1}{R_2}\right) -(\alpha-1)g(x_2-x_3)\\
0&=f_3&&= &&C_2x'_3 - g(x_2-x_3) + \frac{x_3}{R_3}\\
0&=f_4&&= &&C_3(x'_4-x'_5)+\frac{x_4-U_b}{R_4}+\alpha g(x_2-x_3)\\
0&=f_5&&= -&&C_3(x'_4-x'_5)- \frac{U_b}{R_5}+x_5\left(\frac{1}{R_5}+\frac{1}{R_6}\right) -(\alpha-1)g(x_5-x_6)\\
0&=f_6&&= &&C_4x'_6 - g(x_5-x_6) + \frac{x_6}{R_7}\\
0&=f_7&&= &&C_5(x'_7-x'_8)+\frac{x_7-U_b}{R_8}+\alpha g(x_5-x_6)\\
0&=f_8&&= -&&C_5(x'_7-x'_8)+\frac{x_8}{R_9},
\end{alignedat}
\end{equation}
where 
\[
\begin{alignedat}{9}
g(y) &= \beta \bigl( \exp(y/U_F)-1\bigr) &\qquad\qquad& &U_e(t) &= 0.1\sin (200\pi t)
\\ U_b &=6.0                &&  &R_0 &= 1000
\\ U_F &=0.026         && &R_k&=9000 &\quad& \text{for $k=1,\ldots,9$}
\\ \alpha&=0.99         && &C_k&=k\times 10^{-6} &\quad& \text{for $k=1,\ldots,5$}
\\ \beta&=10^{-6}.
\end{alignedat}
\]

\begin{align*}
\hspace{20mm}\Sigma &=
\begin{alignedat}{3}
\tgnstretch
 \begin{blockarray}{rcccccccccc}
 &x_{1} &x_{2} &x_{3} &x_{4} &x_{5} &x_{6} &x_{7} &x_{8} & \s{c_i} \\
\begin{block}{r @{\hspace{10pt}}[cccccccc]cc}
 f_{1}&1&1^\bullet&&&&&&&\s{0}\\ 
 f_{2}&1^\bullet&1&\OK{0}&&&&&&\s{0}\\ 
 f_{3}&&\OK{0}&1^\bullet&&&&&&\s{0}\\ 
 f_{4}&&\OK{0}&\OK{0}&1&1^\bullet&&&&\s{0}\\ 
 f_{5}&&&&1^\bullet&1&\OK{0}&&&\s{0}\\ 
 f_{6}&&&&&\OK{0}&1^\bullet&&&\s{0}\\ 
 f_{7}&&&&&\OK{0}&\OK{0}&1&1^\bullet&\s{0}\\ 
 f_{8}&&&&&&&1^\bullet&1&\s{0}\\ 
\end{block}
\s{d_j} &\s{1}&\s{1}&\s{1}&\s{1}&\s{1}&\s{1}&\s{1}&\s{1}&\mc{2}{c}{$\valSig{\Sig}{8}$} \\
 \end{blockarray}
\end{alignedat} \\
\Jac &=
\begin{alignedat}{3}
\tgnstretch
 \begin{blockarray}{rcccccccccc}
 & x_{1} & x_{2} & x_{3} & x_{4} & x_{5} & x_{6} & x_{7} & x_{8} \\
\begin{block}{r @{\hspace{10pt}}[cccccccc]cc}
 f_{1}&C_1&-C_1&&&&&&\\ 
 f_{2}&-C_1&C_1&&&&&&\\ 
 f_{3}&&&C_2&&&&&\\ 
 f_{4}&&&&C_3&-C_3&&&\\ 
 f_{5}&&&&-C_3&C_3&&&\\ 
 f_{6}&&&&&&C_4&&\\ 
 f_{7}&&&&&&&C_5&-C_5\\ 
 f_{8}&&&&&&&-C_5&C_5\\ 
\end{block}
\\[-3ex]
\mcdetJac{10}{\detJac{\Jac}{0}}
\end{blockarray}.
\end{alignedat}
\end{align*}

SA reports index 1, but produces an identically singular $\Jac$. Observing its structure, we
\[
  \tgnstretch
  \begin{array}{ll}
\text{replace} & \text{by}   \\ \hline
f_1 & \newf_1 = f_1+f_2 \\
f_4 & \newf_4 = f_4+f_5 \\
f_7 & \newf_7 = f_7+f_8.
  \end{array}
\]

The new equations in the converted DAE are
\begin{equation*}
\begin{aligned}
0=\newf_1&= \frac{x_1-U_e(t)}{R_0}- \frac{U_b}{R_2}+x_2\left(\frac{1}{R_1}+\frac{1}{R_2}\right) -(\alpha-1)g(x_2-x_3) \\
0=\newf_4&= \frac{x_4-U_b}{R_4}+\alpha g(x_2-x_3)- \frac{U_b}{R_5}+x_5\left(\frac{1}{R_5}+\frac{1}{R_6}\right) -(\alpha-1)g(x_5-x_6)\\
0=\newf_7&= \frac{x_7-U_b}{R_8}+\alpha g(x_5-x_6)+\frac{x_8}{R_9}.
\end{aligned}
\end{equation*}

\begin{align*}
\hspace{20mm}\newSig &=
\tgnstretch
\begin{alignedat}{3}
 \begin{blockarray}{rccccccccll}
 &x_{1} &x_{2} &x_{3} &x_{4} &x_{5} &x_{6} &x_{7} &x_{8} & \s{c_i} \\
\begin{block}{r @{\hspace{10pt}}[cccccccc]ll}
 \newf_{1}&0^\bullet&0&0&&&&&&\s{1}\\ 
 f_{2}&1&1^\bullet&\OK{0}&&&&&&\s{0}\\ 
 f_{3}&&\OK{0}&1^\bullet&&&&&&\s{0}\\ 
 \newf_{4}&&0&0&0^\bullet&0&0&&&\s{1}\\ 
 f_{5}&&&&1&1^\bullet&\OK{0}&&&\s{0}\\ 
 f_{6}&&&&&\OK{0}&1^\bullet&&&\s{0}\\ 
 \newf_{7}&&&&&0&0&0^\bullet&0&\s{1}\\ 
 f_{8}&&&&&&&1&1^\bullet&\s{0}\\ 
\end{block}
\s{d_j} &\s{1}&\s{1}&\s{1}&\s{1}&\s{1}&\s{1}&\s{1}&\s{1}&\valSig{\newSig}{5} \\
 \end{blockarray}
\end{alignedat} \\
\newJ &= 
\tgnstretch
\begin{alignedat}{3}
\begin{blockarray}{rcccccccccc}
 & x_{1} & x_{2} & x_{3} & x_{4} & x_{5} & x_{6} & x_{7} & x_{8} \\
\begin{block}{r @{\hspace{10pt}}[cccccccc]cc}
 \newf_{1}&R_0^{-1}&\pp{f_1}{x_2}&\pp{f_1}{x_3}&&&&&\\ 
 f_{2}&-C_1&C_1&&&&&&\\ 
 f_{3}&&&C_2&&&&&\\ 
 \newf_{4}&&\pp{f_4}{f_2} &\pp{f_4}{x_3}&R_4^{-1}&\pp{f_4}{x_5}&\pp{f_4}{x_6}&&\\ 
 f_{5}&&&&-C_3&C_3&&&\\ 
 f_{6}&&&&&&C_4&&\\ 
 \newf_{7}&&&&&\pp{f_7}{x_5}&\pp{f_7}{x_6}&R_8^{-1}&R_9^{-1}\\ 
 f_{8}&&&&&&&-C_5&C_5\\ 
\end{block}
\\[-3ex]
\mcdetJac{10}{\detJac{\newJ}{C_1C_2C_3C_4C_5
\left(R_0^{-1}+\pp{f_1}{x_2}\right)
\left(R_4^{-1}+\pp{f_4}{x_5}\right)
\left(R_8^{-1}+R_9^{-1}\right)
}}
\end{blockarray}
\end{alignedat}
\end{align*}

Note that only two partial derivatives shown in $\newJ$ contribute to $\det(\newJ)$:
\begin{align*}
\ppin{f_4}{x_5}&=R_5^{-1}+R_6^{-1}
\\ \ppin{f_1}{x_2}&=R_1^{-1}+R_2^{-1}.
\end{align*}
SA still reports index 1. Since $\newJ$ is not identically singular, SA succeeds if $\det(\newJ)\neq 0$.

\section{Ring modulator}\label{sc:ringmod}
Following is a ring modulator problem originated from electrical circuit analysis \cite{TestSetIVP}. By setting $C_s=0$ in the original problem formulation, we obtain an index-2 DAE consisting of 11 differential and 4 algebraic equations:
\begin{equation}\label{eq:ringmod}
\begin{alignedat}{7}
0&= f_1 &&= -&&y'_1 + C^{-1}\bigl( y_8-0.5y_{10}+0.5y_{11}+y_{14}-R^{-1}y_1 \bigr)\\
0&= f_2 &&= -&&y'_2 + C^{-1}\bigl( y_9-0.5y_{11}+0.5y_{12}+y_{15}-R^{-1}y_2\bigr)\\
0&= f_3 &&= &&y_{10}-q(U_{D1})+q(U_{D4})\\
0&= f_4 &&= -&&y_{11}+q(U_{D2})-q(U_{D3}) \\
0&= f_5 &&= &&y_{12}+q(U_{D1})-q(U_{D3})\\
0&= f_6 &&= -&&y_{13}-q(U_{D2})+q(U_{D4})\\
0&= f_7 &&= -&&y'_7 + C_p^{-1}\bigl(-R_p^{-1}y_7+q(U_{D1})+q(U_{D2})-q(U_{D3})-q(U_{D4}) \bigr)\\
0&= f_8 &&= -&&y'_8 + -L_h^{-1}y_1\\
0&= f_9 &&= -&&y'_9 + -L_h^{-1}y_2\\
0&= f_{10} &&= -&&y'_{10} + L_{s2}^{-1}(0.5y_1-y_3-R_{g2}y_{10}) \\
0&= f_{11} &&= -&&y'_{11} + L_{s3}^{-1}(-0.5y_1+y_4-R_{g3}y_{11}) \\
0&= f_{12} &&= -&&y'_{12} + L_{s2}^{-1}(0.5y_2-y_5-R_{g2}y_{12})\\ 
0&= f_{13} &&= -&&y'_{13} + L_{s3}^{-1}(-0.5y_2+y_6-R_{g3}y_{13})\\
0&= f_{14} &&= -&&y'_{14} + L_{s1}^{-1}(-y_1+U_{in1}(t)-(R_i+R_{g1})y_{14})\\
0&= f_{15} &&= -&&y'_{15} + L_{s1}^{-1}(-y_2-(R_c+R_{g1})y_{15}),
\end{alignedat}
\end{equation}
where
\[
\tgnstretch
\begin{alignedat}{5}
U_{D1} &= &&y_3-y_5-y_7-U_{in2}(t)
&\qquad\qquad&  &q(U)&= \gamma (e^{\delta U}-1)
\\
U_{D2} &= -&&y_4+y_6-y_7-U_{in2}(t)
&& &U_{in1}(t)&= 0.5 \sin (2000\pi t)
\\
U_{D3} &= &&y_4+y_5+y_7+U_{in2}(t)
&&  &U_{in2}(t)&= 2 \sin (20000\pi t)
\\
U_{D4} &= -&&y_3-y_6+y_7+U_{in2}(t).
\end{alignedat}
\]
The parameters are
\begin{alignat*}{11}
C&=1.6\times 10^{-8} &\qquad\qquad\qquad R_{g1}&=36.3
\\ C_p&=10^{-8}         & R_{g2}&=17.3
\\ R&=25\times 10^3  & R_{g3} &=17.3
\\ R_p&=50                & R_i &= 5\times 10
\\ L_h&=4.45              & R_c&=6\times 10^2
\\ L_{s1}&=2\times 10^{-3}  & \gamma&=40.67286402\times 10^{-9}
\\ L_{s2}&=5\times 10^{-4}  & \delta&=17.7493332
\\ L_{s3}&=5\times 10^{-4}.
\end{alignat*}

SA reports index 1 and produces the following $\Sigma$ with $\val\Sig=11$.
{\small
\begin{align*}
\tgnstretch
\Sigma = \begin{blockarray}{rccccccccccccccccc}
 &y_{1} &y_{2} &y_{3} &y_{4} &y_{5} &y_{6} &y_{7} &y_{8} &y_{9} &y_{10} &y_{11} &y_{12} &y_{13} &y_{14} &y_{15} & \s{c_i} \\
\begin{block}{r @{\hspace{10pt}}[ccccccccccccccc]cc}
 f_{1}&1^\bullet&&&&&&&\OK{0}&&\OK{0}&\OK{0}&&&\OK{0}&&\s{0}\\ 
 f_{2}&&1^\bullet&&&&&&&\OK{0}&&&\OK{0}&\OK{0}&&\OK{0}&\s{0}\\ 
 f_{3}&&&0&&0&0^\bullet&\OK{0}&&&\OK{0}&&&&&&\s{0}\\ 
 f_{4}&&&&0&0^\bullet&0&\OK{0}&&&&\OK{0}&&&&&\s{0}\\ 
 f_{5}&&&0^\bullet&0&0&&\OK{0}&&&&&\OK{0}&&&&\s{0}\\ 
 f_{6}&&&0&0^\bullet&&0&\OK{0}&&&&&&\OK{0}&&&\s{0}\\ 
 f_{7}&&&0&0&0&0&1^\bullet&&&&&&&&&\s{0}\\ 
 f_{8}&\OK{0}&&&&&&&1^\bullet&&&&&&&&\s{0}\\ 
 f_{9}&&\OK{0}&&&&&&&1^\bullet&&&&&&&\s{0}\\ 
 f_{10}&\OK{0}&&0&&&&&&&1^\bullet&&&&&&\s{0}\\ 
 f_{11}&\OK{0}&&&0&&&&&&&1^\bullet&&&&&\s{0}\\ 
 f_{12}&&\OK{0}&&&0&&&&&&&1^\bullet&&&&\s{0}\\ 
 f_{13}&&\OK{0}&&&&0&&&&&&&1^\bullet&&&\s{0}\\ 
 f_{14}&\OK{0}&&&&&&&&&&&&&1^\bullet&&\s{0}\\ 
 f_{15}&&\OK{0}&&&&&&&&&&&&&1^\bullet&\s{0}\\ 
\end{block}
\s{d_j} &\s{1}&\s{1}&\s{0}&\s{0}&\s{0}&\s{0}&\s{1}&\s{1}&\s{1}&\s{1}&\s{1}&\s{1}&\s{1}&\s{1}&\s{1} \\
 \end{blockarray}
\end{align*}
}
We do not present $\Jac$ here. The entries that contribute to its determinant are positions $(i,i)$ for $i=1,2,\rnge{7}{15}$, where $\ppin{f_i}{y'_i}=-1$, and the submatrix
\begin{align*}
\Jac(3:6,3:6) = 
\tgnstretch
\begin{blockarray}{rcccccc}
 & y_{3} &y_{4} &y_{5} &y_{6}  \\
\begin{block}{r @{\hspace{10pt}}[cccc]cc}
f_3 &-s_1-s_4 &  & s_1 & -s_4 \\
f_4 & &-s_2-s_3 & -s_3 & s_2\\
f_5 & s_1 & -s_3&-s_1-s_3&\\
f_6 & -s_4 & s_2 & & -s_2-s_4\\
\end{block}
\end{blockarray},\quad \text{where $s_i = \gamma\delta e^{\delta U_{Di}}$}.
\end{align*}
Since this submatrix of $\Jac$ is identically singular, so is $\Jac$.

To remedy this DAE, we replace one equation from $f_3, f_4, f_5, f_6$  by
\[
\newf = f_3-f_4+f_5-f_6 = y_{10} + y_{11} + y_{12} + y_{13}.
\]
Consider the case $f_3$. SA produces $\newSig$ with $\val\newSig=10$:
{\small
\begin{align*}
\tgnstretch
\newSig = \begin{blockarray}{rccccccccccccccccc}
 &y_{1} &y_{2} &y_{3} &y_{4} &y_{5} &y_{6} &y_{7} &y_{8} &y_{9} &y_{10} &y_{11} &y_{12} &y_{13} &y_{14} &y_{15} & \s{c_i} \\
\begin{block}{r @{\hspace{10pt}}[ccccccccccccccc]cc}
 f_{1}&1^\bullet&&&&&&&\OK{0}&&\OK{0}&\OK{0}&&&\OK{0}&&\s{0}\\ 
 f_{2}&&1^\bullet&&&&&&&\OK{0}&&&\OK{0}&\OK{0}&&\OK{0}&\s{0}\\ 
 \newf_{3}&&&&&&&&&&0&0&0&0^\bullet&&&\s{1}\\ 
 f_{4}&&&&0^\bullet&0&0&\OK{0}&&&&\OK{0}&&&&&\s{0}\\ 
 f_{5}&&&0&0&0^\bullet&&\OK{0}&&&&&\OK{0}&&&&\s{0}\\ 
 f_{6}&&&0^\bullet&0&&0&\OK{0}&&&&&&\OK{0}&&&\s{0}\\ 
 f_{7}&&&0&0&0&0&1^\bullet&&&&&&&&&\s{0}\\ 
 f_{8}&\OK{0}&&&&&&&1^\bullet&&&&&&&&\s{0}\\ 
 f_{9}&&\OK{0}&&&&&&&1^\bullet&&&&&&&\s{0}\\ 
 f_{10}&\OK{0}&&0&&&&&&&1^\bullet&&&&&&\s{0}\\ 
 f_{11}&\OK{0}&&&0&&&&&&&1^\bullet&&&&&\s{0}\\ 
 f_{12}&&\OK{0}&&&0&&&&&&&1^\bullet&&&&\s{0}\\ 
 f_{13}&&\OK{0}&&&&0^\bullet&&&&&&&1&&&\s{0}\\ 
 f_{14}&\OK{0}&&&&&&&&&&&&&1^\bullet&&\s{0}\\ 
 f_{15}&&\OK{0}&&&&&&&&&&&&&1^\bullet&\s{0}\\ 
\end{block}
\s{d_j} &\s{1}&\s{1}&\s{0}&\s{0}&\s{0}&\s{0}&\s{1}&\s{1}&\s{1}&\s{1}&\s{1}&\s{1}&\s{1}&\s{1}&\s{1} \\
\end{blockarray}
\end{align*}
}

The consistent initial values for solving this converted DAE are
\begin{alignat*}{3}
y_i &=0 &\qquad\qquad&\text{for $i=\rnge{1}{15}$,}\\
y'_i &=0 &&\text{for $i=1,2,\rnge{7}{15}$,}
\end{alignat*}
which satisfy $f_i$ for all $i$ and $f'_3$. At this consistent point, $\det(\Jac)=-1.2040\times 10^{-14}$. Hence SA succeeds and the DAE is index-2.

\section{An index-overestimated DAE}\label{sc:arbitrary}
Rei\ss ig et al.\cite{Reissig1999a} construct a family of linear constant coefficient DAEs
\begin{equation}\label{eq:arbitrary}
Ax'(t) + Bx(t) - q(t) = 0
\end{equation}
for which SA finds an arbitrarily high structural index $\nu_S>1$, though the d-index is 1.
Here $B$ is an $n\times n$ identity matrix, where $n=2k+1$ and $k\ge 1$; $A$ has the form
\[
A=
\left(\begin{array}{ccccccccc}
0&1&1\\[.5ex] 
&1&1&0  \\[.5ex] 
&&0&1&1\\[.5ex] 
&&&1&1&0  \\[.5ex] 
&&&&0&\ddots&\ddots \\[.5ex] 
&&&&&\ddots&\ddots&0  \\[.5ex] 
&&&&&&0&1&1\\[.5ex] 
&&&&&&&1&1  \\[.5ex] 
&&&&&&&&0
\end{array}.
\right)
\]
That is, for $i=\rnge{1}{k}$, equations $f_{2i-1}$ and $f_{2i}$ have the common expression $x'_{2i}+x'_{2i+1}$.

\medskip

Pryce \cite{Pryce2001a} applies the \Sigmeth on \rf{arbitrary} with $n=5$ and $k=2$:
\begin{equation}\label{eq:arbitraryk=2}
\begin{alignedat}{3}
0 &= f_1 &&= x'_2+x'_3+&&x_1-q_1(t) \\
0 &= f_2 &&= x'_2+x'_3+&&x_2-q_2(t) \\
0 &= f_3 &&= x'_4+x'_5+&&x_3-q_3(t) \\
0 &= f_4 &&= x'_4+x'_5+&&x_4-q_4(t) \\
0 &= f_5 &&= &&x_5-q_5(t).
\end{alignedat}
\end{equation}
\begin{align*}
\Sigma =
\renewcommand{\arraystretch}{1.5}
\begin{alignedat}{3}
\begin{blockarray}{rcccccll}
 &x_{1} &x_{2} &x_{3} &x_{4} &x_{5} & \s{c_i} \\
\begin{block}{r @{\hspace{10pt}}[ccccc]ll}
 f_{1}&0^\bullet&1&1&&&\s{0}\\ 
 f_{2}&&1^\bullet&1&&&\s{0}\\ 
 f_{3}&&&0^\bullet&1&1&\s{1}\\ 
 f_{4}&&&&1^\bullet&1&\s{1}\\ 
 f_{5}&&&&&0^\bullet&\s{2}\\ 
\end{block}
\s{d_j} &\s{0}&\s{1}&\s{1}&\s{2}&\s{2} &\valSig{\Sigma}{2}\\
\end{blockarray}
\end{alignedat}
\Jac =
\renewcommand{\arraystretch}{1.5}
\begin{alignedat}{3} 
\begin{blockarray}{rccccccc}
 & x_{1} & x_{2} & x_{3} & x_{4} & x_{5} \\
\begin{block}{r @{\hspace{10pt}}[ccccc]cc}
 f_{1}&1&1&1&&\\ 
 f_{2}&&1&1&&\\ 
 f_{3}&&&1&1&1\\ 
 f_{4}&&&&1&1\\ 
 f_{5}&&&&&1\\ 
\end{block}
&\mcdetJac{5}{\detJac{\Jac}{1}}
\end{blockarray}
\end{alignedat}
\end{align*}
The method succeeds with $\det(\Jac)=1$ but reports $\nu_S=3$ different from $\nu_d=1$; this occurs in Pantelides's SA as well. 
We illustrate below how to fix this index overestimation problem for \rf{arbitraryk=2}.

Observing the structure of $A$, we
\[
  \tgnstretch
  \begin{array}{ll}
\text{replace} & \text{by}   \\ \hline
f_1 & \newf_1 = f_1-f_2 \\
f_3 & \newf_3 = f_3-f_4.
  \end{array}
\]
The converted DAE is
\begin{equation}\label{eq:arbitraryconv2}
\begin{alignedat}{3}
0 &= \newf_1 &&= x_1-x_2-q_1(t)+q_2(t) \\
0 &= f_2 &&= x'_2+x'_3+x_2-q_2(t) \\
0 &= \newf_3 &&= x_3-x_4-q_3(t)+q_4(t) \\
0 &= f_4 &&= x'_4+x'_5+x_4-q_4(t) \\
0 &= f_5 &&= x_5-q_5(t).
\end{alignedat}
\end{equation}

\begin{align*}
\newSig =
\renewcommand{\arraystretch}{1.5}
\begin{alignedat}{3}
\begin{blockarray}{rcccccll}
 &x_{1} &x_{2} &x_{3} &x_{4} &x_{5} & \s{c_i} \\
\begin{block}{r @{\hspace{10pt}}[ccccc]ll}
 \newf_{1}&0^\bullet&0&&&&\s1\\ 
 f_{2}&&1^\bullet&1&&&\s{0}\\ 
 \newf_{3}&&&0^\bullet&0&&\s{1}\\ 
 f_{4}&&&&1^\bullet&1&\s{0}\\ 
 f_{5}&&&&&0^\bullet&\s{1}\\ 
\end{block}
\s{d_j} &\s1&\s{1}&\s{1}&\s{1}&\s{1}&\valSig{\newSig}{2}\\
\end{blockarray}
\end{alignedat}
\newJ =
\renewcommand{\arraystretch}{1.5}
\begin{alignedat}{3} 
\begin{blockarray}{rcccccll}
 & x_{1} & x_{2} & x_{3} & x_{4} & x_{5} \\
\begin{block}{r @{\hspace{10pt}}[ccccc]ll}
 \newf_{1}&1&-1&&&\\ 
 f_{2}&&1&1&&\\ 
 \newf_{3}&&&1&-1&\\ 
 f_{4}&&&&1&1\\ 
 f_{5}&&&&&1\\ 
\end{block}
&\mcdetJac{5}{\detJac{\newJac}{1}}
\end{blockarray}
\end{alignedat}
\end{align*}
Since no $d_j=0$, SA reports index $\nu_S=\max_i c_i=1$. Note that here we {\em do not} choose the canonical offsets
\[
\~c = (0,0,1,0,1) \quad\text{and}\quad \~d=(0,1,1,1,1),
\]
which still give an overestimated structural index $\nu_S=2$ as $d_1=0$ and $c_3=c_5=1$.

\bigskip

Consider for general case $k\ge 1$. The DAE is
\begin{equation}\label{eq:arbitraryk}
\begin{alignedat}{5}
0 &= &f_{2i-1} &= x'_{2i} + x'_{2i+1} + &x_{2i-1}& - q_{2i-1}(t) &\qquad\qquad &i=\rnge{1}{k} \\
0 &= &f_{2i} &= x'_{2i} + x'_{2i+1} + &x_{2i}& - q_{2i}(t) &\qquad\qquad &i=\rnge{1}{k} \\
0 &= &f_{2k+1} &= &x_{2k+1}& - q_{2k+1}(t).
\end{alignedat}
\end{equation}

\begin{align*}
\hspace{20mm}\Sigma &=
\renewcommand{\arraystretch}{1.5}
\begin{alignedat}{3}
\begin{blockarray}{rccccccccll}
 &x_{1} &x_{2} &x_{3} &x_{4} &x_{5} & \cdots &x_{2k} &x_{2k+1}  & \s{c_i} \\
\begin{block}{r @{\hspace{10pt}}[cccccccc]ll}
 f_{1}&0^\bullet&1&1&&&&&&\s{0}\\ 
 f_{2}&&1^\bullet&1&&&&&&\s{0}\\ 
 f_{3}&&&0^\bullet&1&1&&&&\s{1}\\ 
 f_{4}&&&&1^\bullet&1&&&&\s{1}\\ 
 \vdots&&&&&0^\bullet&\ddots&&&\vdots\\ 
 f_{2k-1}&&&&&&\ddots&1&1&\s{k-1}\\ 
 f_{2k}&&&&&&&1^\bullet&1&\s{k-1} \\ 
 f_{2k+1}&&&&&&&&0^\bullet &\s{k}\\
\end{block}
\s{d_j} &\s{0}&\s{1}&\s{1}&\s{2}&\s{2}&\cdots&\s{k-1} &\s{k}&\valSig{\Sigma}{k}\\\\
\end{blockarray}
\end{alignedat}
\end{align*}
\begin{align*}
\hspace{20mm}\Jac &=
\renewcommand{\arraystretch}{1.5}
\begin{alignedat}{3} 
\begin{blockarray}{rccccccccll}
 &x_{1} &x_{2} &x_{3} &x_{4} &x_{5} & \cdots &x_{2k} &x_{2k+1}\\
\begin{block}{r @{\hspace{10pt}}[cccccccc]ll}
 f_{1}&1&1&1&&&&&\\ 
 f_{2}&&1&1&&&&&\\ 
 f_{3}&&&1&1&1&&&\\ 
 f_{4}&&&&1&1&&&\\ 
\vdots&&&&&\ddots&\ddots&&\\ 
 f_{2k-1}&&&&&&1&1&1\\ 
 f_{2k}&&&&&&&1&1\\ 
 f_{2k+1}&&&&&&&&1\\ 
\end{block}
&\mcdetJac{8}{\detJac{\Jac}{1}}
\end{blockarray}
\end{alignedat}
\end{align*}
SA succeeds and reports structural index $k+1$.

To remedy this index overestimation, we repeat the same strategy used above: for $i=\rnge{1}{k}$, we replace $f_{2i-1}$ by $\newf_{2i-1}=f_{2i-1}-f_{2i}$. The converted DAE is
\begin{equation}\label{eq:arbitraryconv}
\begin{alignedat}{5}
0 &= &\newf_{2i-1} &= x_{2i-1}-x_{2i} - q_{2i-1}(t)+q_{2i}(t) &\qquad\qquad &i=\rnge{1}{k} \\
0 &= &f_{2i} &= x'_{2i} + x'_{2i+1} + x_{2i} - q_{2i}(t) &\qquad\qquad &i=\rnge{1}{k} \\
0 &= &f_{2k+1} &= x_{2k+1} - q_{2k+1}(t).
\end{alignedat}
\end{equation}

\begin{align*}
\hspace{20mm}\newSig &=
\tgnstretch
\begin{alignedat}{3}
\begin{blockarray}{rccccccccll}
 &x_{1} &x_{2} &x_{3} &x_{4} &x_{5} & \cdots &x_{2k} &x_{2k+1}  & \s{c_i} \\
\begin{block}{r @{\hspace{10pt}}[cccccccc]ll}
 \newf_{1}&0^\bullet&0&&&&&&&\s{1}\\ 
 f_{2}&&1^\bullet&1&&&&&&\s{0}\\ 
 \newf_{3}&&&0^\bullet&0&&&&&\s{1}\\ 
 f_{4}&&&&1^\bullet&1&&&&\s{0}\\ 
 \vdots&&&&&0^\bullet&\ddots&&&\vdots\\ 
 \newf_{2k-1}&&&&&&\ddots&0&&\s{1}\\ 
 f_{2k}&&&&&&&1^\bullet&1&\s{0} \\ 
 f_{2k+1}&&&&&&&&0^\bullet &\s{1}\\
\end{block}
\s{d_j} &\s{1}&\s{1}&\s{1}&\s{1}&\s{1}&\cdots&\s{1} &\s{1}&\valSig{\newSig}{k}\\\\
\end{blockarray}
\end{alignedat}\\[-5ex]
\end{align*}
\begin{align*}
\hspace{20mm}\newJ &=
\tgnstretch
\begin{alignedat}{3} 
\begin{blockarray}{rccccccccll}
 &x_{1} &x_{2} &x_{3} &x_{4} &x_{5} & \cdots &x_{2k} &x_{2k+1}\\
\begin{block}{r @{\hspace{10pt}}[cccccccc]ll}
 \newf_{1}&1&-1&&&&&&\\ 
 f_{2}&&1&1&&&&&\\ 
 \newf_{3}&&&1&-1&&&&\\ 
 f_{4}&&&&1&1&&&\\ 
\vdots&&&&&\ddots&\ddots&&\\ 
 \newf_{2k-1}&&&&&&1&-1&\\ 
 f_{2k}&&&&&&&1&1\\ 
 f_{2k+1}&&&&&&&&1\\ 
\end{block}
&\mcdetJac{8}{\detJac{\newJ}{1}}
\end{blockarray}
\end{alignedat}
\end{align*}

Now SA reports the correct $\nu_S=1$ on the converted DAE \rf{arbitraryconv}.
Again, we use non-canonical offsets in $\newSig$, while $d_1=c_1=0$ in the canonical case. 
\end{appendices}

\ifbool{ACM}{
\bibliographystyle{acmsmall}}{
\bibliographystyle{siam}}
\bibliography{../../Articles/NedBib,../TGNBib}
\end{document}